\documentclass[10pt]{article}
\usepackage[utf8]{inputenc}
\usepackage[T1]{fontenc}
\usepackage{ebgaramond}
\usepackage[cmintegrals,cmbraces]{newtxmath}
\usepackage[light,medium]{josefin}
\usepackage{ebgaramond-maths}
\usepackage{mathrsfs}  
\usepackage[outline]{contour}
\usepackage[mathcal]{eucal}
\usepackage[letterpaper, top = 1.25 in, left = 1 in, bottom = 1 in, right = 1 in]{geometry}
\usepackage[font={sf}]{caption}
\usepackage{sidecap}

\makeatletter
  \DeclareSymbolFont{ntxletters}{OML}{ntxmi}{m}{it}
  \SetSymbolFont{ntxletters}{bold}{OML}{ntxmi}{b}{it}
  \re@DeclareMathSymbol{\leftharpoonup}{\mathrel}{ntxletters}{"28}
  \re@DeclareMathSymbol{\leftharpoondown}{\mathrel}{ntxletters}{"29}
  \re@DeclareMathSymbol{\rightharpoonup}{\mathrel}{ntxletters}{"2A}
  \re@DeclareMathSymbol{\rightharpoondown}{\mathrel}{ntxletters}{"2B}
  \re@DeclareMathSymbol{\triangleleft}{\mathbin}{ntxletters}{"2F}
  \re@DeclareMathSymbol{\triangleright}{\mathbin}{ntxletters}{"2E}
  \re@DeclareMathSymbol{\partial}{\mathord}{ntxletters}{"40}
  \re@DeclareMathSymbol{\flat}{\mathord}{ntxletters}{"5B}
  \re@DeclareMathSymbol{\natural}{\mathord}{ntxletters}{"5C}
  \re@DeclareMathSymbol{\star}{\mathbin}{ntxletters}{"3F}
  \re@DeclareMathSymbol{\smile}{\mathrel}{ntxletters}{"5E}
  \re@DeclareMathSymbol{\frown}{\mathrel}{ntxletters}{"5F}
  \re@DeclareMathSymbol{\sharp}{\mathord}{ntxletters}{"5D}
  \re@DeclareMathAccent{\vec}{\mathord}{ntxletters}{"7E}
  \re@DeclareMathSymbol{\epsilon}{\mathord}{ntxletters}{15}
  \re@DeclareMathSymbol{\mu}{\mathord}{ntxletters}{22}
\makeatother

\usepackage[backend=biber, sortcites=true, style=numeric-comp, sorting = none, maxnames = 999, mcite, subentry]{biblatex}
\usepackage[dvipsnames]{xcolor}
\usepackage[colorlinks=true,linkcolor=MidnightBlue,citecolor=blue!70!black,filecolor=green,urlcolor=PineGreen]{hyperref}
\usepackage{graphicx}
\usepackage{tikz}

\title{Optimal Zeno Dragging for Quantum Control: \\ A Shortcut to Zeno with Action--Based Scheduling Optimization}
\author{Philippe Lewalle\thanks{\href{mailto:plewalle@berkeley.edu}{plewalle@berkeley.edu}},~~Yipei Zhang,~~and~K.~Birgitta Whaley \\ 
{\small Berkeley Center for Quantum Information and Computation, and Department of Chemistry,} \\ {\small University of California, Berkeley $94720$}}
\date{\today}

\usepackage{fancyhdr}
\def\be{\begin{equation}}
\def\ee{\end{equation}}
\newcommand{\ensavg}[1]{\left\langle #1 \right\rangle}
\newcommand{\tr}[1]{\mathrm{tr} \left( #1 \right)}
\newcommand{\ket}[1]{| #1 \rangle}

\newcommand{\kett}[1]{| \,#1\, \rangle\kern-0.2em\rangle}
\newcommand{\braa}[1]{\langle\kern-0.2em\langle \,#1\, |}
\newcommand{\bra}[1]{\langle #1 |}

\newcommand{\ip}[2]{\langle #1 | #2 \rangle}

\renewcommand{\eqref}[1]{\textsc{eq.}~(\ref{#1})}
\newcommand{\secref}[1]{\S\,\ref{#1}}

\newcommand{\partl}[3]{\frac{\partial^{#3}#1}{\partial#2^{#3}}}

\newcommand{\contlim}[0]{\underset{\substack{dt\rightarrow 0 \\ N\rightarrow \infty}}{\mathrm{lim}} \;}
\newcommand{\bmath}[1]{{\contour{black}{\text{$#1$}}}}

\begin{document}

\definecolor{amber}{RGB}{255,191,0}
\definecolor{satinsheengold}{rgb}{0.8, 0.63, 0.21}
\definecolor{caribbeangreen}{rgb}{0.0, 0.8, 0.6}
\definecolor{chromeyellow}{rgb}{1.0, 0.65, 0.0}
\definecolor{deepfuchsia}{rgb}{0.76, 0.33, 0.76}
\definecolor{coquelicot}{rgb}{1.0, 0.22, 0.0}
\definecolor{crimsonglory}{rgb}{0.75, 0.0, 0.2}
\definecolor{deeppink}{rgb}{1.0, 0.08, 0.58}
\definecolor{electricviolet}{rgb}{0.56, 0.0, 1.0}
\definecolor{electricgreen}{rgb}{0.0, 1.0, 0.0}
\definecolor{mint}{rgb}{0.24, 0.71, 0.54}
\definecolor{dodgerblue}{rgb}{0.12, 0.56, 1.0}
\definecolor{lincolngreen}{rgb}{0.11, 0.35, 0.02}
\definecolor{persianblue}{rgb}{0.11, 0.22, 0.73}
\definecolor{patriarch}{rgb}{0.5, 0.0, 0.5}
\definecolor{charcoal}{rgb}{0.21, 0.27, 0.31}
\definecolor{darkblue}{rgb}{0.0, 0.0, 0.55}
\definecolor{darkred}{rgb}{0.55, 0.0, 0.0}
\definecolor{ceruleanblue}{rgb}{0.16, 0.32, 0.75}
\definecolor{darkspringgreen}{rgb}{0.09, 0.45, 0.27}
\definecolor{deeplilac}{rgb}{0.6, 0.33, 0.73}
\definecolor{goldenyellow}{rgb}{1.0, 0.87, 0.0}
\definecolor{lightgray}{rgb}{0.8 0.8 0.8}
\definecolor{oceanboatblue}{rgb}{0.0, 0.47, 0.75}
\definecolor{lightseagreen}{rgb}{0.13, 0.7, 0.67}
\definecolor{aqua}{rgb}{0.0, 1.0, 1.0}
\definecolor{mikadoyellow}{rgb}{1.0, 0.77, 0.05}
\definecolor{richcarmine}{rgb}{0.84, 0.0, 0.25}
\definecolor{xkcd_purple}{RGB}{126, 30, 156}
\definecolor{xkcd_lilac}{RGB}{206, 162, 253}
\definecolor{xkcd_sky}{RGB}{117, 187, 253}
\definecolor{xkcd_cobalt}{RGB}{3, 10, 167}
\definecolor{xkcd_pine}{RGB}{10, 72, 30}
\definecolor{xkcd_teal}{RGB}{2, 147, 134}
\definecolor{xkcd_aquamarine}{RGB}{4, 216, 178}
\definecolor{xkcd_chartreuse}{RGB}{193, 248, 10}
\definecolor{xkcd_goldenrod}{RGB}{250, 194, 5}
\definecolor{xkcd_pumpkin}{RGB}{225, 119, 1}
\definecolor{xkcd_brightred}{RGB}{255, 0, 13}
\definecolor{xkcd_deeppink}{RGB}{203, 1, 98}
\definecolor{xkcd_wine}{RGB}{128, 1, 63}
\definecolor{xkcd_charcoal}{RGB}{52, 56, 55}
\definecolor{xkcd_greypurple}{RGB}{130, 109, 140}

\newcommand{\revisn}[1]{{\color{richcarmine}#1}}

%%%%%%%%%%%%%%%%%%%%%%%%%%%%%%%%%%%%%%%%%%%%%%%%%%%%%%%%%%%%%%%%%%%
%%%%%%%%%%%%%%%%%%%%%%%%%%%%%%%%%%%%%%%%%%%%%%%%%%%%%%%%%%%%%%%%%%%
%%%%%%%%%%%%%%%%%%%%%%%%%%%%%%%%%%%%%%%%%%%%%%%%%%%%%%%%%%%%%%%%%%%
%%%%%%%%%%%%%%%%%%%%%%%%%%%%%%%%%%%%%%%%%%%%%%%%%%%%%%%%%%%%%%%%%%%
%%%%%%%%%%%%%%%%%%%%%%%%%%%%%%%%%%%%%%%%%%%%%%%%%%%%%%%%%%%%%%%%%%%

\maketitle
\vspace{-5pt}
\begin{abstract}
The quantum Zeno effect asserts that quantum measurements inhibit simultaneous unitary dynamics when the ``collapse'' events are sufficiently strong and frequent. 
This applies in the limit of strong continuous measurement or dissipation. 
It is possible to implement a dissipative control 
that is known as ``Zeno Dragging'', by dynamically varying the monitored observable, and hence also the eigenstates which are attractors under Zeno dynamics. 
This is similar to adiabatic processes, in that the Zeno dragging fidelity is highest when the rate of eigenstate change is slow compared to the measurement rate. 
We demonstrate here two theoretical methods for using such dynamics to achieve control of quantum systems.
The first, which we shall refer to as `shortcut to Zeno' (STZ), is
analogous to the shortcuts to adiabaticity (counterdiabatic driving) that are frequently used to accelerate unitary adiabatic evolution.
In the second approach we apply the Chantasri Dressel Jordan (2013, CDJ) stochastic action, and demonstrate that the extremal--probability readout paths derived from this are well suited to setting up a Pontryagin--style optimization of the Zeno dragging schedule. 
A fundamental contribution of the latter approach is to show that an action suitable for measurement–-driven control optimization can be derived quite generally from statistical arguments.
Implementing these methods on the Zeno dragging of a qubit, we find that both approaches yield the same solution, namely, that the optimal control is a unitary that matches the motion of the Zeno--monitored eigenstate. 
We then show that such a solution can be more robust than a unitary--only operation, and comment on solvable generalizations of our qubit example embedded in larger systems.
These methods open up new pathways toward systematically developing dynamic control of Zeno subspaces to realize dissipatively--stabilized quantum operations. 
\end{abstract}

\thispagestyle{empty}

%\vspace{10pt} 
%\begin{center}
%\tableofcontents 
%\end{center} 
%\newpage 

%%%%%%%%%%%%%%%%%%%%%%%%%%%%%%%%%%%%%%%%%%%%%%%%%%%%%%%%%%%%%%%%%
%%%%%%%%%%%%%%%%%%%%%%%%%%%%%%%%%%%%%%%%%%%%%%%%%%%%%%%%%%%%%%%%%
%%%%%%%%%%%%%%%%%%%%%%%%%%%%%%%%%%%%%%%%%%%%%%%%%%%%%%%%%%%%%%%%%
\section{Introduction}

Throughout the development of quantum computing to date, the measurement and control of quantum systems have been recognized as fundamental primitives underlying any possibility of realizing quantum information processing. 
In this work we draw on recent advances in the understanding of the interplay between measurement and unitary dynamics exemplified by the quantum Zeno effect to propose new approaches to measurement--based optimal quantum control. 

A large portion of the quantum control literature has focused on the design of unitary controls in both open and closed quantum systems. 
Optimal control theory has been applied in many quantum contexts, usually (although not exclusively) emphasizing the design of unitary operations or pulse shapes to realize optimal quantum dynamics.  
For example, direct implementation of Pontryagin's principle has found widespread use \cite{DAllessandroBook, Boscain2021, zhang2013optimal, yang2017optimizing, brady2021optimal, Lewalle_Pontryagin_2022}, as have related numerical algorithms detailed in e.g.,~Refs.~\cite{Koch_2022, Petruhanov_2023, Muller_2022, Morzhin_2019} and references therein. 
A complementary quantum control strategy that has gained widespread traction is adiabatic control, whereby slowly changing the quantum system's Hamiltonian, a closed system will follow a changing eigenstate with high fidelity. 
Such an evolution $\hat{H}(t)$ can be accelerated via counterdiabatic driving 
(or shortcuts to adiabaticity, \cite{Guery_STA-RMP, Nakahara_2022}), where $\hat{H}(t)$ is driven on a fast time-scale and an additional control Hamiltonian is introduced to suppress the diabatic transitions away from the desired subspace that may occur from varying $\hat{H}(t)$ too rapidly.

A different subset of the control literature has emphasized measurement-- and/or dissipation--based quantum control. 
For example, methods of direct measurement--based feedback have been developed \cite{Jacobs_Shabani_2008, Gough_2012, Zhang2017} and applied to numerous settings \cite{Tanaka_2012, martin2015deterministic, martin2017optimal, Minev2018, Leigh_Phase, zhang2020locally, martin2019single, chen2020quantum, jiang2020optimality, sundaresan2023demonstrating}, including state stabilization \cite{Mirrahimi_2007, amini2011stability, Ticozzi_2013, Benoist_2017, Cardona_2018, liang2019exponential, Cardona_2020, amini2023exponential, liang2023exploring, Ticozzi_2008}.
Advances in measurement--based feedback necessarily draw on the substantial progress made in realizing continuous monitoring of quantum systems over the past few decades, both in the theoretical domain \cite{BookCarmichael, BookWiseman, BookBarchielli, BookJacobs, BookJordan} and in experiments \cite{BookJordan, Gambetta2008, Murch2013, ShayLeigh2016, Ficheux2018, Blais_CQED}. 
Another more recent measurement--based approach has made use of the well--known quantum Zeno effect \cite{misra1977zeno} for control applications. 
The Zeno effect is well understood in the context of generalized measurements \cite{Presilla_1996, Benoist_2017, Burgarth2020quantumzenodynamics, Snizhko2020Zeno, Kumar2020Zeno, walls2022stochastic}, and control via the Zeno effect may be broadly understood as a form of dissipation engineering \cite{Harrington_2022}. 
Changing a measurement observable in time while continuously monitoring the outcomes has been shown to be a viable method for control \cite{ZenoDragging} that shares some important features with adiabatic evolution \cite{Aharonov_1980, Sarandy_2005, Campos_2016, Burgarth2022oneboundtorulethem}. 
Furthermore, one can also use the Zeno effect to stabilize a subspace in a larger quantum system so as to implement useful quantum operations \cite{Burgarth2020quantumzenodynamics, Facchi_2002, facchi2008quantum, Albert_2016, Ticozzi_2013, Benoist_2017}. 
Experimentally--accessible examples of this strategy are increasingly being investigated today \cite{blumenthal2021, Touzard_2018, Guillaud_2019, Lescanne_2020, Gautier_2022, Sellem_2022, gautier2023designing, Regent2023heisenberg, Reglade24, ZenoGateTheory, Mirrahimi_2014}. 
Other related strategies for measurement--based control have also been considered \cite{Roa_2006, Pechen_2006, Shuang_2008, Ashhab_2010, Sthitadhi_2020, Kumar_2022, Herasymenko_2023, Medina-Guerra_2024, Volya_2024}.

In the current paper we develop a theoretical approach to quantum control that builds on the above literature, while combining and unifying select tools in a new way. 
Our focus will be on ``Zeno dragging'' in the spirit of the 2018 experiment by \textcite{ZenoDragging}. 
Specifically, we focus on situations where a continuously--monitored (or dissipated) observable is slowly changed, in a manner that induces the quantum system to follow a measurement eigenstate along a desired trajectory with high probability. 
This Zeno dragging process may equivalently be described as a quasi-adiabatic change in the Liouvillian $\mathscr{L}$ that is engineered to bind the quantum state to an element of the kernel of $\mathscr{L}$ throughout the evolution with high probability \cite{Sarandy_2005, Campos_2016}. 
Our specific contributions here are i) an extension of counterdiabatic driving to the conditional diffusive evolution of quantum trajectories, ii) an extension of the optimal control protocol by \textcite{Kokaew_2022} to incoherent (dissipative) control knobs, and iii) the development of a class of simple examples of Zeno dragging in which the counterdiabatic driving solution and optimal control solution are equivalent. 

We develop these ideas here with the following exposition.
After a brief review of some necessary results from the theory of continuous quantum measurement in \secref{sec-qtrajs}, we then present our two distinct methodologies for quantum control by optimal Zeno dragging in \secref{sec-stz} and \secref{sec-CDJ-P}.  
In \secref{sec-stz} we first develop the ``Shortcut to Zeno dragging'' (STZ) approach. 
This is formally similar to the now well--known shortcuts to adiabaticity (STA)~\cite{Guery_STA-RMP}. 
Just as STA seeks to find a control unitary that suppresses diabatic transitions due to a rapidly changing base Hamiltonian, here we adapt the same machinery to the conditional quantum evolution under continuous monitoring and derive instead the STZ control unitary which suppresses failure of the Zeno pinning under finite--strength continuous measurement of a time--dependent observable. 
In \secref{sec-CDJ-P} we then develop an alternative approach that uses the stochastic action functional of \textcite{Chantasri2013} (CDJ) for Pontryagin--style optimization of the schedule on which a measurement observable is changed, drawing on the same framework as was recently applied to unitary controls by \textcite{Kokaew_2022}.  
We refer to this as the ``CDJ--Pontryagin (CDJ--P)'' approach.
In \secref{sec-example} we then apply both the STZ and CDJ--P methods to a simple example of Zeno dragging a qubit, an illustrative example for which the two methods give equivalent solutions. 
In \secref{sec-beyond-qubit} we elaborate on our single--qubit example, defining a similarly--solvable class of larger problems, including a two--qubit example.
We conclude in \secref{sec-conclude} with a discussion and prognosis for applications and further work. 
%%%%%%%%%%%%%%%%%%%%%%%%%%%%%%%%%%%%%%%%%%%%%%%%%%%%%%%%%%%%%%%%%
%%%%%%%%%%%%%%%%%%%%%%%%%%%%%%%%%%%%%%%%%%%%%%%%%%%%%%%%%%%%%%%%%
%%%%%%%%%%%%%%%%%%%%%%%%%%%%%%%%%%%%%%%%%%%%%%%%%%%%%%%%%%%%%%%%%
\section{Diffusive Quantum Trajectories \label{sec-qtrajs}}

We remind the reader here of some equations relevant to the conditional quantum dynamics accessed via continuous monitoring of a quantum system and introduce some notation that will be used throughout the next sections. 
For further reading about quantum trajectories at a general or introductory level, we recommend consulting the following texts \cite{BookWiseman, BookBarchielli, BookJacobs, BookJordan} or pedagogical/review articles \cite{Wiseman1996, Brun2001Teach, vanHandel_2005, JacobsSteck, Gross_2018, FlorTeach2019, LeighShay2020, Lewalle:21, Chantasri_2021}. 

Consider a set of Kraus \cite{Kraus_1983} operators $\{ \hat{\mathcal{M}}_{s,\ell} \}$ that describe the indirect monitoring of a quantum system via an auxiliary optical degree of freedom, over a short timestep $\Delta t$.  
A common example in experiments is a qubit monitored via an optical mode.\footnote{For example, longitudinal or dispersive coupling as described by \textcite{Blais_CQED}, offer concrete systems grounded in circuit--QED experiments that fit this paradigm. It is then natural to consider the monitoring to be realized by homodyne detection, heterodyne detection, photodetection, or equivalent amplification circuits to obtain a readout signal. \label{foot-Blais-Dyne}}  
Our Kraus operator set will depend on two output channels, the signal channel ($s$) and loss channel ($\ell$), where we suppose that after interacting with the system of interest, our auxiliary / pointer degree of freedom has a probability $\eta$ of going to our detector and becoming signal, and a probability $1-\eta$ of instead being lost in transit.\footnote{See e.g.~\cite{FlorTeach2019, Lewalle:21} for explicit examples detailing how the inefficiency may be modeled by a beamsplitter that diverts some optical signal away from a detector and into a lossy channel.\label{foot-BS-eta}}
Together, detection of these channels should form a complete set of measurement outcomes, in the sense that the Kraus operators should form elements of a positive operator valued measure (POVM)
\be \label{povm}
\sum_{s,\ell} \hat{\mathcal{M}}^\dag_{s,\ell}\, \hat{\mathcal{M}}_{s,\ell} = \hat{\mathbb{I}}.
\ee 
The Kraus operators may then be used to describe the partially--conditioned dynamics
\be \label{discrete-state-up}
\rho(t+\Delta t) = \frac{\sum_\ell \hat{\mathcal{M}}_{s,\ell}(r,\zeta)\,\rho(t)\,\hat{\mathcal{M}}_{s,\ell}^\dag(r,\zeta)}{\sum_\ell \tr{\hat{\mathcal{M}}_{s,\ell}(r,\zeta)\,\rho(t)\,\hat{\mathcal{M}}_{s,\ell}^\dag(r,\zeta)}}, 
\ee 
Here $r$ represents a continuous--valued measurement record/outcome
obtained from detection of the signal channel $s$, while the sum over $\ell$ denotes an average over all lost information. 
In the event of perfectly efficient measurements ($\eta = 1$), the loss channel is never needed, and the state update above then describes a process in which pure states remain pure. 
On the other hand, state mixing will generically occur if $\eta < 1$, due to averaging over losses.

A time--continuous form of \eqref{discrete-state-up} may be derived by e.g.,~performing the trace over the loss channel in the Fock basis, such that to $\mathcal{O}(\Delta t)$ only the terms
\begin{subequations} \label{Kraus-expand} \begin{align}
\label{Kraus-expand-0}\hat{\mathcal{M}}_{s,0} &\approx \CMcal{N}\,e^{-r^2\,\Delta t/4} \left\lbrace \hat{1} + \hat{\mathfrak{Z}}\,\Delta t + \mathcal{O}(\Delta t^2) \right\rbrace \quad\text{with}\quad \hat{\mathfrak{Z}} = \sqrt{\eta}\,r\,\hat{L} - \tfrac{1}{2}\left( \hat{L}^\dag\hat{L} + \eta\,\hat{L}^2\right), \quad \\ \label{Kraus-expand-1}
\hat{\mathcal{M}}_{s,1} &\approx \CMcal{N}\,e^{-r^2\,\Delta t/4} \left\lbrace
\sqrt{1-\eta}\,\sqrt{\Delta t}\,\hat{L} + \mathcal{O}(\Delta t^\frac{3}{2}) 
\right\rbrace,
\end{align}\end{subequations}
survive (see, e.g.,~Refs.~\cite{BookWiseman, Wiseman1996, Chantasri_2021, Rouchon_2015} for more rigorous justification, and further comments below). 
Such an expansion then results in the continuous--time dynamics
\be \label{continuous-state-up}
\dot{\rho} = \mathcal{F}(\rho,r,\zeta) = \hat{\mathfrak{Z}}(r,\zeta)\,\rho + \rho\,\hat{\mathfrak{Z}}^\dag(r,\zeta) + (1-\eta)\,\hat{L}(\zeta)\,\rho\,\hat{L}^\dag(\zeta) - \rho\,\tr{\hat{\mathfrak{Z}}(r,\zeta)\,\rho + \rho\,\hat{\mathfrak{Z}}^\dag(r,\zeta) + (1-\eta)\,\hat{L}(\zeta)\,\rho\,\hat{L}^\dag(\zeta)}.
\ee
In the following, the dependence of $\hat{L}$, $\hat{L}^\dag$ and related operators on the control parameter $\zeta$ will not be explicitly shown until it is required again in \secref{sec-CDJ-P}. 
We have set $\hbar=1$ here and throughout this paper. 

We shall associate the readout $r$ in the signal port with the outcome of a homodyne measurement (quadrature basis outcome) of the auxiliary optics in the timestep of interest $^{\text{\ref{foot-Blais-Dyne},\ref{foot-BS-eta}}}$ and have chosen units such that the readout may be expressed in the time-continuum limit as
\be \label{readout-dW}
r\,\Delta t \approx \sqrt{\eta}\,\tr{\hat{L}\,\rho+\rho\,\hat{L}^\dag}\,\Delta t + \Delta W = \sqrt{\eta}\,S\,\Delta t + \Delta W,
\ee
where $S$ is the ideal expected signal, which is equal to 
\be \label{signal}
S = \tr{\hat{L}\,\rho+\rho\,\hat{L}^\dag} = \ensavg{\hat{L} + \hat{L}^\dag},
\ee
and $\Delta W$ is the mean zero, Gaussian-distributed, intrinsic measurement uncertainty or ``noise'' that has variance $\Delta t$ \cite{BookWiseman, BookBarchielli, BookJacobs}. 
The device readout $r\,\Delta t$ is thus a Gaussian stochastic variable of mean $\sqrt{\eta}\, S$ and variance $\Delta t^{-1}$, which is a
sum of signal and noise, where the effect of inefficient (lossy) measurement is seen to be attenuation of the ideal average signal $S$ relative to the measurement noise $\Delta W$.
Concrete examples have been developed using a similar style and conventions in, e.g., ~Refs.~\cite{FlorTeach2019, Lewalle:21, Steinmetz_2022}. 
The ``ostensible'' probability distribution $\CMcal{N}\,e^{-r^2\,\Delta t/4}$ \cite{BookWiseman} may be eliminated from the dynamics in \eqref{discrete-state-up} or \eqref{continuous-state-up} because it appears on every term. 
We note that the definition of $\hat{\mathfrak{Z}}$ in terms of $\hat{L}$ used above is consistent with the Stratonovich\footnote{For further reading on stochastic processes see e.g.~\cite{GardinerStochastic, bookVanKampen, BookKloedenPlaten}.} Stochastic Master Equation (SME; see e.g.~\cite{Szigeti_2014}). Similar expansions suited to It\^{o} calculus can also be found in the literature \cite{Chantasri_2021, Rouchon_2015, amini2011stability, Guevara_2020, Wonglakhon_2021}. 

We now highlight two features that will be used in the subsequent analysis. 
The first is that we will emphasize the monitoring of a Hermitian observable ($\hat{L} = \hat{L}^\dag$), and the second is that any analysis we perform using conditional quantum dynamics will rely on pure--state trajectories corresponding to the case of perfect measurement efficiency ($\eta = 1$).\footnote{Initially pure states will remain pure under ideal continuous monitoring.} 
For $\hat{L} = \hat{L}^\dag$ the time--continuous conditional dynamics read
\begin{subequations}\label{all-dynamics} \be \label{gen-hermitian}
\dot{\rho} = (1-\eta)\,\hat{L}\,\rho\,\hat{L} + \sqrt{\eta}\,r\,\left(\hat{L}\,\rho+\rho\,\hat{L} \right) - \tfrac{\eta + 1}{2}\left( \hat{L}^2\,\rho + \rho\,\hat{L}^2 \right) - 2\,\rho\,\sqrt{\eta}\,r\,\tr{\rho\,\hat{L}} + 2\,\eta\,\rho\,\tr{\hat{L}\,\rho\,\hat{L}},
\ee
and with our further assumption that $\eta = 1$, this reduces to 
\be \label{ideal-hermitian}
\dot{\rho} = r \, \left( \hat{L}\,\rho+\rho\,\hat{L} \right) - \hat{L}^2\,\rho - \rho\,\hat{L}^2 - 2\,\rho \left( r \,\ensavg{\hat{L}} - \ensavg{\hat{L}^2} \right).
\ee 
Substitution of \eqref{readout-dW} into \eqref{gen-hermitian} or \eqref{ideal-hermitian} will result in Stratonovich equations of motion, rather than It\^{o} equations, due to our use of regular calculus and in accordance with the Wong--Zakai theorem \cite{WongZakai_1965a, WongZakai_1965b, Gough_2006}. 
We also point out that substituting $\eta = 0$ instead of $\eta = 1$ into \eqref{gen-hermitian} just reduces the dynamics to the usual Lindblad Master Equation (LME), i.e., to
\be \label{lme}
\dot{\rho} = \hat{L}\,\rho\,\hat{L}^\dag - \tfrac{1}{2}\,\hat{L}^\dag\hat{L}\,\rho - \tfrac{1}{2}\,\rho\,\hat{L}^\dag\hat{L}.
\ee \end{subequations}
This corresponds to the evolution averaged over many measurement realizations, or equivalently to dissipation without detection, and will be useful later for characterizing the average fidelity of our control processes. 
Note that unitary evolution $i[\rho,\hat{H}]$ may generically be added to any of \eqref{all-dynamics}.

The probability density to obtain the measurement outcome $r$ in a given timestep is governed by
\begin{subequations} \label{statistics-G} \be \label{pre-G}
\CMcal{P}(r|\rho) = \tr{\hat{\mathcal{M}}_{s,0}\,\rho(t)\,\hat{\mathcal{M}}_{s,0}^\dag + \hat{\mathcal{M}}_{s,1}\,\rho(t)\,\hat{\mathcal{M}}_{s,1}^\dag} \approx \CMcal{N}\,\tr{\rho + \Delta t\left( \hat{\mathfrak{Z}}\,\rho + \rho\,\hat{\mathfrak{Z}}^\dag + (1-\eta)\,\hat{L}\,\rho\,\hat{L}^\dag - \tfrac{1}{2}\,r^2\,\rho\right) + \mathcal{O}(\Delta t^2)}.
\ee
Expansion of the log--probability density to $\mathcal{O}(\Delta t)$ gives
\be \label{G-full}
\log\,\CMcal{P}(r|\rho) = \log \CMcal{N} + \mathcal{G}\,\Delta t + \mathcal{O}(\Delta t^2) \quad\text{with}\quad \mathcal{G} = -\tfrac{1}{2}\left( r - \sqrt{\eta}\,S \right)^2 - \mathsf{g},
\ee \be \label{g-hermitian}
\text{and} \quad \mathsf{g}(\rho,\hat{L}(\zeta)) \equiv 2\,\eta \left( \ensavg{\hat{L}^2} - \ensavg{\hat{L}}^2 \right) = \tfrac{\eta}{2}\,\text{var}(S) \quad\text{for}\quad \hat{L} = \hat{L}^\dag.
\ee \end{subequations}
This defines a function $\mathcal{G}$ that contains all of the features of the un-normalized probability density relevant to the time--continuum limit.
Recall that for $\hat{L} = \hat{L}^\dag$, the ideal expected signal $S$ in \eqref{readout-dW} is $S = 2\ensavg{L}$, so that the first term in $\mathcal{G}$ summarizes the basic statistics expected of $r$ mentioned above, namely that the readout is a Gaussian stochastic variable of mean $\sqrt{\eta}\, S$ and variance $\Delta t^{-1}$. 
The remaining term $\mathsf{g}$ is $r$--independent: for $\hat{L} = \hat{L}^\dag$ it takes the form \eqref{g-hermitian}, where $\text{var}(S)$ denotes the variance in the signal,\footnote{Note that the ``variance in the signal'' here is \emph{not at all} the same as the variance in the noisy measurement records.} and we have highlighted the dependence of $\hat{L}$ on $\zeta$.
The instantaneous measurement statistics $\mathcal{G}$ remain unchanged in the event that unitary evolution is present alongside the measurement dynamics; unitary evolution will add a term $i[\rho,\hat{H}]\,\Delta t$ inside the trace on the RHS of \eqref{pre-G}, which then immediately cancels out of the trace.

The function $\mathsf{g}(\rho,\hat{L})$ has elsewhere been interpreted as providing a measure of the rate of information acquisition due to continuously monitoring $\hat{L}$, or as a naturally--arising measure of distance from the eigenstates of $\hat{L}$, which vanishes at the Zeno points, i.e.,~at the eigenstates of $\hat{L}$ \cite{Philippe_Thesis, Karmakar_2022, Cylke_2022inprep, BookBarchielli}. 
In the current context, it is important to appreciate that this distance measure $\mathsf{g}(\rho,\hat{L})$ is now implicitly dependent on the control parameter $\zeta$.

This summarizes the main aspects of continuous quantum monitoring that we will need going forward in this work. We now turn our attention to quantum control based on the Zeno Dragging demonstrated in \cite{ZenoDragging}. 
Such a process goes as follows: 
the system is first initialized in an eigenstate of $\hat{L}$ at the initial time. 
Incoherent control is then realized by varying the system measurement operator $\hat{L}(t)$ as time evolves. 
If this is changed slowly compared to the accompanying dissipation rate $|| \hat{L}(t) ||^2$, then the system state will be ``Zeno dragged'' with high probability along the trajectory imposed on the measurement eigenstate. This may be regarded as a special case of the adiabatic Liouvillian evolution described in \cite{Campos_2016}. 
The dynamics summarized in \eqref{all-dynamics} may be used to model both individual realizations and ensemble averages of such a process. 
The ideally--realized Zeno dragging process would begin at the pure eigenstate of $\hat{L}(0)$, and then maintain purity throughout its evolution to terminate at the eigenstate of $\hat{L}(T)$ as $t$ passes from $0\rightarrow T$, not only in its individual realizations, but also on average (in the Lindbladian dynamics). 
Such an ideal process, that maintains perfect purity in the Lindblad dynamics by exactly following an instantaneous eigenstate of $\hat{L}(t)$ at all times, necessarily corresponds to the control becoming deterministic.
 
%%%%%%%%%%%%%%%%%%%%%%%%%%%%%%%%%%%%%%%%%%%%%%%%%%%%%%%%%%%%%%%%%
%%%%%%%%%%%%%%%%%%%%%%%%%%%%%%%%%%%%%%%%%%%%%%%%%%%%%%%%%%%%%%%%%
%%%%%%%%%%%%%%%%%%%%%%%%%%%%%%%%%%%%%%%%%%%%%%%%%%%%%%%%%%%%%%%%%
\section{Shortcut to Zeno Dragging \label{sec-stz}}

The first of the two methods we develop to control such Zeno dragging is a ``Shortcut To Zeno'' (STZ). 
This is directly analogous to the counterdiabatic driving, often referred to as a shortcut to adiabaticity, that has been extensively used in adiabatic dynamics of closed quantum systems \cite{Guery_STA-RMP, Nakahara_2022}. 
Such methods have recently been considered in the context of Lindbladian evolution of open quantum systems \cite{Vacanti_2014, Dann_2019, Alipour2020shortcutsto, Dupays_2020, Touil_2021, Menu_2022}. 
Here we show that such approaches may be further extended to control of the conditional evolution describing diffusive monitoring of a Hermitian observable. 

Let us add a control Hamiltonian $\hat{H}_\mathrm{STZ}$ to our ideal dynamics \eqref{ideal-hermitian}, such that
\begin{subequations} \label{eq-eqmo-for-stz} \begin{align}
\dot{\rho} &= i[\rho,\hat{H}_\mathrm{STZ}] + \left(r\,\hat{L} - \hat{L}^2 \right) \rho + \rho\left( r\,\hat{L} - \hat{L}^2 \right) + 2\,\rho \left( \tr{\rho\,\hat{L}^2} - r \,\tr{\rho\,\hat{L}} \right) \\
&= i[\rho,\hat{H}_\mathrm{STZ}] + \hat{\mathfrak{Z}}\,\rho + \rho\, \hat{\mathfrak{Z}} - 2\,\rho\,\tr{\rho\,\hat{\mathfrak{Z}}} \quad\text{for}\quad \hat{\mathfrak{Z}} = r\,\hat{L} - \hat{L}^2,
\end{align} \end{subequations}
where $\hat{H}_\mathrm{STZ}$, $\hat{\mathfrak{Z}}$, and $\hat{L}$ are all assumed to be Hermitian. 
Our aim now is to find the $\hat{H}_\mathrm{STZ}$ that minimizes the probability of escape from the rotating Zeno pinning implemented by the measurement operator $\hat{\mathfrak{Z}}(t)$ \eqref{Kraus-expand} 
that depends on a parameter $\zeta$ which has continuous and time--differentiable temporal dependence (i.e.,~$\zeta$ is not stochastic). 
To this end, we implement a change of frame on these dynamics, namely $\varrho \equiv \hat{Q}^\dag\,\rho\,\hat{Q}$, where the unitary transformation $\hat{Q}$ will be chosen so as to diagonalize $\hat{\mathfrak{Z}}$ (i.e.,~we define $\hat{Q}$ such that $\hat{D}_\mathfrak{Z} = \hat{Q}^\dag\,\hat{\mathfrak{Z}}\,\hat{Q}$ where $\hat{D}_\mathfrak{Z}$ is the diagonal matrix containing the real eigenvalues of $\hat{\mathfrak{Z}}$). 
Transforming \eqref{eq-eqmo-for-stz}, we find that in this new frame that we shall refer to as the ``Zeno frame'', the dynamics become
\be \label{Zeno-frame-dynamics}
\dot{\varrho} = \varrho \left\lbrace \hat{Q}^\dag\dot{Q} + i \,\hat{Q}^\dag\,\hat{H}_\mathrm{STZ}\,\hat{Q} + \hat{D}_\mathfrak{Z} \right\rbrace + \left\lbrace \dot{Q}^\dag\hat{Q} - i \,\hat{Q}^\dag\,\hat{H}_\mathrm{STZ}\,\hat{Q} + \hat{D}_\mathfrak{Z} \right\rbrace \varrho - 2\,\varrho\,\tr{\hat{D}_\mathfrak{Z}\,\varrho}.
\ee 
Now if the dynamics of $\dot{\varrho}$ in the Zeno frame are completely diagonal, then we have suppressed transitions out of the Zeno subspace (or target eigenstate) that we are interested in staying in. 
We then may think of this Zeno frame as directly analogous to an adiabatic frame.
The condition of interest for successful Zeno dragging is therefore that we choose the control Hamiltonian $\hat{H}_\mathrm{STZ}$ such that
\be \label{STZ_condition_1}
\dot{Q}^\dag\hat{Q} - i\,\hat{Q}^\dag\,\hat{H}_\mathrm{STZ}\,\hat{Q} + \hat{D}_\mathfrak{Z} = \hat{\lambda}, 
\ee
where $\hat{\lambda}$ can be any diagonal matrix. 

Notice that $\hat{\mathfrak{Z}}$ in \eqref{eq-eqmo-for-stz} is quadratic in $\hat{L}$ because of our choice of a Hermitian observable. (We note here that while we are working with a Stratonovich--like SME, this would still be true if we did a corresponding derivation compatible with the It\^{o} SME). 
It follows that the unitary frame--change operator $\hat{Q}$ diagonalizing $\hat{L}$ also diagonalizes $\hat{\mathfrak{Z}}$, i.e.,~for $\hat{D}_L = \hat{Q}^\dag\,\hat{L}\,\hat{Q}$ we necessarily have $\hat{D}_\mathfrak{Z} = r\,\hat{D}_L - \hat{D}_L^2$, which is also diagonal. 
This makes finding $\hat{Q}$ significantly easier in practice. 
Importantly, since the measurement observable $\hat{L}$ does not itself depend on the measurement outcome $r$, we can then also see that for measurement of a single Hermitian observable neither the frame change $\hat{Q}$ nor the STZ control Hamiltonian $\hat{H}_\mathrm{STZ}$ will depend on the readout $r$.
Thus, for a single Zeno dragging channel, we necessarily have an open--loop control rather than a closed--loop feedback control. 

With this in mind, we solve \eqref{STZ_condition_1} to find 
\be \label{STZ_condition_2}
\hat{H}_\mathrm{STZ} = i\,\hat{Q} \left(\hat{\lambda} - r\,\hat{D}_L + \hat{D}_L^2 \right) \hat{Q}^\dag - i\,\hat{Q}\,\dot{Q}^\dag = i\,\hat{Q}\,\hat{\lambda}\,\hat{Q}^\dag - i\,r\,\hat{L} + i\,\hat{L}^2 - i\,\hat{Q}\,\dot{Q}^\dag. 
\ee
Parameterizing the unitary $\hat{Q} = e^{-i\,\hat{\mathsf{h}}(t)}$ with some Hermitian $\hat{\mathsf{h}}(t) = \hat{\mathsf{h}}^\dag(t)$ for all time, yields $\dot{Q}^\dag = i\,\dot{\mathsf{h}}\,e^{i\,\hat{\mathsf{h}}}$, and $\hat{Q}\,\dot{Q}^\dag = i\,\dot{\mathsf{h}}$.
The terms $ - i\,r\,\hat{L} + i\,\hat{L}^2$ in \eqref{STZ_condition_2} are anti-Hermitian and are therefore unphysical for our present purposes. 
However, since $\hat{\lambda}$ is a completely arbitrary diagonal matrix, we may simply choose $\hat{\lambda} = \hat{D}_\mathfrak{Z}$ to cancel these unwanted terms, to arrive at the explicit solution
\be \label{STZ_solution}
\hat{H}_\mathrm{STZ} = -i\,\hat{Q}\,\dot{Q}^\dag = \dot{\mathsf{h}}. 
\ee
In short then, we may implement a STZ that supports a dissipative dragging operation by implementing a counterdiabatic unitary control that matches the rate at which we rotate our measurement eigenstates. 
This will be made clearer through examples in \secref{sec-example} and \secref{sec-beyond-qubit}.  
Additional technical details are given in the Appendices. 

%%%%%%%%%%%%%%%%%%%%%%%%%%%%%%%%%%%%%%%%%%%%%%%%%%%%%%%%%%%%%%%%%
%%%%%%%%%%%%%%%%%%%%%%%%%%%%%%%%%%%%%%%%%%%%%%%%%%%%%%%%%%%%%%%%%
%%%%%%%%%%%%%%%%%%%%%%%%%%%%%%%%%%%%%%%%%%%%%%%%%%%%%%%%%%%%%%%%%
\section{CDJ--Pontryagin Optimal Zeno Dragging \label{sec-CDJ-P}}

We now re-consider the above Zeno dragging problem in an alternative framework. 
Here we will optimize the CDJ \cite{Chantasri2013, Chantasri2015} stochastic action, deriving controls for the measurement dynamics that are conditioned on the extremal--probability measurement records in a manner similar to that recently proposed by \textcite{Kokaew_2022} for unitary control of open quantum system dynamics. 
One may regard this section as an application of their method to Zeno dragging.

%%%%%%%%%%%%%%%%%%%%%%%%%%%%%%%%%%%%%%%%%%%%%%%%%%%%%%%%%%%%%%%%%
%%%%%%%%%%%%%%%%%%%%%%%%%%%%%%%%%%%%%%%%%%%%%%%%%%%%%%%%%%%%%%%%%
\subsection{Review of the CDJ Stochastic Path Integral}

We first briefly summarize the main ideas needed to construct the \textcite{Chantasri2013} stochastic action. 
We begin with a Chapman--Kolmogorov equation that expresses the joint probability density to obtain a sequence of continuous--valued measurement outcomes together with the corresponding conditional state dynamics:
\be \label{cdj-CK} \begin{split}
\CMcal{P}(\lbrace \rho \rbrace, \lbrace r \rbrace) = \prod_{k = 0}^{N-1} \CMcal{P}(\rho_{k+1}|\rho_{k},r_k) \CMcal{P}(r_k|\rho_k) &= \prod_{k=0}^{N-1} \delta \left(\rho_{k+1} -  \frac{\sum_\ell\hat{\mathcal{M}}_{s,\ell}\,\rho_k\,\hat{\mathcal{M}}_{s,\ell}^\dag}{\sum_\ell\tr{\hat{\mathcal{M}}_{s,\ell}\,\rho_k\,\hat{\mathcal{M}}_{s,\ell}^\dag}} \right) \,\tr{ \sum_\ell \hat{\mathcal{M}}_{s,\ell}\,\rho_k\,\hat{\mathcal{M}}_{s,\ell}^\dag} \\
& \approx \prod_{k=0}^{N-1} \delta \left(\rho_{k+1} - \rho_k - \Delta t\,\mathcal{F}(\rho_k,r_k,\zeta_k) \right) \,\CMcal{N}_k\,\exp\left[\mathcal{G}(\rho_k,r_k,\zeta_k)\,\Delta t \right].
\end{split} \ee 
Using the notations of \eqref{discrete-state-up}, \eqref{Kraus-expand}, and \eqref{continuous-state-up}, and using $k$ to index the timestep, we then have a deterministic state update conditioned on the stochastic readout \eqref{readout-dW}, which has the statistics described by \eqref{statistics-G}. We have used \eqref{continuous-state-up} and \eqref{statistics-G} in moving from the first to the second line in \eqref{cdj-CK}, where the second line gives an approximation to $\mathcal{O}(\Delta t)$.
Boundary terms may be added to the above expression as needed \cite{Chantasri2013, Chantasri2015}, including e.g.,~the typical constraint to an initial state $\rho_i$ as per $\delta(\rho_i - \rho_0)$ or an exact post-selection to $\rho_f$ as per $\delta(\rho_f-\rho_N)$.\footnote{More general constraints to enforce an approximate terminal condition may also be used, if for instance, this is needed in connection with numerical optimization methods.}
Any unitary evolution $i[\rho,\hat{H}]$ will appear in $\mathcal{F}$ (from \eqref{continuous-state-up} or \eqref{ideal-hermitian}), but \emph{not} in $\mathcal{G}$ (from \eqref{statistics-G}).

It will be convenient to express the $d\times d$ density matrix $\rho$ in terms of at most $d^2-1$ real coordinates $\mathbf{q}$.
We write the equation of motion in the $\mathbf{q}$ coordinates as $\dot{\mathbf{q}} = \boldsymbol{\mathcal{F}}(\mathbf{q},r,\zeta) = \tr{\dot{\rho}\,\hat{\bmath{\sigma}}}$, where $\hat{\bmath{\sigma}}$ is the vector of generalized Gell--Mann matrices defining the $\mathbf{q}$ coordinate system \cite{Bertlmann2008}. 
Taking the continuous--time limit, and using the Fourier definition of the $\delta$--functions (which introduces the co-state variables $\boldsymbol{\Lambda}$, conjugate to $\mathbf{q}$), one arrives at the stochastic path integral \cite{Chantasri2013, Chantasri2015}
\begin{subequations}\label{path_integral}\be 
\CMcal{P}(\lbrace \rho \rbrace, \lbrace r \rbrace) \underset{\contlim}{\longmapsto} \CMcal{P}(\mathbf{q}(t),r(t)) = \int \mathcal{D}[\boldsymbol{\Lambda}] e^{-\int dt \left( \boldsymbol{\Lambda}\cdot\dot{\mathbf{q}} - \boldsymbol{\Lambda}\cdot\boldsymbol{\mathcal{F}} - \mathcal{G}\right)} = \int \mathcal{D}[\boldsymbol{\Lambda}] e^{-\mathcal{S}} 
\ee\be
\text{for}\quad \mathcal{S} = \int_0^T dt\left\lbrace \boldsymbol{\Lambda}\cdot\dot{\mathbf{q}} - \mathcal{H} \right\rbrace \quad \& \quad \mathcal{H} = \boldsymbol{\Lambda}^\top\cdot\boldsymbol{\mathcal{F}}(\mathbf{q},r,\zeta) + \mathcal{G}(\mathbf{q},r,\zeta).
\ee\end{subequations}
For a single measurement, the CDJ stochastic action $\mathcal{S}$ is characterized by the stochastic Hamiltonian 
\be \label{HCDJ-general}
\mathcal{H}(\mathbf{q},\boldsymbol{\Lambda},r,\zeta) = \boldsymbol{\Lambda}^\top\cdot\boldsymbol{\mathcal{F}}(\mathbf{q},r,\zeta)  -\tfrac{1}{2}\left[r-\sqrt{\eta}\,S(\mathbf{q},\zeta)\right]^2 - \mathsf{g}(\mathbf{q},\zeta),
\ee
using the definition of $\mathcal{G}$ from \eqref{statistics-G}. 

The action that we have derived above represents the trajectory probability density for the conditional quantum dynamics.
As such, extremization of the action leads to extremal--probability paths (trajectories following the extremal--probability measurement record) \cite{Chantasri2013, Weber2014}. 
Specifically, we may use $\delta \mathcal{S} = 0$ to obtain the following equations of motion for these optimal--readout paths:
\be 
\dot{\mathbf{q}} = \partl{\mathcal{H}}{\boldsymbol{\Lambda}}{}, \quad \dot{\boldsymbol{\Lambda}} = - \partl{\mathcal{H}}{\mathbf{q}}{},\quad \partl{\mathcal{H}}{r}{} = 0.
\ee
By then solving $\partial_r \mathcal{H} = 0$ for an optimal value of the readout $r^\star(\mathbf{q},\boldsymbol{\Lambda},
\zeta)$, we essentially find the extremal--probability measurement record (subject to boundary conditions). 
The quantum trajectories conditioned on $r^\star$ are generated by $\mathcal{H}^\star(\mathbf{q},\boldsymbol{\Lambda},\zeta) = \mathcal{H}(\mathbf{q},\boldsymbol{\Lambda},\zeta,r^\star(\mathbf{q},\boldsymbol{\Lambda},\zeta))$, using Hamilton's equations.

Extensive work has been done to investigate the dynamics of the optimal readout trajectories
\cite{Chantasri2013, Karmakar_2022, Chantasri2015, Weber2014, Jordan2015flor, Lewalle_Caustic-thy, Naghiloo_Caustic, Lewalle_chaos}.
Since the integrand in the stochastic path integral is Gaussian in $r$ (recall the form of \eqref{statistics-G}), one may analytically marginalize $r$ out of the stochastic path integral by integration; this turns out to leave exactly the same Hamiltonian $\mathcal{H}^\star$ as that obtained from the readout optimization above \cite{Chantasri2013, Philippe_Thesis, Karmakar_2022, Chantasri2015}. 
If we write the corresponding Stratonovich equation of motion for the conditional dynamics \eqref{continuous-state-up} or \eqref{all-dynamics} as 
\be 
\dot{\mathbf{q}} = \boldsymbol{\mathcal{F}}(\mathbf{q},r,\zeta) = \bmath{\mathsf{A}}(\mathbf{q},\zeta) + \bmath{\mathsf{b}}(\mathbf{q},\zeta)(dW/dt) = \bmath{\mathsf{A}}(\mathbf{q},\zeta) + \bmath{\mathsf{b}}(\mathbf{q},\zeta)(r-\sqrt{\eta}\,S(\mathbf{q},\zeta)),
\ee
where $\bmath{\mathsf{A}}$ is the Stratonovich drift vector (which includes any contribution from a unitary drive $\hat{H}$) and $\bmath{\mathsf{b}}$ is the diffusion tensor,
then the $r$--optimized CDJ stochastic Hamiltonian reads 
\begin{subequations}\label{HStar_CDJ_General}\begin{align} 
\mathcal{H}^\star &= \boldsymbol{\Lambda}^\top \cdot \boldsymbol{\mathcal{F}} [\text{$\mathbf{q}$},r^\star(\text{$\mathbf{q}$},\boldsymbol{\Lambda}),\zeta] - \tfrac{1}{2}\,\boldsymbol{\Lambda}^\top\cdot \bmath{\mathsf{B}}(\mathbf{q},\zeta)\cdot\boldsymbol{\Lambda} - \mathsf{g}( \mathbf{q},\zeta ) \\ 
&= \tfrac{1}{2}\,\boldsymbol{\Lambda}^\top\cdot\bmath{\mathsf{B}}(\mathbf{q},\zeta)\cdot\boldsymbol{\Lambda} + \boldsymbol{\Lambda}^\top\cdot\bmath{\mathsf{A}}(\mathbf{q},\zeta) - \mathsf{g}(\mathbf{q},\zeta),
\end{align}\end{subequations}
where $\bmath{\mathsf{B}}$ denotes $\bmath{\mathsf{b}}\,\bmath{\mathsf{b}}^\top$ \cite{Philippe_Thesis, Karmakar_2022}.
This form is quite general, and holds for arbitrary measurement efficiency $\eta > 0$, as well as in the event that there are multiple channels that are monitored simultaneously. In the latter case $\bmath{\mathsf{b}}$ is a tensor rather than simply a column vector. 
Note that in general the stochastic readout(s) reads $\mathbf{r} = \sqrt{\bmath{\eta}}\,\mathbf{S} + d\mathbf{W}/dt$, while the corresponding optimal readout(s) is $\mathbf{r}^\star = \sqrt{\bmath{\eta}}\,\mathbf{S} + \bmath{\mathsf{b}}^\top\cdot\boldsymbol{\Lambda}$. 
We can understand $\bmath{\mathsf{b}}^\top\cdot\boldsymbol{\Lambda}$ to represent the optimal noise, and $\boldsymbol{\Lambda}^\top\cdot\bmath{\mathsf{B}}(\mathbf{q},\zeta)\cdot\boldsymbol{\Lambda}$ to represent the quadratic form of the optimal noise.
In particular, it is clear that for each readout $(r^\star - \sqrt{\eta}\,S)^2 = (\bmath{\mathsf{b}}^\top\,\boldsymbol{\Lambda})^2 = \boldsymbol{\Lambda}^\top\,\bmath{\mathsf{B}}\,\boldsymbol{\Lambda}$ must be non-negative.

More recently, joint optimization over the readout $r$ and unitary control parameters $\boldsymbol{\Omega}$ has been performed to find optimal control solutions $\boldsymbol{\Omega}^\star(t)$ based on the optimal readouts \cite{Kokaew_2022}. 
That work by Kokaew et.~al demonstrated the utility of the CDJ stochastic path integral, and its $r$--optimal paths in particular, for finding optimal control solutions. 
Below, we will employ the same conceptual framework, but apply it now to the optimal control of Zeno dragging. 
We shall show that the CDJ stochastic action is exceptionally well--suited to deriving optimal Zeno--based controls. 

%%%%%%%%%%%%%%%%%%%%%%%%%%%%%%%%%%%%%%%%%%%%%%%%%%%%%%%%%%%%%%%%%
%%%%%%%%%%%%%%%%%%%%%%%%%%%%%%%%%%%%%%%%%%%%%%%%%%%%%%%%%%%%%%%%%
\subsection{General Zeno Dragging Optimization: CDJ--Pontryagin optimization \label{sec-CDJP}}

We consider the following Zeno dragging scenario. 
Suppose that we have a single measurement characterized by $\hat{L}(\zeta) = \hat{L}^\dag(\zeta)$, where the control function $\zeta(t)$ specifies a fixed (open-loop) schedule over which $\hat{L}$ is varied, i.e., $\zeta$ specifies how the measurement eigenspace that is used for the Zeno dragging is varied in time. 
$\zeta(t)$ is assumed continuous and time--differentiable. 
A unitary drive $\hat{H}$ can optionally be included in the dynamics.
For the purposes of CDJ--Pontryagin (CDJ--P) optimization, we may apply the general form of the readout-optimized CDJ Hamiltonian \eqref{HStar_CDJ_General} to the Zeno dragging scenario. 
This Hamiltonian is already in a form suitable for Pontryagin optimization, since we have Lagrange multipliers $\boldsymbol{\Lambda}$ that constrain us to dynamics $\boldsymbol{\mathcal{F}}^\star$ (here the conditional dynamics following the extremal--probability readout $r^\star$), with the remaining term $\tfrac{1}{2}\,\boldsymbol{\Lambda}\cdot \bmath{\mathsf{B}}(\mathbf{q},\zeta)\cdot\boldsymbol{\Lambda} + \mathsf{g}(\mathbf{q},\zeta)$ acting essentially as a cost function for the optimization.

It turns out that $\mathsf{g}(\mathbf{q},\zeta) = \tfrac{1}{2}\,\text{var}(\hat{S})$ \eqref{g-hermitian} is an ideal cost function for Zeno dragging optimization, for the following reasons: First, it is non-negative everywhere by virtue of being a variance, and second, it only has roots at the eigenstates of $\hat{L}(\zeta)$, i.e.~at the Zeno points. 
$\bmath{\mathsf{B}}$ also has roots at these Zeno points, and is similarly non-negative as noted above.
Minimization of the two terms in $\tfrac{1}{2}\,\boldsymbol{\Lambda}\cdot \bmath{\mathsf{B}}(\mathbf{q},\zeta)\cdot\boldsymbol{\Lambda} + \mathsf{g}(\mathbf{q},\zeta)$ goes hand in hand: minimizing $\mathsf{g}$ means keeping the system as close as possible to a measurement eigenstate throughout the evolution, which implicitly also minimizes the diffusive noise $\tfrac{1}{2}\,\boldsymbol{\Lambda}\cdot \bmath{\mathsf{B}}(\mathbf{q},\zeta)\cdot\boldsymbol{\Lambda}$, and is exactly what we aim to accomplish by successful Zeno dragging. 

Implementing the Pontryagin optimization $\delta \mathcal{S} = 0$ with $\zeta$ as the control to be optimized then leads to 
\be \label{zeta-star-condition} 
\dot{\mathbf{q}} = \partl{\mathcal{H}^\star}{\boldsymbol{\Lambda}}{}, \quad
\dot{\boldsymbol{\Lambda}} = -\partl{\mathcal{H}^\star}{\mathbf{q}}{}, \quad
\partl{\mathcal{H}^\star}{\zeta}{} = 0 \quad\rightarrow\quad \partial_\zeta\mathsf{g}
= \tfrac{1}{2} \boldsymbol{\Lambda}^\top \cdot \left(\partial_\zeta \bmath{\mathsf{B}} \right) \cdot \boldsymbol{\Lambda} + \boldsymbol{\Lambda}^\top\cdot \left(\partial_\zeta \bmath{\mathsf{A}}\right),
\ee
where the second second expression is the optimality condition that $\zeta^\star$ must satisfy at all times, given by the Pontryagin Maximum Principle $\partial_\zeta \mathcal{H}^\star = 0$.
One may think of \eqref{zeta-star-condition} as an application of Pontryagin's principle, where the measurement schedule $\zeta(t)$ is the control function to be optimized, and the action 
\begin{subequations}\begin{align}
\mathcal{S}^\star &= \int_0^T dt \left\lbrace \dot{\mathbf{q}}\cdot\boldsymbol{\Lambda} - \tfrac{1}{2}\boldsymbol{\Lambda}^\top \,\bmath{\mathsf{B}}\, \boldsymbol{\Lambda} - \boldsymbol{\Lambda}^\top \bmath{\mathsf{A}} + \mathsf{g} \right\rbrace \label{SHstar} \\ \label{SLstar}
&= \int_0^T dt\left\lbrace \tfrac{1}{2}(\dot{\mathbf{q}}-\bmath{\mathsf{A}})^\top \bmath{\mathsf{B}}^{-1} (\dot{\mathbf{q}} - \bmath{\mathsf{A}}) + \mathsf{g} \right\rbrace 
\end{align}\end{subequations}
has been constructed from the CDJ stochastic path integral with dynamics constrained to its $\mathbf{r}$--optimal paths. 
(In particular, one may apply $\int \mathcal{D}[\mathbf{r}]$ to \eqref{path_integral} to obtain the first line, and then additionally perform the $\int\mathcal{D}[\boldsymbol{\Lambda}]$ integration to obtain the Lagrangian form in the second line \cite{Philippe_Thesis, Cylke_2022inprep}.
Both are straightforward Gaussian integrations when the path integral is written in discrete time.)

We make a brief digression here to remark on the contrast between the STZ approach of \secref{sec-stz} and the CDJ--P approach of \secref{sec-CDJP}. 
First, note that of these two, only the CDJ--P method explicitly seeks \emph{optimal} control; counterdiabatic controls do not generally come with any kind of cost function or optimality guarantees attached.
Second, these two methods approach the Zeno dragging protocol from complementary viewpoints. 
In deriving the STZ, we first provide a measurement and then search for a unitary control satisfying some specific properties (namely, suppression of diabatic transitions in the Zeno frame). 
In the CDJ--P approach, we first specify any unitary that may be part of the system dynamics, and then perform an optimization of the measurement axis within the context of those dynamics.
These qualitative differences suggest that quite different solutions might be found when using the two approaches. 
However, in \secref{sec-example} we shall show that for at least one example the two approaches yield equivalent solutions, and further connections between the methods will be elucidated in \secref{sec-beyond-qubit}.

Finally, we highlight what we have done in this section: \textsc{eqs}. (\ref{cdj-CK} -- \ref{zeta-star-condition}) show that a cost function to optimize measurement--driven quantum control can be derived from the measurement statistics. Thus, in deriving the CDJ--P optimal controls via $\delta \mathcal{S} = 0$, one is also optimizing the likelihood that the controlled evolution occurs.

%%%%%%%%%%%%%%%%%%%%%%%%%%%%%%%%%%%%%%%%%%%%%%%%%%%%%%%%%%%%%%%%%
%%%%%%%%%%%%%%%%%%%%%%%%%%%%%%%%%%%%%%%%%%%%%%%%%%%%%%%%%%%%%%%%%
%%%%%%%%%%%%%%%%%%%%%%%%%%%%%%%%%%%%%%%%%%%%%%%%%%%%%%%%%%%%%%%%%
\section{Example: Zeno Dragging a Qubit \label{sec-example}}

To demonstrate the methods presented above, we will now consider the example of Zeno dragging a single qubit, similar to what was experimentally demonstrated by \textcite{ZenoDragging}.
For continuity with the presentation of the previous sections, in \secref{sec-example_CDJP} we first continue our CDJ--P analysis in the context of this example, and then in \secref{sec-example_STZ} we demonstrate the application of our STZ method in the same context. 
In \secref{sec-example_noSTZ} we consider the special case of Zeno dragging without unitary assistance.
The average fidelity and other indicators of the performance of the optimal protocol derived in the preceding sections are detailed in \secref{sec-avg-Zeno-performance}. 
We revisit our example in the Zeno frame in \secref{sec-ExZenoFrame}, which yields some additional insights about the optimal schedule, and then prepares us for a discussion of the robustness of STZ against controller errors, appearing in \secref{sec-ExRobustness}.

%%%%%%%%%%%%%%%%%%%%%%%%%%%%%%%%%%%%%%%%%%%%%%%%%%%%%%%%%%%%%%%%%
%%%%%%%%%%%%%%%%%%%%%%%%%%%%%%%%%%%%%%%%%%%%%%%%%%%%%%%%%%%%%%%%%
\subsection{CDJ--P Zeno Dragging for a Qubit}
\label{sec-example_CDJP}

Let us suppose that we can monitor any qubit observable in the $xz$--Bloch plane. 
The Kraus operator
\begin{equation} \label{Kraus-Op-Zeta}
\hat{\mathcal{M}}_r^{(\zeta)} =  \left( \frac{\Delta t}{2 \pi} \right)^\frac{1}{4} \exp
\left[- \Gamma\,\Delta t - \frac{r^2\Delta t}{4} \right] \bigg\lbrace \cosh\left[ r\:\Delta t\:\sqrt{\Gamma} \right] \hat{\mathbb{I}}  +
\sinh\left[ r\:\Delta t\:\sqrt{\Gamma} \right] \left( \hat{\sigma}_z\cos\zeta + \hat{\sigma}_x \sin\zeta \right)\bigg\rbrace
\end{equation}
is appropriate for this \cite{Lewalle_chaos}. 
The measurement strength is given by the characteristic collapse time $\tau$, or alternatively as the corresponding rate $\Gamma = 1/4\tau$. 
The Kraus operator \eqref{Kraus-Op-Zeta} may be expanded following the methods of \secref{sec-qtrajs}, leading to the equivalent Lindblad operator $\hat{L} = \sqrt{\Gamma}\left( \hat{\sigma}_x\,\sin\zeta + \hat{\sigma}_z\,\cos\zeta \right)$.
See \secref{sec-get-L} for a more general recipe for constructing $\hat{L}$ in similar settings.

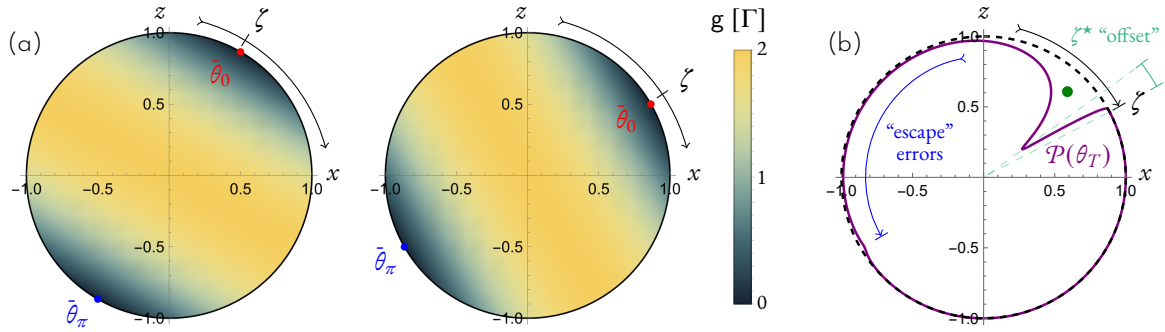
\begin{figure}[t]
\centering 
\includegraphics[width = .25\textwidth]{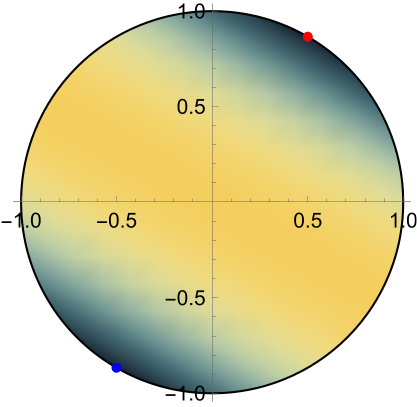} \hspace{0.5cm}
\includegraphics[width = .25\textwidth]{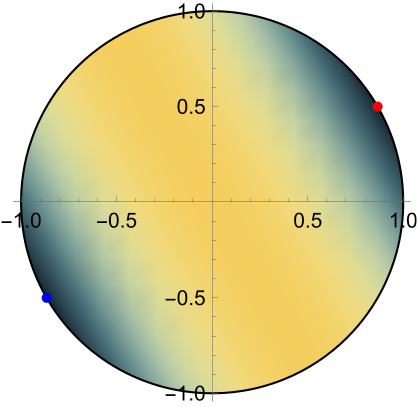} \hspace{0.5cm}
\includegraphics[height = .17\textheight]{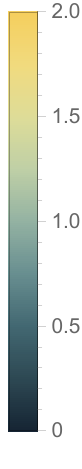} \hspace{0.5cm}
\includegraphics[width = .25\textwidth]{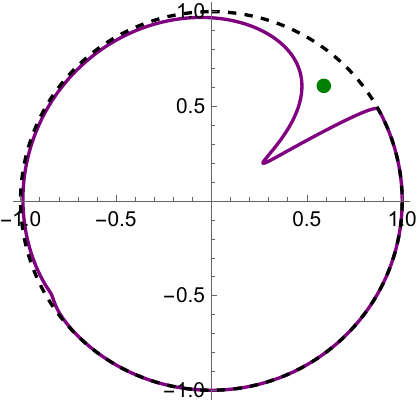} \\
\begin{tikzpicture}[overlay]
%%%%%%%%%%%%%%%%%%%%
\node[] at (-7.3,4.2) {\sf (a)};
\node[] at (3.6,4.2) {\sf (b)};
%%%%%%%%%%%%%%%%%%%%
\node[] at (-3.2,2.45) {$x$};
\node[] at (1.6,2.45) {$x$};
\node[] at (-5.5,4.6) {$z$};
\node[] at (-0.7,4.6) {$z$};
\node[] at (2.2,4.5) {$\mathsf{g}~[\Gamma]$};
\draw[draw = white, fill = white] (2.4,0.7) rectangle (2.8,4.2);
\node[] at (2.5,0.75) {\footnotesize $0$};
\node[] at (2.5,2.42) {\footnotesize $1$};
\node[] at (2.5,4.13) {\footnotesize $2$};
%%%%%%%%%%%%%%%%%%%%
\node[] at (-6.6,0.6) {\color{blue} $\bar{\theta}_\pi$};
\node[] at (-2.5,1.3) {\color{blue} $\bar{\theta}_\pi$};
\node[] at (-4.7,3.8) {\color{red} $\bar{\theta}_0$};
\node[] at (0.7,3.2) {\color{red} $\bar{\theta}_0$};
\draw[<-<] (-3.3,2.8) arc(15:75:2.4cm);
\draw[<-<] (1.5,2.8) arc(15:75:2.4cm);
\draw[] (-4.42,4.16) -- (-4.32,4.32);
\node[] at (-4.15,4.5) {$\zeta$};
\draw[] (1.1,3.47) -- (1.25,3.57); 
\node[] at (1.5,3.7) {$\zeta$};
%%%%%%%%%%%%%%%%%%%%
\node[] at (7.6,2.45) {$x$};
\node[] at (5.45,4.6) {$z$};
\draw[|<-<] (7.25,3.35) arc(30:75:2.25cm);
\node[] at (7.5,3.4) {$\zeta$};
\node[] at (6.7,2.7) {\color{patriarch} $\CMcal{P}(\theta_T)$};
\draw[|-|,mint] (7.75,3.62) arc(30:37.2:3cm);
\draw[dashed, opacity = 0.5, mint] (5.5,2.45) -- (7.25,3.35);
\draw[dashed, opacity = 0.5, mint] (5.5,2.45) -- (7.45,3.85);
\node[] at (7.2,4.3) {\color{mint}\footnotesize $\zeta^\star$ ``offset''};
\draw[>->|,blue] (5.2,4) arc(100:210:1.6cm);
\node[] at (4.6,3) {\color{blue}\footnotesize ``escape''};
\node[] at (4.6,2.7) {\color{blue}\footnotesize errors};
\end{tikzpicture} \\ \vspace{-12pt}
\caption{
(a) Density plots of $\mathsf{g}(\mathbf{q})$ in the $xz$ Bloch plane for two different values of $\zeta$, in our single qubit example. 
Color denotes $\mathsf{g}$ in units of $\Gamma$. 
Recall that the function $\mathsf{g}$ (\eqref{g-hermitian} in general, and \eqref{g-qubit} in the current example) plays the role of a cost function for our CDJ--P Zeno dragging optimization. 
Panel (b) offers a qualitative illustration of how diffusion of the conditional state in different measurement realizations leads on average to loss of purity: The purple curve represents the probability density of pure states after some diabatic Zeno dragging (distance inside the Bloch sphere represents higher probability density at some particular time $T$; this was obtained as a solution to the Fokker--Planck equation \cite{GardinerStochastic}). The green point represents the corresponding Lindblad estimate at this time, which is the average of pure states weighted by the purple conditional distribution. Loss of purity occurs when the distribution is no longer well--localized, i.e.~when the conditional dynamics have been allowed to diffuse more widely.
A perfect control scheme, either using STZ or in the adiabatic limit, prevents any diffusion and follows the root of $\mathsf{g}(\mathbf{q})$ exactly for all time, and thereby both retains purity, even on average, and becomes deterministic evolution. 
We also observe in panel (b) that the peak of the distribution $\CMcal{P}(\theta_T)$ lags behind the diabatically--moving measurement axis. 
Optimal controls will explicitly compensate for this kind of ``offset'' (see \eqref{zeta-star} or \eqref{zeta-star-Om0}). 
The tail of the distribution $\CMcal{P}(\theta_T)$ extending to $\bar{\theta}_\pi$ illustrates how ``escape'' errors can occur: Trajectories left behind by diabatic motion of the measurement axis can have overlap with unwanted eigenstates, and may then eventually ``collapse'' towards such unwanted states instead of the intended one.
}\label{fig-g-rotation}
\end{figure}

The time--continuous conditional dynamics based on this single-qubit Kraus operator (with $\eta = 1$ and $y = 0$) and in the presence of 
a unitary drive $\hat{H} = \tfrac{1}{2}\,\Omega\,\hat{\sigma}_y$ may be expressed as
\begin{subequations} \be \label{qubit_F-OM}
\boldsymbol{\mathcal{F}} = \left( \begin{array}{c} \dot{x} \\ \dot{z} \end{array} \right) = \left( \begin{array}{c}
z\,\Omega-2\,r\,\sqrt{\Gamma}\,[(x^2-1)\sin\zeta + x\,z\,\cos\zeta] \\
-x\,\Omega-2\,r\,\sqrt{\Gamma}\,[(z^2-1)\cos\zeta + x\,z\,\sin\zeta]
\end{array} \right).
\ee 
The readout statistics may be characterized by $\mathcal{G} =  - \tfrac{1}{2}\left(r-S \right)^2 - \mathsf{g}$, with
\be \label{g-qubit}
S = 2\sqrt{\Gamma}\left(x\,\sin\zeta + z\,\cos\zeta \right), \quad \mathsf{g} = 2\Gamma\left\lbrace 1 - (x\,\sin\zeta+z\,\cos\zeta)^2 \right\rbrace. 
\ee \end{subequations}
Based on these expressions, we can immediately construct the CDJ stochastic Hamiltonian $\mathcal{H}$ in Cartesian Bloch coordinates $\mathbf{q} = (x,z)^\top$ 
as
\be \label{HCDJ-xz-general}
\mathcal{H}_\zeta = \Lambda_x\,\mathcal{F}_x + \Lambda_z\,\mathcal{F}_z + \mathcal{G}. 
\ee
Recall from the preceding section that $\mathsf{g}(\mathbf{q})$ will play the role of a cost function in our optimization procedure. 
Fig.~\ref{fig-g-rotation} shows a visual representation of the specific $\mathsf{g}(\mathbf{q})$  in \eqref{g-qubit}, that  underlies the calculations described below.

Two further manipulations are helpful as we move towards an analytic solution to this problem:
\begin{enumerate}
    \item We convert $\mathcal{H}_\zeta$ to polar coordinates in the $xz$--Bloch plane. This can be accomplished via a canonical transformation \cite{Lewalle_Pontryagin_2022, FlorTeach2019}:
    \be \begin{split}
    x \rightarrow R\,\sin\theta, ~~&~~ z \rightarrow R\,\cos\theta, \\ \Lambda_x \rightarrow \Lambda_R\,\sin\theta + \Lambda_\theta \,\cos\theta/R, ~~&~~ \Lambda_z \rightarrow \Lambda_R\, \cos\theta - \Lambda_\theta\, \sin\theta/R.
    \end{split} \ee
    \item We observe that $\dot{R}= 0$ for $R = 1$, i.e.,~pure states remain pure since our dynamics are conditioned on complete measurement information ($\eta = 1$). We can consequently set $R = 1$, which also decouples $\Lambda_R$ from the dynamics. 
\end{enumerate}
By implementing this transformation, we will re-write our Hamiltonian, originally in terms of two canonically--conjugate coordinate/co-state pairs $\mathbf{q} = (x,z)$ and $\boldsymbol{\Lambda} = (\Lambda_x,\Lambda_z)$, in terms of a single pair $q\rightarrow \theta$ and $\Lambda \rightarrow \Lambda_\theta$.

Implementing this coordinate change, eliminating the radial coordinate, and optimizing or marginalizing away 
the readout $r$ (recall the process leading from \eqref{HCDJ-general} to \eqref{HStar_CDJ_General}), transforms the CDJ stochastic Hamiltonian to
\be \label{H_star_zeta}
\mathcal{H}_\zeta^\star = 2 \Gamma  (\Lambda_\theta ^2-1) \sin ^2(\zeta -\theta )+\Lambda_\theta  \left[2 \Gamma  \sin (2 \zeta -2\theta)+\Omega \right].
\ee
In the notation of \textsc{eqs.} (\ref{HStar_CDJ_General},\ref{zeta-star-condition}), we here have $\mathsf{g}(\theta,\zeta) = 2\Gamma\,\sin^2(\zeta-\theta) = \tfrac{1}{2}\mathsf{B}(\theta,\zeta)$ and $\mathsf{A} = 2 \Gamma  \sin (2 \zeta -2\theta)+\Omega$,
where $\mathsf{A}$ and $\mathsf{B} = \mathsf{b}^2$ are all scalar functions instead of vectors or tensors, because we have reduced the problem to a single coordinate and a single noise source.
The fact that the cost function $\mathsf{g}(\theta,\zeta)$ and diffusion coefficient $\mathsf{b}(\theta,\zeta)$ vanish at the Zeno points $\bar{\theta}_0 = \zeta$ and $\bar{\theta}_\pi = \zeta + \pi$ is central to the functioning of the Zeno dragging control we analyze here. 
We denote $\bar{\theta}_0$ as the target eigenstate, while escape to the ``wrong'' $\bar{\theta}_\pi$ eigenstate implies failure of the control.
These two options are the only roots (minima) of $\mathsf{g}(\theta,\zeta)$ in this one qubit scenario, and are the instantaneous attractors of the conditional dynamics. 
See Fig.~\ref{fig-g-rotation} for an illustration.
Note that because \eqref{H_star_zeta} is a function of $\zeta - \theta$ only, we have $\partial_\zeta \mathcal{H}^\star_\zeta = - \partial_\theta \mathcal{H}^\star_\zeta$. 
By \eqref{zeta-star-condition}, we may then immediately understand that when we solve the optimization condition $\partial_\zeta\mathcal{H}^\star_\zeta = 0$, we must also subsequently have $\Lambda_\theta$ a conserved quantity in the optimal dynamics.

We are then in a position to find the optimal measurement axis control $\zeta^\star$ by solving $\partial_\zeta \mathcal{H}_\zeta^\star = 0$. 
This reveals that
\be \label{zeta-star}
\zeta^\star = \theta + \frac{1}{2}\,\arctan\left( \frac{2\,\Lambda_\theta}{1-\Lambda_\theta^2} \right) = \theta + \arctan(\Lambda_\theta),
\ee
where the first and second forms of $\zeta^\star$ are equivalent up to piecewise additions of integer multiples of $\pi/2$.
Substitution of this solution into the stochastic Hamiltonian gives us the generator of the optimized dynamics as
\be \label{H-2opt-OM}
\mathcal{H}_{\zeta^\star}^{r^\star} = 2\Gamma\,\Lambda_\theta^2 + \Lambda_\theta\,\Omega.
\ee
Since \eqref{H-2opt-OM} is independent of $\theta$, we may immediately understand that $\Lambda_\theta$ is conserved in the optimized dynamics (as expected), with state dynamics that are linear in time.
In particular, applying $\partial_{\Lambda_\theta} \mathcal{H}^{r^\star}_{\zeta^\star}|_{\Lambda_\theta^\star} = \dot{\theta}^\star = 4\Gamma\,\Lambda_\theta^\star + \Omega$ (recall \eqref{zeta-star-condition}), with $\Lambda_\theta^\star$ a constant of motion due to $\partial_\theta\mathcal{H}^{r^\star}_{\zeta^\star} = 0$, leads to 
\begin{subequations}\label{linear-omega-dynamics} \be \label{linear_theta}
\theta(t) = \theta_i + (\Omega + 4\Gamma\,\Lambda_\theta^\star)t = \theta_i + (\theta_f - \theta_i)\tfrac{t}{T}.
\ee 
This implicitly includes the optimal co-state 
\be \label{linear_Lambda}
\Lambda^\star_\theta = \frac{\theta_f - \theta_i}{4\Gamma\,T}- \frac{\Omega}{4\Gamma},
\ee \end{subequations}
which maps the boundary value problem in the state $\theta_i$ and $\theta_f$ to an initial value problem in the co-state $\Lambda_\theta$. 
Collecting these expressions into \eqref{zeta-star}, we then have the schedule
\be \label{linear-zeta-star-general}
\zeta^\star = \theta_i + (\theta_f - \theta_i)\tfrac{t}{T} + \arctan\left( \frac{\theta_f - \theta_i}{4\Gamma\,T} - \frac{\Omega}{4\Gamma} \right). 
\ee

We now remark that global optima of the action $\mathcal{S}$ are given either by $\mathcal{H} = 0$ or the path with $\Lambda_\theta = 0$ at the final time \cite{Lewalle_Caustic-thy, BookArnoldClassical, Cline_CMech}. \footnote{Specifically, the optimized dynamics with $\mathcal{H}^\star = 0$ identify the particular solution that traverses the distance between boundary states $\theta_i \rightarrow \theta_f$ in the optimal time. 
Because $\partial_{\mathbf{q}_f} \mathcal{S} = \boldsymbol{\Lambda}_f$, the dynamics terminating at $\boldsymbol{\Lambda}_f = 0$ are by contrast the dynamics leading to the action--extremal final state $\theta_f$ given $\theta_i$ and the evolution time $T$. See the cited references \cite{Lewalle_Caustic-thy, BookArnoldClassical, Cline_CMech} for further justification and discussion. Since these extrema are the same in the present instance, we will not interrogate these subtleties further here.} 
Setting \eqref{H-2opt-OM} to zero is equivalent to setting $\Lambda_\theta^\star = 0$, such that here these two optimal conditions coincide. 
Solving \eqref{linear_Lambda}$~=0$ in the presence of a finite unitary drive ${\Omega} \neq 0$ (see \secref{sec-example_noSTZ} for the optimal solution in the absence of a unitary drive, i.e., $\Omega=0$) then gives us
\be \label{STZ_CDJ-P}
\Omega^\star = \frac{\theta_f - \theta_i}{T} = \dot{\theta}^\star = \dot{\zeta}^\star. 
\ee 
This optimal schedule solution illustrates two useful results: i) The optimal schedule for the single qubit measurement axis $\zeta^\star(t)$ in this simplest example of Zeno dragging is linear in time (in the angular $\theta$ coordinates within the $xz$ plane), and ii) the best solution for Zeno dragging assisted by a \emph{unitary drive} has the unitary matched exactly with the linear schedule of the measurement, $\Omega = \dot{\zeta}^\star$. 

%%%%%%%%%%%%%%%%%%%%%%%%%%%%%%%%%%%%%%%%%%%%%%%%%%%%%%%%%%%%%%%%%
%%%%%%%%%%%%%%%%%%%%%%%%%%%%%%%%%%%%%%%%%%%%%%%%%%%%%%%%%%%%%%%%%
\subsection{STZ Dragging for a Qubit}
\label{sec-example_STZ}

Here we compare the CDJ--P optimal solution for drive--assisted Zeno dragging with that obtained from the STZ method that we derived in \secref{sec-stz}. 
Application of the STZ process to the single qubit example requires diagonalization of $\hat{\mathfrak{Z}}$ which is accomplished by the time--dependent Zeno frame transformation
\be 
\hat{Q} = \left(
\begin{array}{cc}
 \cos \left(\frac{\zeta(t)}{2}\right) & -\sin \left(\frac{\zeta(t)}{2}\right) \\
 \sin \left(\frac{\zeta(t)}{2}\right) & \cos \left(\frac{\zeta(t)}{2}\right) \\
\end{array}
\right), \quad\text{to yield}\quad \hat{D}_\mathfrak{Z} = \hat{Q}^\dag\,\hat{\mathfrak{Z}}\,\hat{Q} = \left(
\begin{array}{cc}
-\Gamma + \sqrt{\Gamma}\,r & 0 \\
 0 & -\Gamma - \sqrt{\Gamma}\,r \\
\end{array}
\right).
\ee 
Having determined the Zeno frame transformation $\hat{Q}$, it is then straightforward to apply \eqref{STZ_condition_1} and \eqref{STZ_solution} to find
\be \label{STZ_example}
\dot{Q}^\dag \hat{Q} + \hat{D}_\mathfrak{Z} - i\,\hat{Q}^\dag\, \hat{H}_\mathrm{STZ} \,\hat{Q} =
\frac{1}{2}\left(\begin{array}{cc}
 -2(\Gamma - \sqrt{\Gamma}\,r) & \dot{\zeta}-\Omega \\
\Omega-\dot{\zeta} & -2(\Gamma + \sqrt{\Gamma}\,r) \\
\end{array}\right).
\ee
This suggests that $\hat{H}_\mathrm{STZ} = \tfrac{1}{2}\,\Omega\,\hat{\sigma}_y$, with the choice
\be \label{STZ-optimal}
\dot{\zeta} = \Omega,
\ee
will diagonalize the dynamics in the Zeno frame. 
The condition \eqref{STZ-optimal} is furthermore identical to the optimal schedule solution from the CDJ-P approach in \eqref{STZ_CDJ-P}. 

We have thereby confirmed that our STZ and CDJ--P optimal solutions for the time-dependence of the measurement axis, \eqref{STZ-optimal} and \eqref{STZ_CDJ-P} are identical and that both of these rates are equal to the magnitude of the unitary drive.
Both of these control approaches were analytically tractable for this simple example. 
We remark that the STZ approach was perhaps simpler to implement (and should be expected to scale to larger problems more easily), but that when tractable the CDJ--P approach has the considerable benefit of explicitly providing optimality guarantees with the schedule $\zeta^\star(t)$, rather than merely giving the condition that the driving frequency $\Omega$ should match the rate of change of the measurement axis rotation (or vice versa). 
A third way of deriving this solution is presented in Appendix \ref{sec-feedback}.

%%%%%%%%%%%%%%%%%%%%%%%%%%%%%%%%%%%%%%%%%%%%%%%%%%%%%%%%%%%%%%%%%
%%%%%%%%%%%%%%%%%%%%%%%%%%%%%%%%%%%%%%%%%%%%%%%%%%%%%%%%%%%%%%%%%
\subsection{Optimal Zeno Dragging without Unitary Assistance}
\label{sec-example_noSTZ}

For comparison, we briefly analyze here how the CDJ--P optimal solution changes when the unitary drive $\Omega = 0$, i.e.,~we look at optimal Zeno dragging alone without the possibility of any unitary assistance or a ``shortcut'' to staying in the Zeno subspace. 

In the special case $\Omega = 0$, \eqref{zeta-star} remains correct, i.e.
\be \label{zeta-star-Om0}
\partial_\zeta \mathcal{H}_\zeta^\star = 0 \quad\rightarrow\quad \zeta^\star = \theta + \arctan\Lambda_\theta,
\ee 
with the remaining equations from \eqref{H_star_zeta} through \eqref{linear_theta} also remaining correct upon setting $\Omega = 0$.
Specifically, \eqref{H-2opt-OM} becomes $\mathcal{H}_{\zeta^\star}^{r^\star} = 2\Gamma\,\Lambda_\theta^2$ such that $\Omega = 0$ may be implemented directly into \eqref{linear-omega-dynamics} to obtain $\theta(t) = \theta_i + 4\Gamma\,\Lambda_\theta^\star\,t$ with $\Lambda_\theta^\star = (\theta_f - \theta_i)/(4\Gamma\,T)$.
Thus, in the absence of any unitary shortcut or control, the best measurement schedule that one can implement for finite-time Zeno dragging is
\be \label{pre-compute-linear-gZeno}
\zeta^\star(t) = \theta_i + (\theta_f - \theta_i)\tfrac{t}{T} + \arctan \frac{\theta_f - \theta_i}{4\,\Gamma\,T},
\ee
which is a special case of \eqref{linear-zeta-star-general}.
Interestingly, this implies that one should \emph{not} set the measurement axis exactly on the expected instantaneous state, but that one should rather set it \emph{ahead} of that expected Bloch angle $\theta$ by the factor $\arctan{\Lambda_\theta^\star}$. 
This forward lead in the Zeno dragging gets larger when attempting to do the process quickly ($\Gamma\,T \sim 1$), but shrinks in the adiabatic limit $\Gamma\,T\gg 1$ where the Zeno dragging process performs well without any unitary assistance. 
We note that in the experiment most closely related to our present formulation of the problem, it is clear that the main cluster of conditional states lags behind the measurement axis by a small amount \cite{ZenoDragging}; our optimal control solution here explicitly compensates for this effect.
See Fig.~\ref{fig-g-rotation}(b).
Related ``offset'' phenomena have been observed in other optimal control settings that aim to accelerate an adiabatic process \cite{Stefanatos_2022, Evangelakos_2023}.

Since the Hamiltonian generator of the optimal dynamics reduces to $\mathcal{H}^{r^\star}_{\zeta^\star} = 
\mathcal{H}_\zeta^\star(\zeta = \zeta^\star) = 2\,\Gamma\,\Lambda_\theta^2$ for $\Omega = 0$, the globally optimal solution (corresponding equivalently to $\mathcal{H}^{r^\star}_{\zeta^\star} = 0$ or $\Lambda_\theta^\star(T) = 0$, as above), is then given by $\Lambda_\theta^\star = (\theta_f - \theta_i)/(4\Gamma\,T) = 0$.
This shows that optimal Zeno dragging with measurement alone is obtained only in the limit $\Gamma\,T\rightarrow \infty$ that minimizes $\Lambda_\theta^\star$, i.e.,~by rotating $\zeta$ as slowly as possible from $\theta_i$ to $\theta_f$. This makes physical sense: our solution \eqref{pre-compute-linear-gZeno} encourages us to work in the adiabatic limit for Zeno dragging, while our equivalent solutions \eqref{STZ_CDJ-P} and \eqref{STZ-optimal} point us towards the possibility of a faster operation via cooperative unitary and dissipative evolution. 

%%%%%%%%%%%%%%%%%%%%%%%%%%%%%%%%%%%%%%%%%%%%%%%%%%%%%%%%%%%%%%%%%
%%%%%%%%%%%%%%%%%%%%%%%%%%%%%%%%%%%%%%%%%%%%%%%%%%%%%%%%%%%%%%%%%
\subsection{Evaluating the Performance of our Single--Qubit Optimal Solutions \label{sec-avg-Zeno-performance}}

\begin{figure}[p]
\hspace{2.9cm}\includegraphics[width = .75\textwidth]{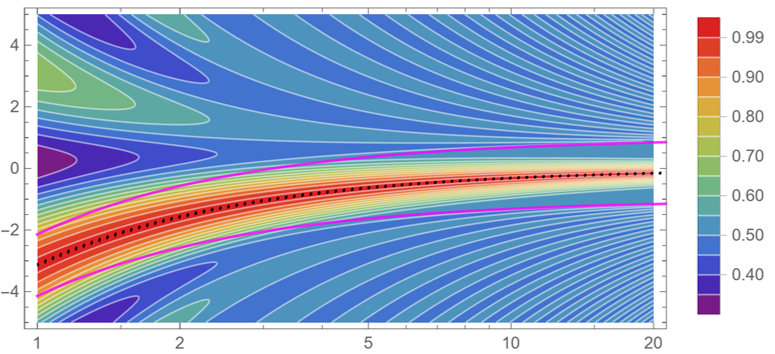} \\ 
\begin{center}
\includegraphics[height = .18\textheight]{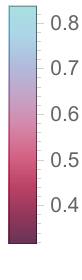} \hspace{0.5cm}
\includegraphics[height = .18\textheight]{dpfid.pdf} 
\hspace{0.5cm}
\includegraphics[height = .18\textheight]{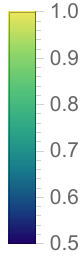} \end{center} 
\begin{tikzpicture}[overlay]
\node[] at (0.5,10) {\sf (a)};
\node[] at (0.5,4) {\sf (b)};
\node[] at (1.3,4.9) {$\CMcal{F}$};
\node[] at (14.7,4.9) {$\CMcal{F}$};
\node[] at (3.2,4.9) {$\Omega~[\Gamma]$};
\node[] at (8.8,0.8) {$T~[\Gamma^{-1}]$};
\node[] at (1.3,0.8) {\small $\Gamma^2 < (\Omega-\dot{\zeta})^2$};
\node[] at (15,0.8) {\small $\Gamma^2 > (\Omega - \dot{\zeta})^2$};
\node[] at (14.8,6.1) {$\CMcal{F}$};
\node[] at (12.25,5.5) {$T~[\Gamma^{-1}]$};
\node[] at (2.6,9.85) {$\Omega~[\Gamma]$};
\end{tikzpicture} \vspace{-15pt}
\caption{
Average fidelity, $\CMcal{F}$ \eqref{fidelity}, for the Zeno dragging protocol that performs a bit-flip $\ket{g}\rightarrow\ket{e}$ using the CDJ--P optimal schedule \eqref{linear-zeta-star-general}, with boundary conditions $\theta_f = 0$ and $\theta_i = \pi$. 
Both panels show the Rabi rate $\Omega$ of the unitary drive (in units of the measurement strength $\Gamma$) on the vertical axis, and the total dragging time $T$ (in units $\Gamma^{-1}$) on the horizontal axis. 
The globally optimal CDJ--P/STZ solution \eqref{STZ_CDJ-P} is marked with a dotted black line in both panels.
Panel (a): contour plot for a relatively short range of dragging times $T$ and a large range of rotation rates $\Omega$. 
The magenta lines denote the crossover point $\Gamma^2 = (\Omega - \dot{\zeta})^2$ between oscillatory Lindblad solutions ($\Gamma^2 < (\Omega - \dot{\zeta})^2$), and decaying solutions ($\Gamma^2 > (\Omega - \dot{\zeta})^2$). 
Our protocol is intended for the regime of decaying solutions, i.e.,~where the Zeno effect stabilizes our target trajectory. 
The globally optimal CDJ--P/STZ solution (dotted black line) sits at the center of this damped regime. 
The low--purity oscillations of the oscillatory Lindblad regime are clearly visible outside the magenta curves. 
Panel (b): density plot illustrating the same dynamics, with an extended horizontal axis that goes deeper into the adiabatic regime where Zeno dragging works increasingly well without unitary assistance. 
In this panel a red--hued colorbar is employed for the oscillatory regime, and a green--hued colorbar for the damped regime. 
The STZ solution (dotted black line) clearly converges towards the unitary drive amplitude $\Omega \rightarrow 0$ in the long--time limit, corresponding to the adiabatic limit where perfect Zeno dragging fidelity can be achieved deterministically without unitary assistance. 
Both panels were obtained by numerical evaluation of the quasi-analytic solutions for the averaged dynamics presented in Appendix \ref{sec-LME-solve}.}
\label{fig-fidelities}
\end{figure}

\begin{figure}[p]
\centering
\includegraphics[width = .4\textwidth]{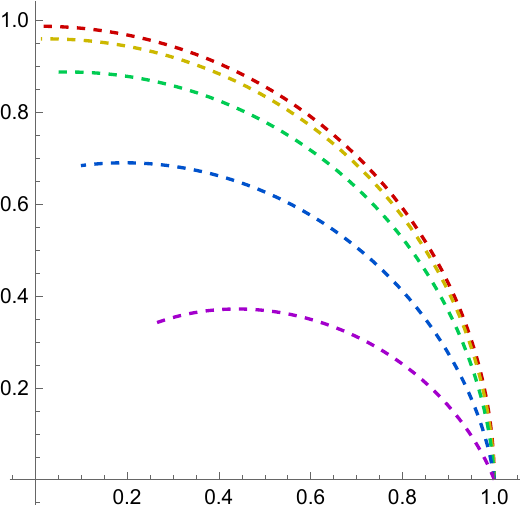}
\hspace{1cm}\includegraphics[width = .5\textwidth]{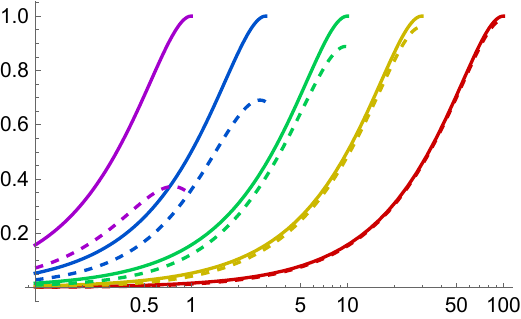} \\
\begin{tikzpicture}[overlay]
\node[] at (6.2,0.4) {$\Gamma\,t$};
\node[] at (-0.3,4.8) {$z$};
\node[] at (-8,6.1) {$z$};
\node[] at (-2.25,0.4) {$x$};
\draw[>->,black!15!white, line width = 0.05cm] (-1.4,1) arc (0:90:6cm);
\node[draw = black!15!white,fill = white, rounded corners = 0.1cm,line width = 0.05cm] at (-2.9,5.2) {\color{black!30!white} $\,\zeta\,$};
%%%%%%%%%%%%%%%%%%%%%%%
\end{tikzpicture} \\ \begin{tikzpicture}[xshift = 2 cm, yshift = 7 cm, overlay]
\node[right] at (4,0) {\color{red!80!black} $\Gamma T = 100$};
\node[right] at (2,0) {\color{mikadoyellow!80!black} $\Gamma T = 30$};
\node[right] at (0,0) {\color{darkspringgreen!50!green} $\Gamma T = 10$};
\node[right] at (-2,0) {\color{dodgerblue!50!blue} $\Gamma T = 3$};
\node[right] at (-4,0) {\color{patriarch!50!magenta} $\Gamma T = 1$};
\draw[draw = black!15!white,line width = 0.05cm, rounded corners = 0.2cm] (-4.3,-0.4) rectangle (5.85,0.4);
\end{tikzpicture} \vspace{-20pt}
\caption{  
Lindblad dynamics for a Zeno dragging process on a single qubit that pulls the initial $x = +1$ eigenstate at $t = 0$ to the $z = +1$ eigenstate at $t = T$, are evaluated using the optimal linear schedule $\zeta^\star$ \eqref{linear-zeta-star-general} and plotted as paths in the $xz$ Bloch plane (left panel), and as the time-dependent $z(t)$ (right panel). 
Dashed lines show the solutions with $\Omega = 0$, i.e.,~Zeno dragging alone, with no corrective unitary control. 
Solid lines (right panel only) show the CDJ--P optimal solutions with the STZ drive $\Omega = \dot{\zeta}$ added to the scheme, revealing effectively perfect solutions from the optimal condition matching the rate of change of the measurement axis and the unitary drive, provided that the measurement axis rotation $\zeta(t)$ and unitary $\Omega$ are both implemented perfectly. 
Comparing the STZ and measurement--only traces, we see how the matched measurement and unitary overcome the adiabatic--like timescale issue inherent in Zeno dragging alone, allowing us to now reproduce the good Zeno dragging dynamics, i.e.,~those normally accessible only for $\Gamma\,T \gg 1$, at arbitrarily faster timescales. 
}\label{fig-STZ-correction}
\end{figure}
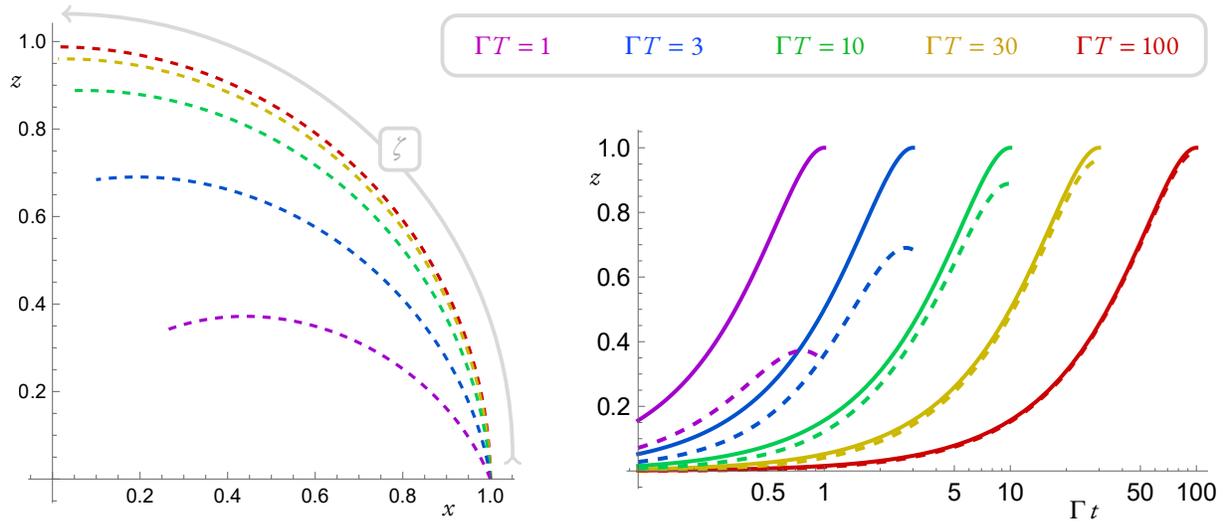

\begin{figure}[p]
    \hfill
    \includegraphics[height = .157 \textheight, trim = {45 30 30 30}, clip]{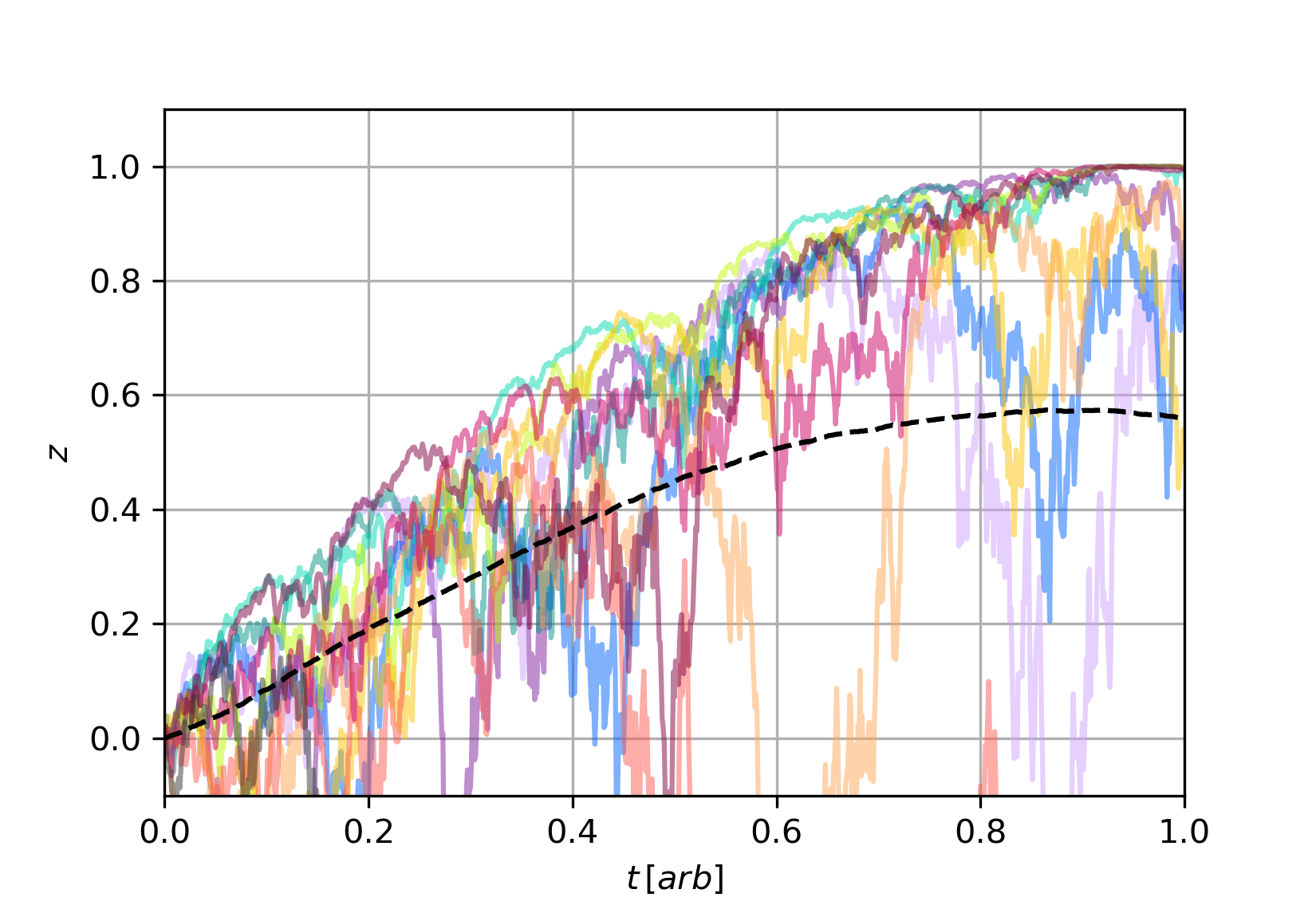}
    \includegraphics[height = .157 \textheight, trim = {45 30 30 30}, clip]{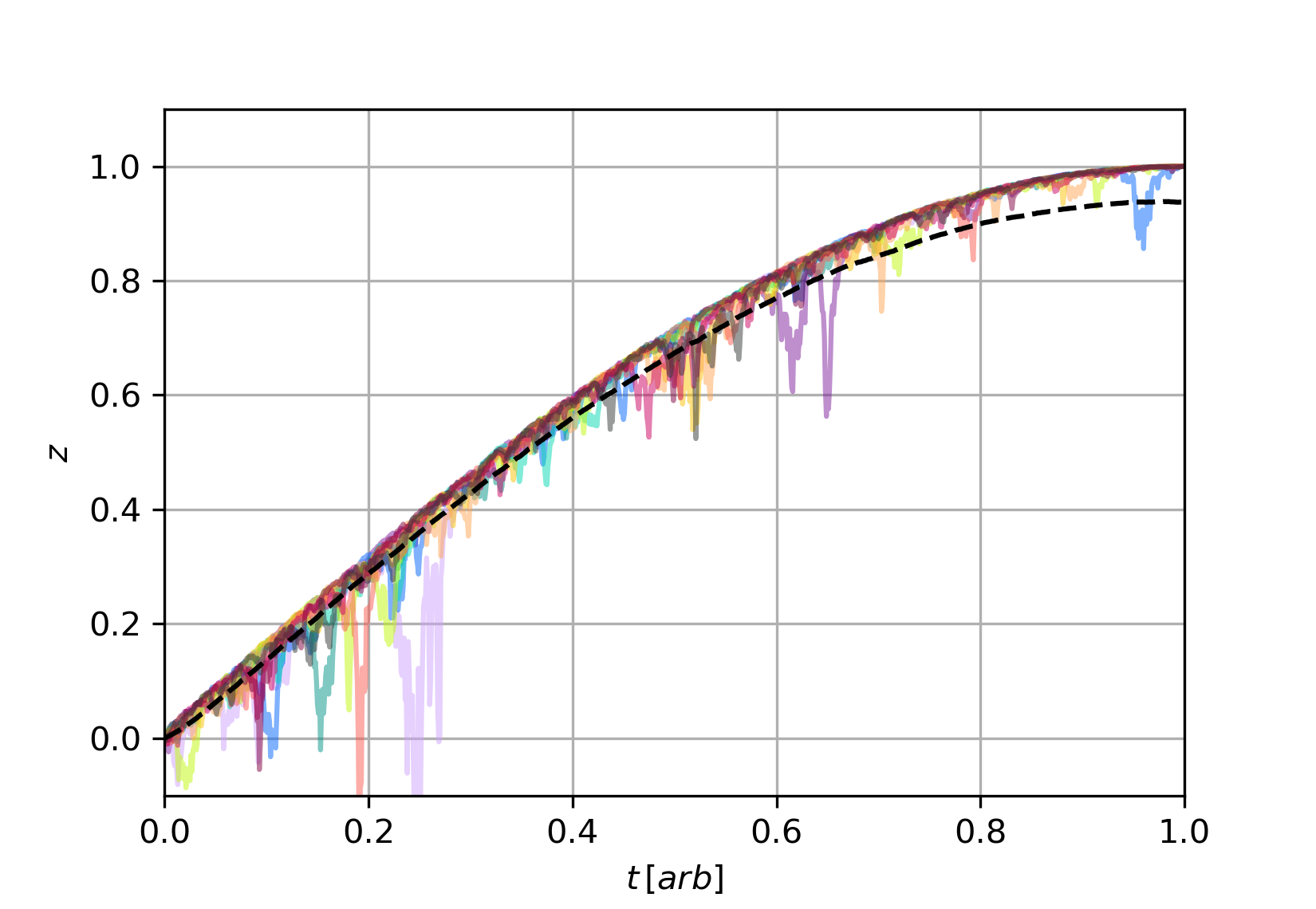}
    \includegraphics[height = .157 \textheight, trim = {45 30 30 30}, clip]{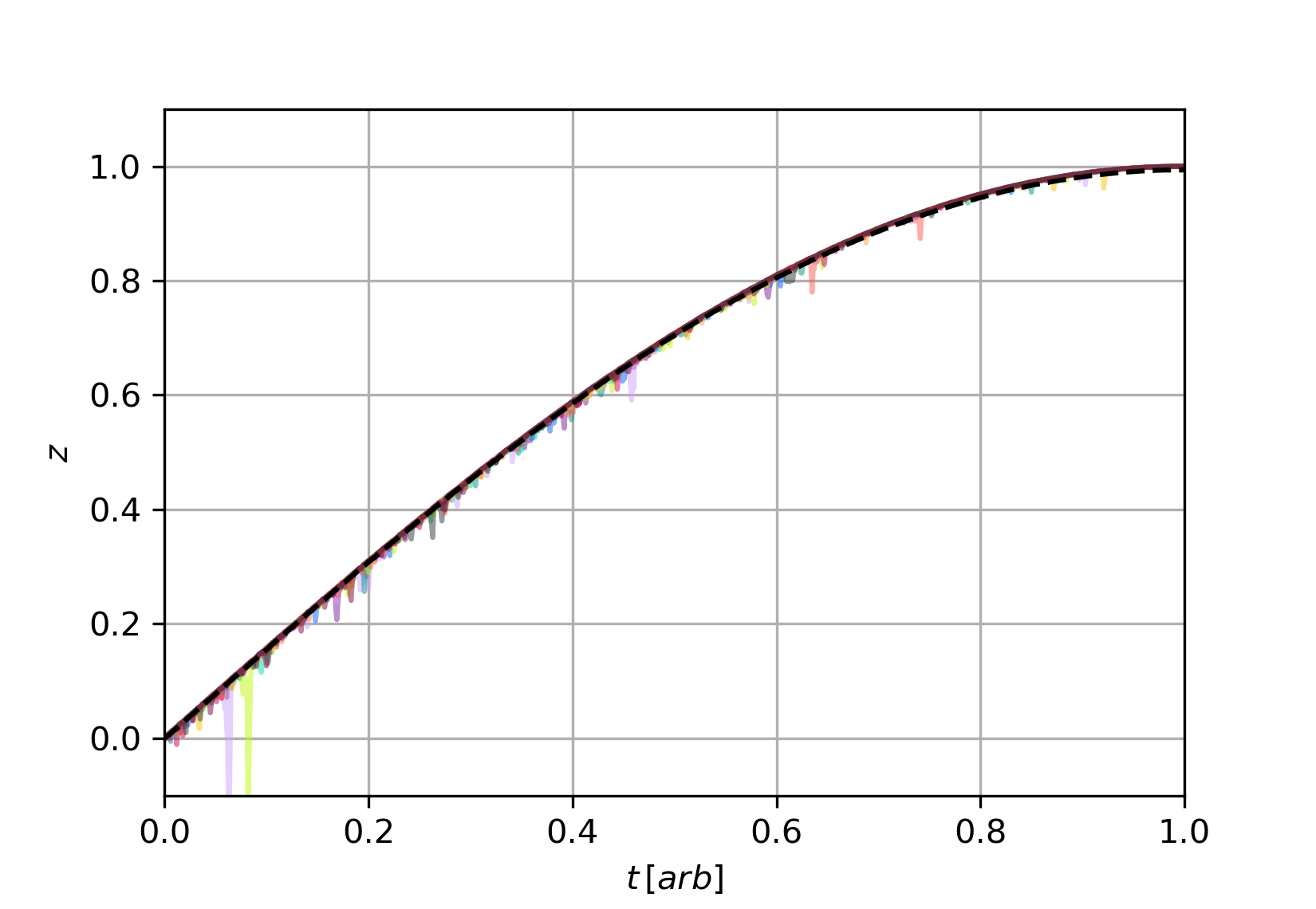} 
    \begin{tikzpicture}[overlay]
    \node[] at (-15.4,3.1) {\sf (a)};
    \node[] at (-10,3.1) {\sf (b)};
    \node[] at (-4.65,3.1) {\sf (c)};
    \node[] at (-16.2,3.2) {\footnotesize $1$};
    \node[] at (-16.2,1.8) {\footnotesize $z$};
    \node[] at (-16.2,0.4) {\footnotesize $0$};
    \node[] at (-15.9,-0.2) {\footnotesize $0$};
    \node[] at (-13.45,-0.2) {\footnotesize $t\,[\mathrm{arb}]$};
    \node[] at (-11,-0.2) {\footnotesize $T$};
    \node[] at (-10.55,-0.2) {\footnotesize $0$};
    \node[] at (-8.1,-0.2) {\footnotesize $t\,[\mathrm{arb}]$};
    \node[] at (-5.65,-0.2) {\footnotesize $T$};
    \node[] at (-5.2,-0.2) {\footnotesize $0$};
    \node[] at (-2.75,-0.2) {\footnotesize $t\,[\mathrm{arb}]$};
    \node[] at (-0.3,-0.2) {\footnotesize $T$};
    \end{tikzpicture} \\ \vspace{-5pt}
    \caption{
    Individual stochastic quantum trajectories for Zeno dragging a single qubit from $\tfrac{1}{\sqrt{2}}\ket{e} + \tfrac{1}{\sqrt{2}}\ket{g}$ to $\ket{e}$, using the pre-computed linear in $\theta$ schedule \eqref{pre-compute-linear-gZeno}. These are computed with a fixed dragging time $T$ (arbitrary units), but with differing measurement strengths $\Gamma$. Specifically we use values $\Gamma\,T = 2$ (a), $\Gamma\,T =20$ (b), and $\Gamma\,T = 200$ (c). 
    The average of an ensemble of $5000$ trajectories is shown in dashed black, together with a dozen individual trajectories within that ensemble (various colors). 
    The distance of each trajectory from $z_f = 1$ at the final time $T$ is a good indicator of the overall error. 
    As expected, the protocol is more robust for large values of $\Gamma\,T$. 
    Interestingly, we also note that for individually weak measurements (we use $\Delta t = 0.001\,[T]$ throughout), fluctuations at shorter times are often corrected by subsequent measurements. 
    }
    \label{fig-Pontryagin-1Q-Sim}
\end{figure}

A successful Zeno dragging process will result in high--purity quantum trajectories that closely follow a measurement eigenstate throughout the time evolution of the system. 
The LME \eqref{lme} describes the average evolution over an ensemble of quantum trajectories (recall also that this emerges from \eqref{gen-hermitian} in the special case where the measurement outcomes are not collected, i.e.,~for zero measurement efficiency, $\eta = 0$). 
As such, we may use the LME to characterize the average effectiveness of our various Zeno dragging solutions; both the purity of the LME solutions and their fidelity to the desired evolution quantify the degree to which a Zeno dragging process (characterized by a choice of $\zeta(t)$ and $\Gamma$) will be effective on average. 

It is generically possible to solve the Lindblad equation quasi-analytically via diagonalization of the Liouvillian if none of the unitary drives or dissipators are time--dependent \cite{Manzano_2020}. 
See Appendix \ref{sec-LME-solve} for details pertinent to our present example. 
We further show there that when the schedule on which the dissipator in \eqref{lme} varies is linear in time (i.e.,~$\dot{\mathsf{h}}$ is a constant, as is the case for our optimal schedule \eqref{linear-zeta-star-general}), then the time dependence of the dissipator is eliminated by transforming to the Zeno frame, so that the diagonalization approach may be applied in that frame.
We may then use such a solution to compute the average fidelity of a Zeno dragging operation. 
In particular, if $\ket{\psi_f}$ is the target final state and $\rho(T)$ is the final average state  at the end of a Zeno dragging operation, then the general mixed state fidelity~\cite{jozsa1994fidelity}
\begin{subequations} \label{fidelity} \be 
\CMcal{F} = \tr{\sqrt{\sqrt{\ket{\psi_f}\bra{\psi_f}}\,\rho(T)\,\sqrt{\ket{\psi_f}\bra{\psi_f}}}}^2
\ee
may be used. 
For a qubit restricted to the $xz$--Bloch plane, as in the example above, this reduces to 
\be 
\CMcal{F} = \tfrac{1}{2}\left(1 + x(T)\,\sin\theta_f + z(T)\,\cos\theta_f \right), 
\ee \end{subequations}
where $\theta_f$ is the target angle, and $x(T)$ and $z(T)$ are the Bloch coordinate representation of the solution to the Lindblad dynamics. 
This fidelity can be evaluated using the solutions from Appendix~\ref{sec-LME-solve}, and is plotted in Fig.~\ref{fig-fidelities} for a Zeno dragging operation that performs a bit-flip $\ket{g}\rightarrow\ket{e}$ using the optimal schedule \eqref{linear-zeta-star-general}. 
Two important and expected features appear in Fig.~\ref{fig-fidelities}. 
First, we confirm that the solution \eqref{STZ_CDJ-P} is globally optimal, allowing a perfect average fidelity to be achieved for all values of dragging time $T$. 
Second, we see how this solution converges towards the un-assisted Zeno dragging of \secref{sec-example_noSTZ} for long dragging times.  

Fig.~\ref{fig-STZ-correction} shows now the dynamics for Zeno dragging of a qubit from the $+x$ eigenstate to the $+z$ eigenstate over a finite time interval $T$. 
Specifically, here we show the average state evolution for CDJ--P optimal Zeno dragging that is designed to generate the target dynamics $\theta(t) = (\pi/2T)(T-t)$.
We compare different measurement strengths, as well as solutions with and without the unitary STZ.
We see once again that working in the adiabatic regime $\Gamma T \gg 1$ can generate a high dragging fidelity even for $\Omega = 0$, i.e., for Zeno dragging without supporting unitary controls, while for smaller values of $\Gamma\,T$ the quality of the Zeno dragging control degrades. Here the solutions are increasingly impure as diabatic effects become more important at smaller $T$ and/or smaller $\Gamma$ values, but the target dynamics are nevertheless still recovered when the STZ protocol is implemented.
All of these features are expected, given that adiabatic dynamics and that our STZ is designed to accelerate those dynamics in a similar manner to the shortcuts to adiabaticity of the more standard all--unitary dynamical situation \cite{Guery_STA-RMP}. 

Fig.~\ref{fig-Pontryagin-1Q-Sim} shows some representative individual trajectories together with the ensemble average. 
It is evident that these plots are consistent with the conclusions above. 
We additionally note that the main cluster of trajectories follows the intended dynamics more closely than does the average dynamics. 
This is because the average dynamics include large fluctuations involving escape of the system to the ``wrong'' measurement eigenstate. 
As noted in \textcite{Kokaew_2022}, part of the motivation for using the CDJ $r$--optimal paths to derive optimal controls instead of the LME solutions is precisely this clustering behavior in the conditional dynamics.  
Furthermore, it is apparent from Fig.~\ref{fig-Pontryagin-1Q-Sim} that in the limit of weak continuous measurements (individual detector integration intervals $\Delta t \ll \Gamma^{-1}$), many trajectories that go off course at short times actually correct themselves at longer times. 
This is because measurement fluctuations away from the target eigenspace are small enough to not constitute a complete ``collapse'' to the ``wrong'' eigenstate, so that subsequent weak measurements may still correct those fluctuations with high probability. 

%%%%%%%%%%%%%%%%%%%%%%%%%%%%%%%%%%%%%%%%%%%%%%%%%%%%%%%%%%%%%%%%%
%%%%%%%%%%%%%%%%%%%%%%%%%%%%%%%%%%%%%%%%%%%%%%%%%%%%%%%%%%%%%%%%%
\subsection{Insights from the Zeno Frame\label{sec-ExZenoFrame}}

\begin{SCfigure}
$\quad$\includegraphics[width = .4\textwidth, trim = {20 20 0 0},clip]{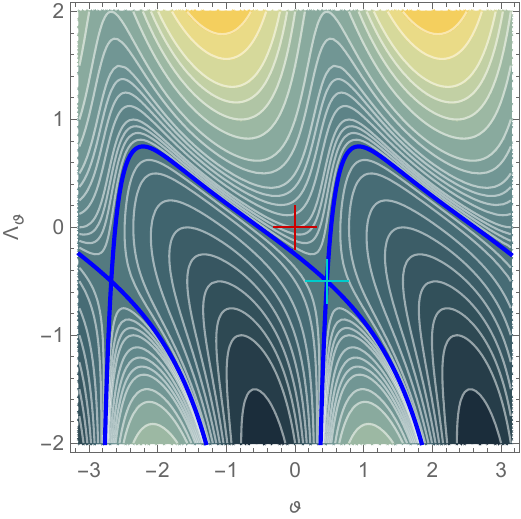}
\includegraphics[height = .27\textheight]{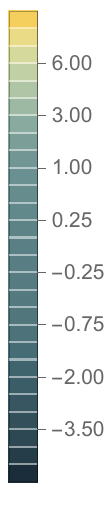} 
\begin{tikzpicture}[overlay,yshift = -0.7cm, xshift = -4cm]
\node[] at (3.5,6.9) {$\mathcal{E} = \mathcal{H}^\star_\zeta$};
\node[] at (1.9,0.6) {$\vartheta$};
\node[] at (-4,6.3) {$\Lambda_\vartheta$};
\end{tikzpicture} 
\caption{
We show the Hamiltonian phase portrait of \eqref{H_star_zeta_zeno}, for $\Gamma = 1$ and $\Omega_{eff} = 2$. This highly diabatic case is shown to improve visual contrast.
The ``energy'' $\mathcal{E} = \mathcal{H}^\star_\zeta$ is conserved for constant $\dot{\zeta}$, such that the dynamical solutions follow lines of constant $\mathcal{E}$. 
The separatrix $\mathcal{E} = -\Omega^2/8\Gamma$ is shown in dark blue, and the fixed point within it, \eqref{ZF-fixedpoint}, is higlighted with a cyan $+$. 
The offset between the origin (red $+$, representing the measurement axis $\zeta = \theta - \vartheta$ in this frame) and fixed point (cyan $+$) gives us the optimal shift needed for a linear schedule, c.f.~\eqref{linear-zeta-star-general}. 
Related analyses appear in e.g.~Refs.~\cite{Chantasri2013, FlorTeach2019, Philippe_Thesis, Jordan2015flor, shea2023action}.
}\label{fig-Hlevels-ZF}
\end{SCfigure}
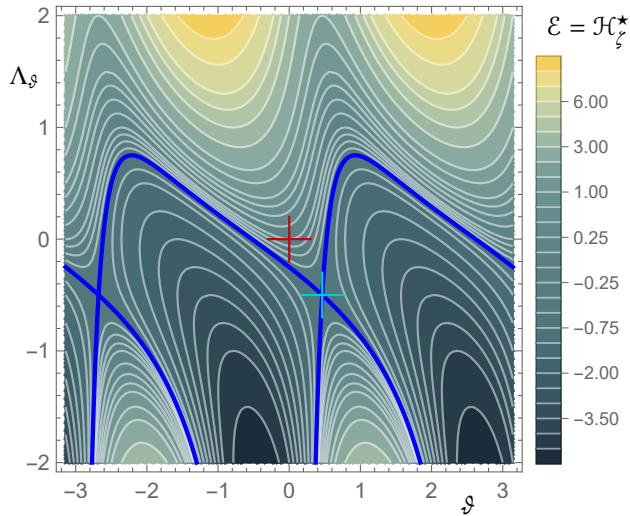

Let us briefly reconsider this simple qubit problem in the Zeno frame defined in the course of deriving our STZ results. 
To put \eqref{H_star_zeta} in the Zeno frame, we may simply define a new coordinate $\vartheta = \theta - \zeta$, and write
\be \label{H_star_zeta_zeno}
\mathcal{H}_\zeta^\star = 2 \Gamma  (\Lambda_\vartheta ^2-1) \sin ^2(\vartheta)+\Lambda_\vartheta  \bigg[\underbrace{\Omega - \dot{\zeta}}_{{\Omega_{eff}}} - 2 \Gamma  \sin ( 2\vartheta) \bigg],
\ee
where the diabatic frame rotation term now appears as an effective unitary rotation.
This transformation was simple due to the previous observation that \eqref{H_star_zeta} was already a function of $\zeta - \theta$ only. 
When $\dot{\zeta}$ is constant (i.e.~for a linear schedule), $\mathcal{H}_\zeta^\star$  \eqref{H_star_zeta_zeno} is time--independent in this frame; thus we have a frame that rotates with the optimal schedule, in which the optimally--controlled dynamics must follow lines of constant ``stochastic energy'' $\mathcal{E} = \mathcal{H}^\star_\zeta$. 
We may now derive the optimal ``offset'' in the schedule solely by examining the fixed points in this Zeno frame \cite{shea2023action}. 
The fixed points $\bar{\vartheta},~\bar{\Lambda}_\vartheta$ satisfy
\begin{subequations} \be 
\dot{\vartheta} = \partl{\mathcal{H}^\star_\zeta}{\Lambda_\vartheta}{} = 0 \quad\&\quad 
\dot{\Lambda}_\vartheta = -\partl{\mathcal{H}^\star_\zeta}{\vartheta}{} = 0, 
\ee
which may be solved to obtain
\be \label{ZF-fixedpoint}
\bar{\Lambda}_\vartheta = \frac{\dot{\zeta}-\Omega}{4\,\Gamma} \quad\&\quad \bar{\vartheta} = -\arctan(\bar{\Lambda}_\vartheta). 
\ee \end{subequations}
Substitution of these fixed point expressions into $\zeta = \theta - \vartheta$ then recovers \eqref{linear-omega-dynamics} and \eqref{linear-zeta-star-general}, i.e.~the fixed points allow us to quickly derive the ``offset'' by which the measurement axis is ahead of the current state for finite dragging times (because the whole idea of Zeno dragging is to remain stationary in the Zeno frame). 
Hamiltonian phase portraits \cite{BookArnoldClassical, Cline_CMech} of \eqref{H_star_zeta_zeno} appear in Fig.~\ref{fig-Hlevels-ZF} to illustrate the idea.

%%%%%%%%%%%%%%%%%%%%%%%%%%%%%%%%%%%%%%%%%%%%%%%%%%%%%%%%%%%%%%%%%
%%%%%%%%%%%%%%%%%%%%%%%%%%%%%%%%%%%%%%%%%%%%%%%%%%%%%%%%%%%%%%%%%
\subsection{Robustness of the Optimal Solutions \label{sec-ExRobustness}}

Given that our STZ solution calls for a redundant measurement and unitary, one may ask: Is there any advantage to ``doubling up'' on our controls in this way, as compared to just performing a unitary by itself?
We offer an analysis in this section centered about this question, and offer some further perspective again in \secref{sec-conclude}. 

We here address the following situation: Suppose that we perform STZ, but both our measurement axis control and unitary drive experience some drift or noise that makes the operations imperfect. 
Suppose $\Omega = \Omega_0 + \Omega_\epsilon$ and $\zeta = \zeta_0 + \zeta_\epsilon$ for the moment, where the $\epsilon$-subscripted terms denote errors relative to the intended  STZ operations $\Omega_0 = \dot{\zeta}_0$.  
We can then ask a more targeted version of our question: For what, if any, types of errors $\Omega_\epsilon$ and $\zeta_\epsilon$, and measurement strength $\Gamma\,T$ do the STZ dynamics remain more tightly clustered about the intended evolution than the corresponding dynamics generated by the unitary evolution alone? 
We here offer some numerical evidence that a suitable Zeno measurement does in fact add robustness; the most--closely related formal work we are aware of in the literature includes Refs.~\cite{Tanaka_2012, liang2023exploring, shtanko2023bounds, Medina-Guerra_2024}.

%%%%%%%%%%%%%%%%%%%%%%%%%%%%%%%%%%%%%%%%%%%%%%%%%%%%%%%%%%%%%%%%%
\subsubsection{Impact of Unitary Noise on STZ}

We begin by moving in the Zeno frame of our \emph{intended} operation, i.e.~we take the Kraus operator \eqref{Kraus-Op-Zeta} and implement $\hat{Q}_0^\dag\, \hat{\mathcal{M}}_r(\zeta)\,\hat{Q}_0 = \hat{\mathcal{M}}_r(\zeta-\zeta_0)$, for $\hat{Q}_0 = e^{-i\,\zeta_0\,\hat{\sigma}_y/2}$. 
Note that this implies that $\hat{L}(\zeta) \rightarrow \hat{L}(\zeta - \zeta_0)$ from this frame change, so that we may use \eqref{Zeno-frame-dynamics} to write the conditional dynamics
\be \begin{split}
\dot{\varrho} = & i[\varrho,\hat{H}_\Omega - \dot{\mathsf{h}}_0] + \hat{\mathfrak{Z}}(\zeta-\zeta_0)\, \varrho + \varrho\,\hat{\mathfrak{Z}}(\zeta-\zeta_0) - 2\,\varrho\,\tr{\varrho\,\hat{\mathfrak{Z}}(\zeta-\zeta_0)},
\end{split} \ee
where $\varrho$ and $\hat{H}_\Omega$ are in the frame defined by $\hat{Q}_0$. 
We will assume that the intended dynamics follow our STZ protocol, such that $\Omega_0 = \dot{\zeta}_0$. 
Our previous results can consequently be adapted straightforwardly, where for a \emph{deterministically} mis-aliged measurement axis we have (still in Stratonovich form)
\be \label{Zeno0_Strato} 
\dot{\vartheta} = \mathsf{A}_\vartheta + \mathsf{b}_\vartheta\,\frac{dW}{dt} \quad\text{for}
\quad
\mathsf{A}_\vartheta = \Omega_\epsilon - 2\Gamma\,\sin(2\vartheta - 2\zeta_\epsilon) \quad\&\quad \mathsf{b}_\vartheta =  2\sqrt{\Gamma}\,\sin(\vartheta - \zeta_\epsilon),
\ee 
with $\Omega = \Omega_0 + \Omega_\epsilon$ and $\zeta = \zeta_0 + \zeta_\epsilon$.

Consider first the following simple scenario: The unitary drive experiences state--uniform Gaussian white noise (independent of the fundamental measurement noise) as per $\Omega_\epsilon = \varepsilon_\Omega\,dW_\Omega/dt$, while $\zeta_\epsilon(t)$ is some deterministic drift away from the intended measurement axis.
We may then re-write
\begin{subequations}\be 
\dot{\vartheta} = \mathsf{A}_\vartheta|_{\Omega_\epsilon = 0} + \mathsf{b}_\vartheta \frac{dW}{dt} + \varepsilon_\Omega \frac{dW_\Omega}{dt},
\ee
as well as the corresponding Fokker--Planck equation (FPE, or forward Kolmogorov equation) \cite{GardinerStochastic} 
\be \label{FPE}
\partl{\CMcal{P}}{t}{} = \partl{}{\vartheta}{} \left\lbrace \tfrac{1}{2} \mathsf{b}_\vartheta \partl{}{\vartheta}{} [\mathsf{b}_\vartheta\,\CMcal{P}] + \frac{\varepsilon_\Omega^2}{2}\partl{\CMcal{P}}{\vartheta}{} - (\mathsf{A}_\vartheta|_{\Omega = 0})\,\CMcal{P} \right\rbrace, 
\ee \end{subequations}
where $\CMcal{P} = \CMcal{P}(\vartheta,t|\CMcal{P}(\vartheta,t=0))$. Given some initial distribution of states $\CMcal{P}(\vartheta,t=0)$, the FPE tells us how to propagate the distribution forward in time.
Note that in the event of unitary operations only (i.e.~without measurement at all, $\Gamma = 0$), the error is analytically solvable for an exactly--prepared initial state, i.e.
\be \label{FPE-basic-nomeas}
\partl{\CMcal{P}}{t}{} = \frac{\varepsilon_\Omega^2}{2}\partl{\CMcal{P}}{\vartheta}{2} \quad\rightarrow\quad \CMcal{P}_t = \CMcal{P}(\vartheta,t|\CMcal{P}_0) = \frac{e^{-\frac{(\vartheta - \vartheta_0)^2}{2\,\varepsilon_\Omega^2\,t}}}{\sqrt{2\pi\,\varepsilon_\Omega^2\,t}} \quad\text{for}\quad \CMcal{P}_0 = \delta(\vartheta - \vartheta_0).
\ee

\begin{figure} \centering
\includegraphics[width = .49\columnwidth]{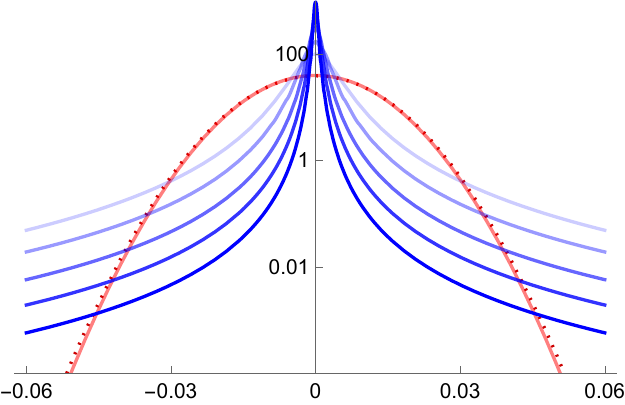}
\includegraphics[width = .49\columnwidth]{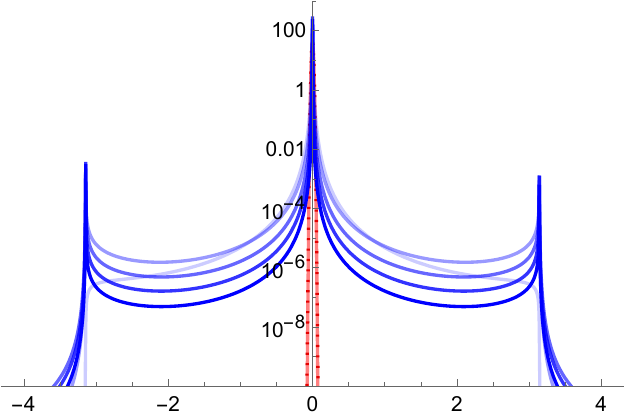} \\ 
\includegraphics[width = .49\columnwidth]{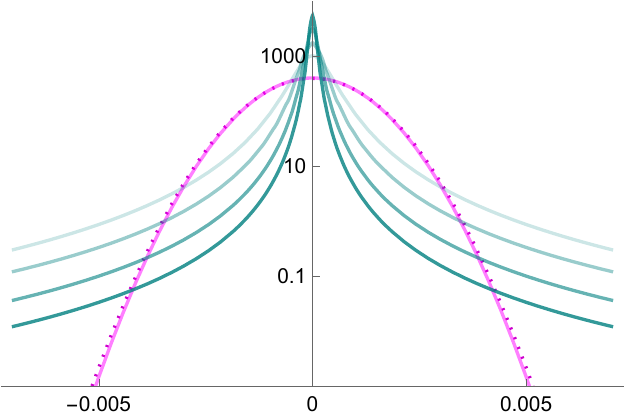}
\includegraphics[width = .49\columnwidth]{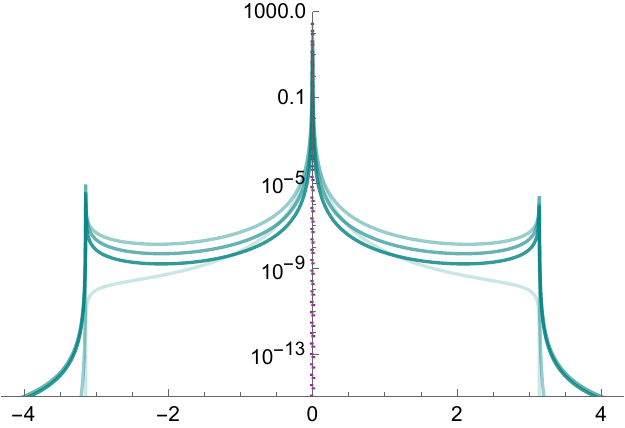} \\ 
\begin{tikzpicture}[overlay,scale = 1.8,yshift = -0.35cm,xshift=-0.2cm]
\node[] at (-3.8,6) {\sf (a)};
\node[] at (0.4,6) {\sf (b)};
\node[] at (-3.8,3) {\sf (c)};
\node[] at (0.4,3) {\sf (d)};
\node[] at (0.25,4.1) {$\vartheta$};
\node[] at (0.25,1) {$\vartheta$};
\node[] at (-1.7,6.2) {$\CMcal{P}_T$};
\node[] at (-1.7,3.2) {$\CMcal{P}_T$};
\node[] at (2.75,6.2) {$\CMcal{P}_T$};
\node[] at (2.75,3.2) {$\CMcal{P}_T$};
\end{tikzpicture} \vspace{-10pt}
\caption{
We plot solutions to the FPE \eqref{FPE}, obtained numerically (via finite element methods). 
In panels (a) and (b) diffusive unitary errors are generated as per $\varepsilon_\Omega = 0.01/\sqrt{T}$, while in (c) and (d) this is reduced to $\varepsilon_\Omega = 0.001/\sqrt{T}$. 
We choose the initial distribution $\CMcal{P}_0 = (\pi/2\,\sqrt{T}\,\varepsilon_\Omega) \cos^2(\pi\,\vartheta/\sqrt{T}\,\varepsilon_\Omega)$ for numerical purposes, which effectively assumes that the initial state is prepared with a similar error rate as appears in our subsequent unitary operations, and is localized with certainty to a small region about the desired initial state. 
Numerical solution of \eqref{FPE-basic-nomeas} with this finite--width initial state is shown in dotted red (a,b) or magenta (c,d), and deviates negligibly from the analytic solution to \eqref{FPE-basic-nomeas} for exact initial state preparation (solid red or magenta lines). These reference lines without measurement may be compared to the solutions including measurement (blue in a,b; teal in c,d). 
FPE solutions with measurement are shown for $\Gamma\,T = 1,3,10,30,100$ (blue or teal, with increasing opacity denoting stronger measurement).
In general, we see that measurement tends to localize the peak of the distribution around the desired state, but \emph{also} grows the tails of the distribution (by sometimes encouraging collapse to ``other'' measurement eigenstate). 
Recall that $\vartheta = 0$ is the target state in the Zeno frame, while $\vartheta = \pm \pi$ is the opposite measurement eigenstate. 
These localization and escape effects become more exaggerated with increased measurement strength. Note that all vertical axes are logarithmic, so that even quite rare escape events can be discerned in these distributions.  
}\label{fig-FPE}
\end{figure}
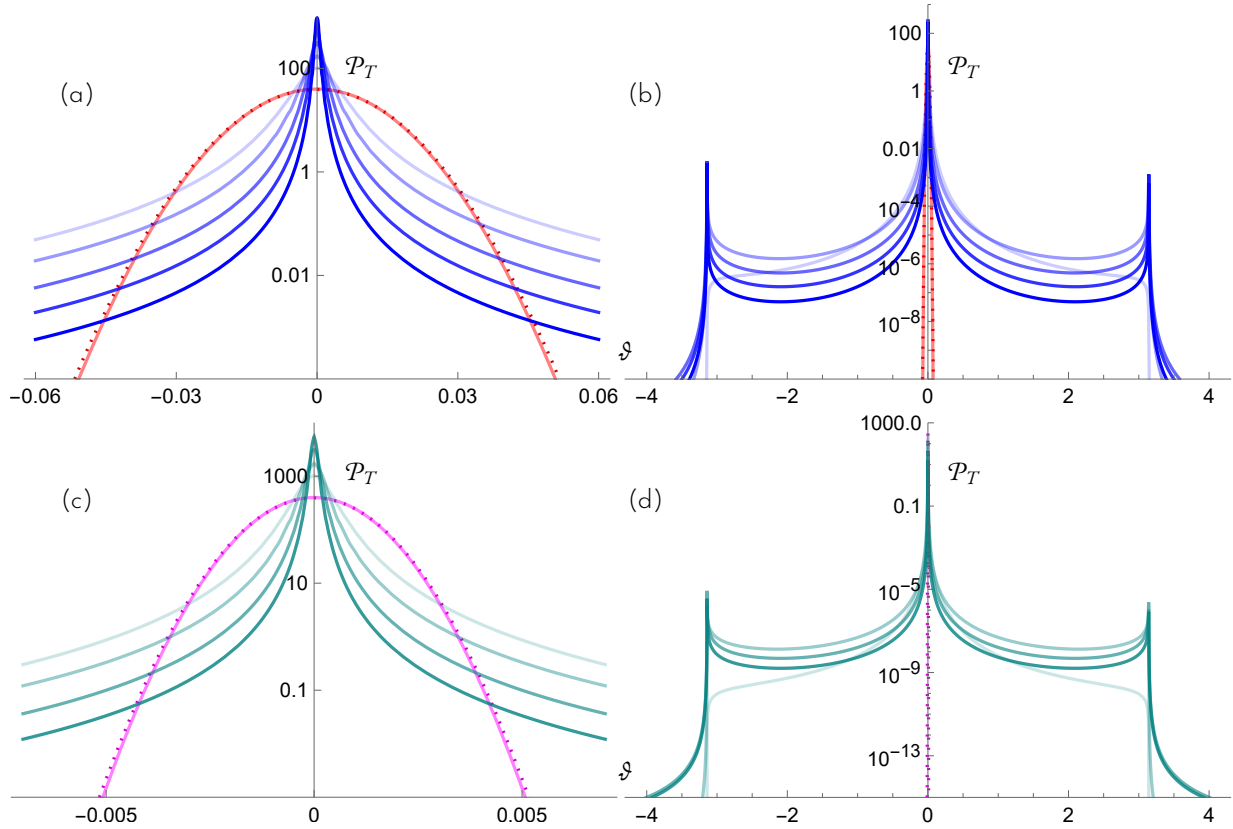

Solutions to the FPE \eqref{FPE} can be obtained numerically, and are shown in Fig.~\ref{fig-FPE}. 
Does perfect Zeno dragging help to mitigate errors in the paired unitary? Fig.~\ref{fig-FPE} says that the answer is both yes and no: Yes, the measurement helps to localize the state distribution about the target state more tightly than if we performed a unitary alone; but also no, because the measurement grows the tails of the distribution by fueling escape towards the orthogonal eigenstate. 
Both the positive and negative effects grow with increased measurement strength. 
As expected however, escape errors become increasingly unlikely when the unitary error rate is smaller. 
In short, the addition of a measurement is a net positive here, so long as the error rate is small enough to make escape events sufficiently rare.

%%%%%%%%%%%%%%%%%%%%%%%%%%%%%%%%%%%%%%%%%%%%%%%%%%%%%%%%%%%%%%%%%
\subsubsection{Impact of Measurement--Axis Noise on STZ}

\begin{figure}[t]\centering
\includegraphics[width = .33\columnwidth, trim = {20 25 10 10}, clip]{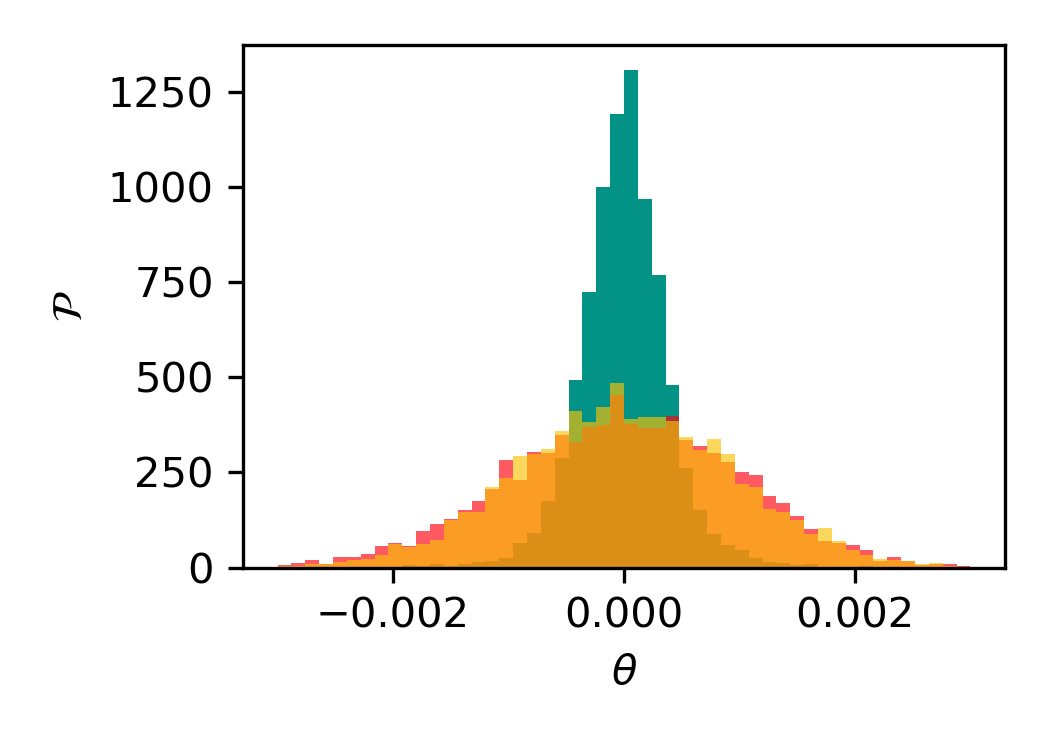}
\includegraphics[width = .33\columnwidth, trim = {20 25 10 10}, clip]{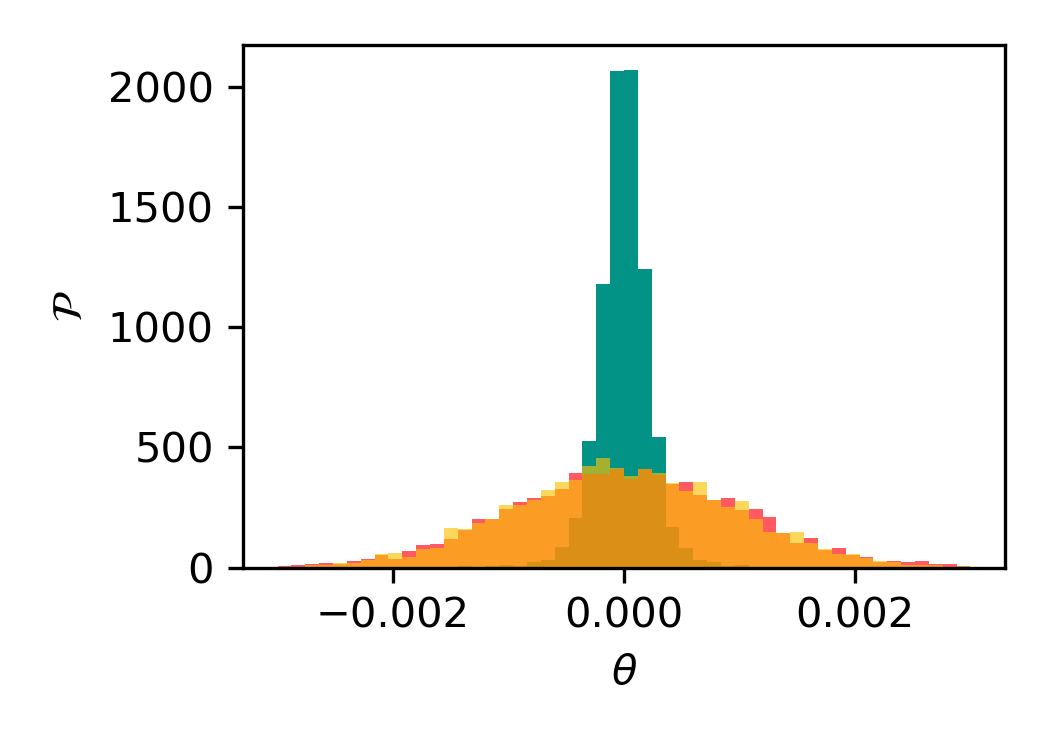}
\includegraphics[width = .33\columnwidth, trim = {20 25 10 10}, clip]{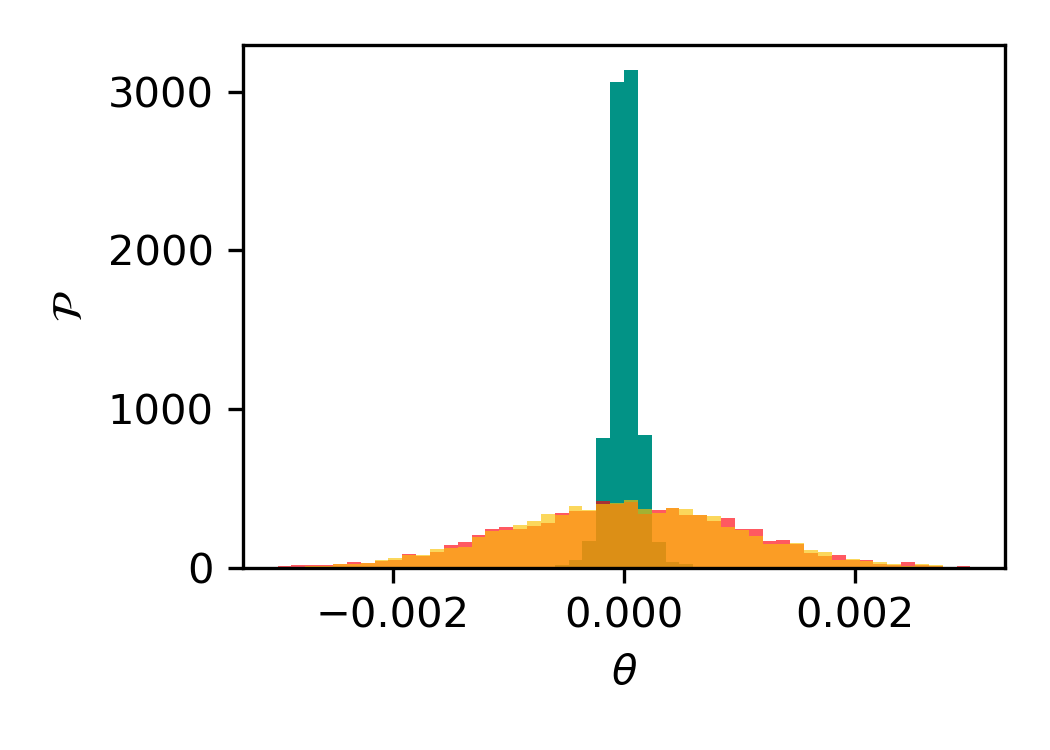} \\
\includegraphics[width = .33\columnwidth, trim = {20 25 10 10}, clip]{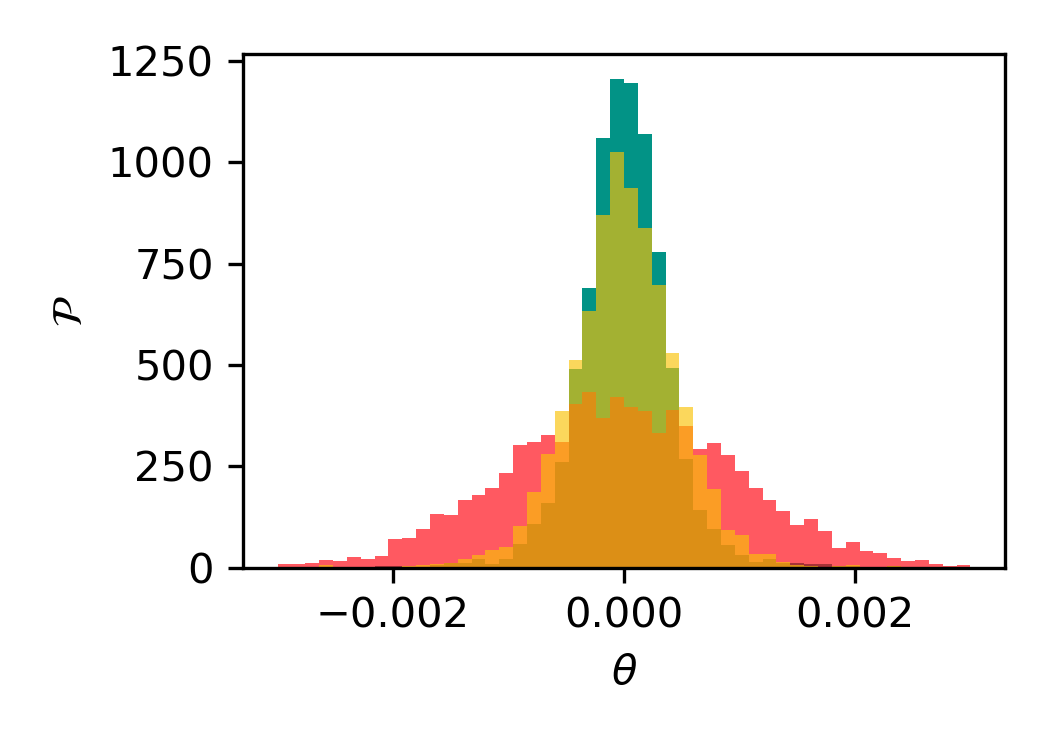} 
\includegraphics[width = .33\columnwidth, trim = {20 25 10 10}, clip]{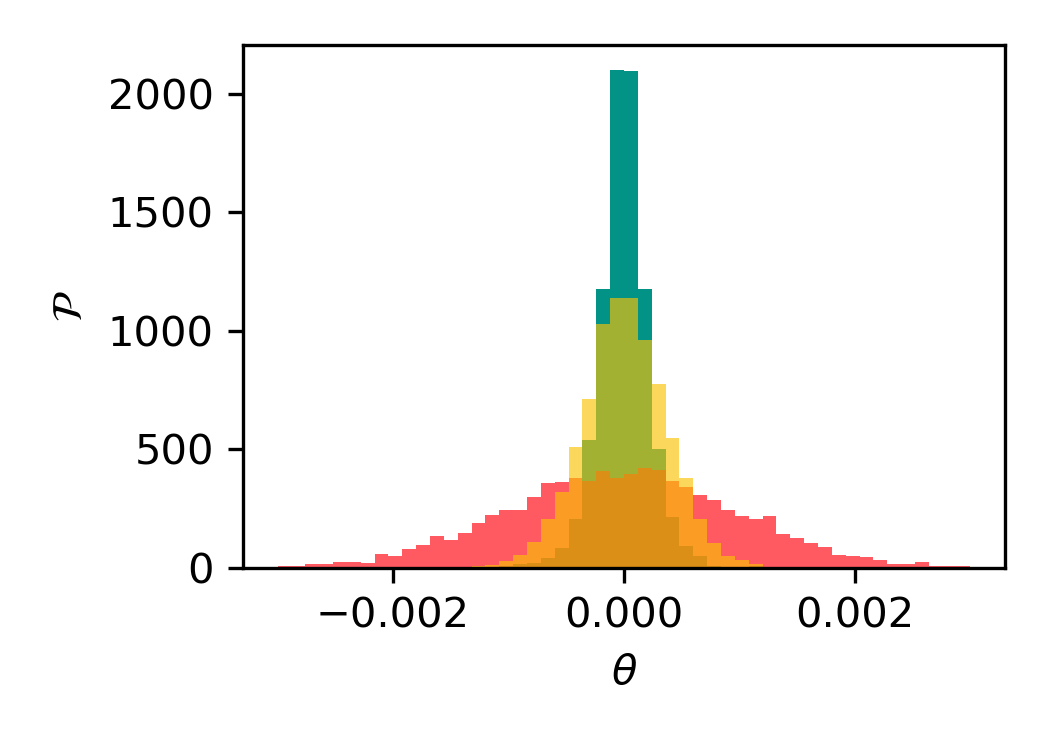} 
\includegraphics[width = .33\columnwidth, trim = {20 25 10 10}, clip]{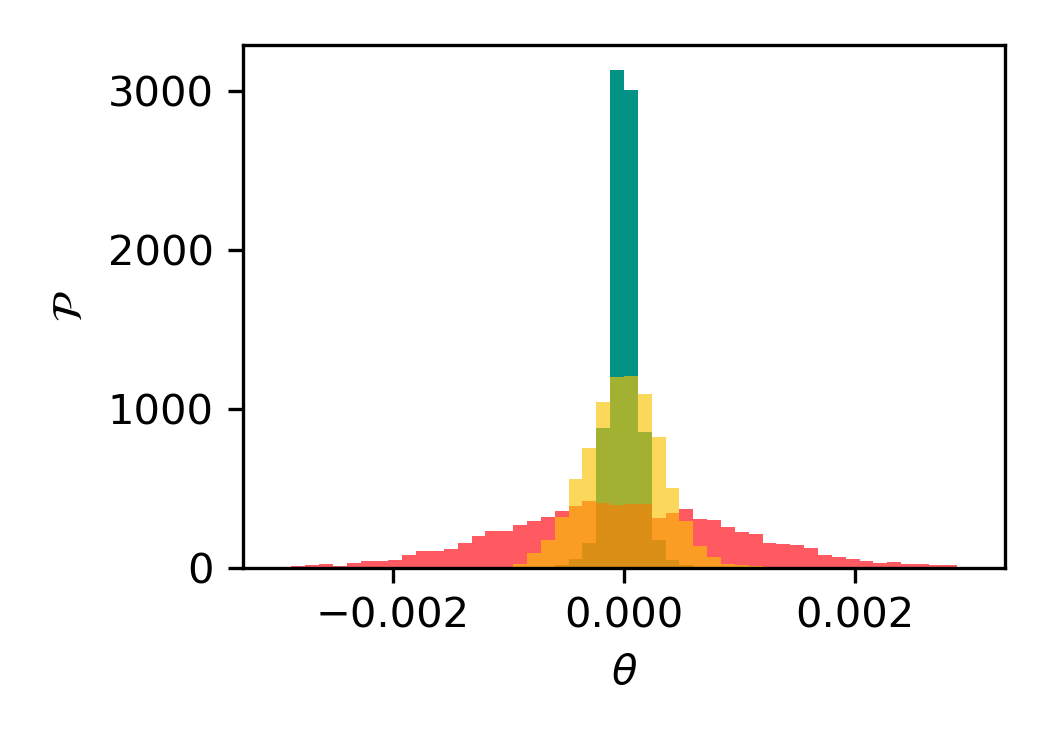} \\
\begin{tikzpicture}[overlay]
\node[] at (-3.2,7.2) {\sf (a)};
\node[] at (-3.2,3.5) {\sf (b)};
\node[] at (2.3,7.2) {\sf (c)};
\node[] at (2.3,3.5) {\sf (d)};
\node[] at (7.8,7.2) {\sf (e)};
\node[] at (7.8,3.5) {\sf (f)};
\end{tikzpicture} \vspace{-10pt}
\caption{
Histograms of the final--time states from simulations of $5000$ trajectories each, including stochastic errors, are shown. 
A bitflip is performed via STZ over time interval $T = 1$, with $\Gamma\,T = 3$ (a,b), $\Gamma\,T = 10$ (c,d), and $\Gamma\,T = 30$ (e,f). 
The teal distributions are for unitary error $\varepsilon_\Omega > 0$ but perfect measurement $\varepsilon_\zeta = 0$ (recall Fig.~\ref{fig-FPE}). 
The red distributions are for unitary error $\varepsilon_\Omega > 0$ with no measurement performed at all ($\Gamma = 0$). 
The yellow distributions are for both unitary mis-control $\varepsilon_\Omega >0$ and measurement axis mis-control $\varepsilon_\zeta > 0$. 
On the top (a,c,e) we use $\varepsilon_\Omega = 0.001/\sqrt{T} = \varepsilon_\zeta$, and observe that the red and yellow distributions more or less overlap (indicating that the addition of a measurement prone to errors of a similar severity to the unitary has only a small effect on the distribution). 
One the bottom (b,d,f) we use $\varepsilon_\zeta = 0.0003/\sqrt{T} = 0.3\,\varepsilon_\Omega$, and observe that despite the small probability of ``escape'' to the wrong eigenstate, the measurement is a net positive for control if it is calibrated better than the unitary. 
The measurement record is not used to perform feedback in any of the above; however, continuous monitoring does offer the possibility of continuously correcting operations, which would further improve control by providing information about rare escape events, such that they could be corrected in real time. 
}\label{fig-err-histo}
\end{figure}

Let us now investigate the case in which our measurement axis is also prone to stochastic errors. To what extent are the results outlined in and around Fig.~\ref{fig-FPE} modified? We imagine that $\zeta = \zeta_0 + \varepsilon_\zeta\,dW_\zeta$. 
We cannot use a FPE in the same way as in the previous subsection, because the coefficients $\mathsf{A}$ and $\mathsf{b}$ depend on $\zeta$ rather than $\dot{\zeta}$, such that the history of the diffusion is involved in setting the next step, making the process non-Markovian. 
We can however still perform pure--state trajectory simulations, and histogram the distributions to obtain the analogous Monte Carlo analysis. 

See Fig.~\ref{fig-err-histo} for results from such an analysis. 
In broad strokes, we draw the following conclusions from those simulations: 
i) For stochastic measurement--axis mis-control at a similar scale to the unitary mis-control, the peak of the distribution widens to about the same width as for the unitary errors without any measurement at all. 
ii) The measurement still encourages some escape to the ``wrong'' eigenstate, but for small--scale errors these escape events remain quite rare. Only for large measurement strength and large mis-control errors are these issues substantial. 
iii) If dissipation (without detection) is used for STZ, the most basic intuition one might have about the problem appears to be about right: Dissipation will help stabilize an operation if the measurement axis control is at least as precise as the unitary.

%%%%%%%%%%%%%%%%%%%%%%%%%%%%%%%%%%%%%%%%%%%%%%%%%%%%%%%%%%%%%%%%%
\subsubsection{Discussion: Monitored versus un-Monitored Errors}

Point iii) above suggests the ways in which a ``measurement'' may have autonomous benefits even without actual detection, provided the dissipation engineering is sufficiently good to avoid creating more errors than it suppresses. 
More work could be done to determine the exact bounds or break--even points for realistic types and scales of errors in STZ dragging operations. 
Some related work about the scaling of autonomous quantum error correction (AQEC) already appears in the literature \cite{shtanko2023bounds}. 
This dissipation--only case is however the weakest case of Zeno dragging: The full power of such an appproach becomes apparent if one actually monitors the unitary, and therefore gives oneself the ability to detect and correct escape errors (escape away from the neighborhood of the target eigenspace) in real time. 
Such a feedback approach would effectively be an instance of continuous quantum error correction \cite{Ahn_2002, Oreshkov_2007, Mascarenhas_2010, Atalaya_CQEC, Mohseninia2020alwaysquantumerror, Atalaya_2021_CQEC, Convy2022logarithmicbayesian, Livingston_2022, Convy_2022, Ahn_2003, Ahn_2004, Sarovar_2004, vanhandel2005optimal} (CQEC instead of AQEC).
Given that continuous measurements can be used to diagnose coherent errors in experiments \cite{Siva_2023}, we speculate that STZ with error detection could potentially also be developed with applications to calibration tasks. 
This connection between AQEC and CQEC is effectively the same relationship that exists between dissipative Zeno stabilization \cite{Facchi_2002, amini2011stability, Ticozzi_2013, Benoist_2017, Burgarth2020quantumzenodynamics, Burgarth2022oneboundtorulethem}, and feedback--assisted Zeno stabilization \cite{Mirrahimi_2007, Ticozzi_2008, amini2011stability, Ticozzi_2013, Benoist_2017, Cardona_2018, liang2019exponential, Cardona_2020, amini2023exponential, liang2023exploring}.

%%%%%%%%%%%%%%%%%%%%%%%%%%%%%%%%%%%%%%%%%%%%%%%%%%%%%%%%%%%%%%%%%
%%%%%%%%%%%%%%%%%%%%%%%%%%%%%%%%%%%%%%%%%%%%%%%%%%%%%%%%%%%%%%%%%
%%%%%%%%%%%%%%%%%%%%%%%%%%%%%%%%%%%%%%%%%%%%%%%%%%%%%%%%%%%%%%%%%
\section{Beyond a Single Qubit \label{sec-beyond-qubit}}

%%%%%%%%%%%%%%%%%%%%%%%%%%%%%%%%%%%%%%%%%%%%%%%%%%%%%%%%%%%%%%%%%
%%%%%%%%%%%%%%%%%%%%%%%%%%%%%%%%%%%%%%%%%%%%%%%%%%%%%%%%%%%%%%%%%
\subsection{Setting up Simple Zeno Dragging Operations \label{sec-get-L}}

We briefly describe how the formulas in \secref{sec-qtrajs} and \secref{sec-stz} may be re-purposed to find a Lindblad operator $\hat{L}$ to perform a Zeno dragging operation from $\ket{\psi_i}$ to $\ket{\psi_f}$, where $\ket{\psi_i}$ and $\ket{\psi_f}$ are states in an arbitrary Hilbert space.

Our first step will be to write down a $\hat{Q} = e^{-i\,\hat{\mathsf{h}}}$, that defines a change of basis into the Zeno frame (in which the $\hat{L}$ we wish to find is diagonal).
Consider 
\be \label{h_planar-rotation}
\hat{\mathsf{h}} = \frac{i\,\zeta}{2} \frac{\ket{\psi_i}\bra{\psi_f} - \ket{\psi_f}\bra{\psi_i}}{\sqrt{1-\left|\ip{\psi_i}{\psi_f}\right|^2}},
\ee
where the factor $i$ and sign between the two terms in the numerator have been chosen so that $\hat{Q}$ will be a real--valued orthogonal matrix. One may generalize this phase if a different convention better suits their needs. 
The angle $\zeta$ swept over the full control operation is $\int \dot{\zeta}\,dt = \int d\zeta = \pi\left(1- \left| \ip{\psi_i}{\psi_f} \right|^2 \right)$. 
We stress that construction of the basis change associated tracing a path between boundary states above is not unique; it is meant only to be a simple and valid formula. 
In writing down this particular expression, we are imposing a specific path over which to realize $\ket{\psi_i}\rightarrow\ket{\psi_f}$. 
Alternatively, the CDJ--P strategy \secref{sec-CDJP}, or another quantum control strategy, might be used to optimize the path taken given a suitable parameterization of the abilities and constraints of a given physical system. 
One could also impose a more complicated continuous trajectory through Hilbert space via a sequence of infinitesimal $\hat{\mathsf{h}}$ like the one above. 
The main goal of this section however will be to illustrate that in the event that we have simple dynamics like the planar rotation \eqref{h_planar-rotation}, then the schedule--optimization problem will remain relatively simple to solve, even if the dynamics in question are embedded in a larger Hilbert space.

We now aim to turn the basis change $\hat{Q}$ into a Lindblad operator $\hat{L}(\zeta)$ that Zeno drags a marked state (or group of states) through the larger Hilbert space. 
It is convenient to use a $\tilde{D}_\mathfrak{Z}$, which differs from the convention of \eqref{Kraus-expand} and/or \eqref{eq-eqmo-for-stz} only in that the ``ostensible'' probability is expanded and retained in $\tilde{\mathfrak{Z}} = \hat{\mathfrak{Z}} - \tfrac{1}{4}\,r^2\,\hat{\mathbb{I}}$. 
In the simplest case, one may then construct a two outcome $\tilde{D}_\mathfrak{Z}$ as per e.g.
\be \label{DZ-example} \tilde{D}_\mathfrak{Z} =
 -\frac{1}{4} \left( \begin{array}{cccc}
(r-2\sqrt{\Gamma})^2 & & & \\
& (r+2\sqrt{\Gamma})^2 & & \\
& & (r+2\sqrt{\Gamma})^2 & \\
& & & ~~\ddots~~
\end{array} \right).
\ee
This may be constructed so that outcomes centered around $r = +2\sqrt{\Gamma}$ herald success in following a marked state or subpace (i.e.~choose $(r-2\sqrt{\Gamma})^2$ on matrix element(s) so as to select the marked state or subspace one wishes to restrict population to), while outcomes centered around $r = -2\sqrt{\Gamma}$ herald failure of the dissipative / Zeno confinement (i.e.~assign this outcome to all state(s) besides those marked). 
With the diagonal matrix $\tilde{D}_\mathfrak{Z}$ in place, one may then simply recover 
\be \label{Zed_from_DQ}
\hat{\mathfrak{Z}} = \hat{Q}\,\tilde{D}_\mathfrak{Z}\,\hat{Q}^\dag + \tfrac{1}{4}\,r^2\,\hat{\mathbb{I}} = r\,\hat{L} -\hat{L}^2,
\ee
and solve for $\hat{L}$. 
Note that the above construction of $\tilde{D}_\mathfrak{Z} = \text{diag}\lbrace -\tfrac{1}{4}(r\mp 2\,\sqrt{\Gamma})^2\rbrace$ implies that $\hat{D}_L = \text{diag}\lbrace \pm\sqrt{\Gamma} \rbrace$ and therefore that $\hat{D}_L^2 = \Gamma\,\hat{\mathbb{I}}$. 
Refs.~\cite{blumenthal2021, ZenoGateTheory} offer a case study in how an outcome structure and Zeno subspace division like \eqref{DZ-example} can be approximately engineered in practice.
Generalization to the case of more than two groups of outcomes is straightforward.

%%%%%%%%%%%%%%%%%%%%%%%%%%%%%%%%%%%%%%%%%%%%%%%%%%%%%%%%%%%%%%%%%
%%%%%%%%%%%%%%%%%%%%%%%%%%%%%%%%%%%%%%%%%%%%%%%%%%%%%%%%%%%%%%%%%
\subsection{A Two-Qubit Example: Dissipative Bell State Generation \label{sec-Bell}}

We now construct an explicit example using the general procedure outlined above.
We here illustrate how one might perform Zeno dragging from a separable two--qubit state to a Bell state, once again using a linear optimal schedule. 
In the next subsection we will then be in a firm position to discuss common features of our one-- and two--qubit examples that define a ``simple'' class of analytically--solvable problems via CDJ--P. 

Let us start by defining $\hat{\mathsf{h}}(\zeta)$ for initial state $\ket{gg}$ and final state $\ket{\Phi^+} \propto \ket{ee} + \ket{gg}$. 
Note that the boundary states in question have $50\%$ overlap, i.e.~$|\ip{gg}{\Phi^+}|^2 = \tfrac{1}{2}$. 
This gives us 
\begin{subequations} \be \label{h_2Qubit}
\hat{\mathsf{h}} = \tfrac{i}{2\sqrt{2}}\,\zeta\,\left( \ket{\Phi^+}\bra{gg} - \ket{gg}\bra{\Phi^+} \right) = \frac{i}{2}\left(
\begin{array}{cccc}
 0 & 0 & 0 & \zeta \\
 0 & 0 & 0 & 0 \\
 0 & 0 & 0 & 0 \\
 -\zeta & 0 & 0 & 0 \\
\end{array}
\right),
\ee \be 
\hat{Q} = e^{-i\,\hat{\mathsf{h}}} = \left(
\begin{array}{cccc}
 \cos \left(\frac{\zeta }{2}\right) & 0 & 0 & \sin \left(\frac{\zeta }{2}\right) \\
 0 & 1 & 0 & 0 \\
 0 & 0 & 1 & 0 \\
 -\sin \left(\frac{\zeta }{2}\right) & 0 & 0 & \cos \left(\frac{\zeta }{2}\right) \\
\end{array}
\right).
\ee \end{subequations}
We mark the bottom element of $\tilde{D}_\mathfrak{Z}$, which initially corresponds to $\ket{gg}$ at $\zeta = 0$. 
Then application of \eqref{Zed_from_DQ} leads us to
\be \label{Bell_lindop}
\hat{L}(\zeta) = \left(
\begin{array}{cccc}
 -\sqrt{\Gamma } \cos (\zeta ) & 0 & 0 & \sqrt{\Gamma } \sin (\zeta ) \\
 0 & -\sqrt{\Gamma } & 0 & 0 \\
 0 & 0 & -\sqrt{\Gamma } & 0 \\
 \sqrt{\Gamma } \sin (\zeta ) & 0 & 0 & \sqrt{\Gamma } \cos (\zeta ) \\
\end{array}
\right).
\ee

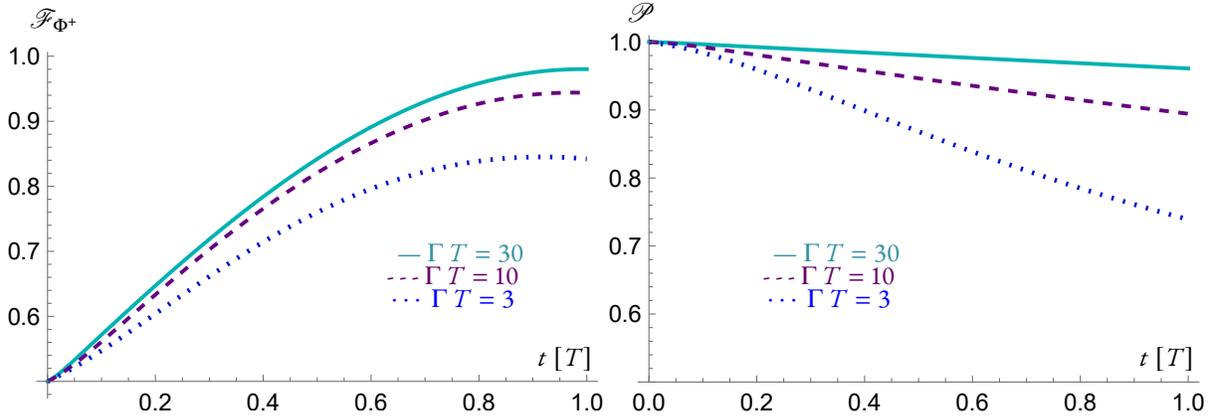
\begin{figure}\centering
\includegraphics[width = .48\textwidth]{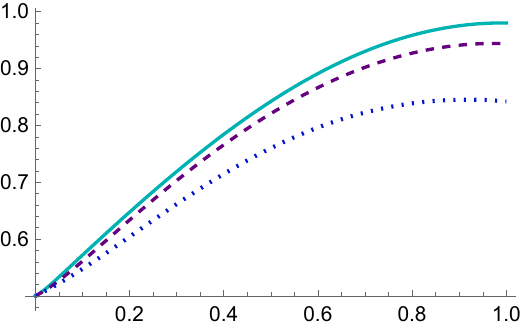}
\includegraphics[width = .48\textwidth]{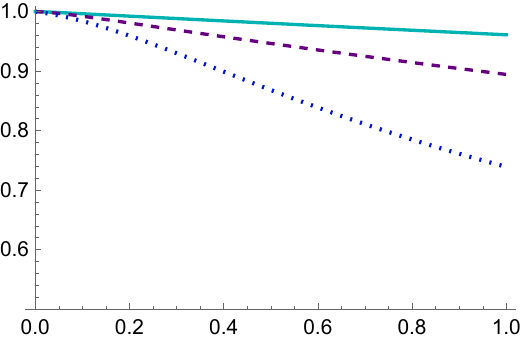} \\ 
\begin{tikzpicture}[overlay]
\node[] at (-7.3,5.7) {$\mathscr{F}_{\Phi^+}$};
\node[] at (0.5,5.8) {$\mathscr{P}$};
\node[] at (-0.5,1.2) {$t~[T]$}; 
\node[] at (7.5,1.2) {$t~[T]$};
\node[] at (-1.9,2.6) {\color{mint!80!blue} --- $\Gamma\,T = 30$};
\node[] at (-2,2.3) {\color{patriarch!80!black} - - - $\Gamma\,T = 10$}; 
\node[] at (-2,2) {\color{blue} $\cdots$ $\Gamma\,T = 3$};
\node[] at (3.1,2.6) {\color{mint!80!blue} --- $\Gamma\,T = 30$};
\node[] at (3,2.3) {\color{patriarch!80!black} - - - $\Gamma\,T = 10$}; 
\node[] at (3,2) {\color{blue} $\cdots$ $\Gamma\,T = 3$};
\end{tikzpicture} \vspace{-10pt}
\caption{
We plot the fidelity to the target state $\mathscr{F}_{\Phi^+}$ and purity $\mathscr{P} = \tr{\rho^2}$ as a function of time, under Lindblad evolution with \eqref{Bell_lindop} and \eqref{zeta_chi_opt}. 
As expected, an entangled state can be created with near unit fidelity and purity (and therefore near unit probability, as well) in the limit $\Gamma\,T \gg 1$. 
Time on the horizontal axes is in units of the dragging interval $T$. 
}\label{fig-BellFid}
\end{figure}

It is helpful now to parameterize our real and pure two--qubit state in terms of three angles: We may use two local angles, where $\theta$ has the same meaning as in our previous example for one qubit, with $\vartheta$ in the analogous local Bloch angle for the other qubit. 
A third non-local angle $\chi$ is used to represent entanglement (such that concurrence $\mathcal{C} = \sin\chi$) \cite{wharton2016natural}. See Appendix \ref{app_2Q} for details. 
This turns out to be ideal for our chosen boundary conditions: We have $\ket{gg} \leftrightarrow \lbrace \theta = \pi, \vartheta = \pi, \chi = 0 \rbrace$ and $\ket{\Phi^+} \leftrightarrow \lbrace \theta = \pi, \vartheta = \pi, \chi = \pi/2 \rbrace$. 
Then we are able to set $\theta = \pi = \vartheta$ and work in the one dimensional sub-manifold where only $\chi$ changes. 
Under these simplifications, our conditional dynamics are again reduced to a one-dimensional problem: The signal and information gain rate (see \eqref{readout-dW} and \eqref{g-hermitian}) given by our $\hat{L}(\zeta)$ are 
\be 
S = 2\,\sqrt{\Gamma}\,\cos(\zeta - \chi) \quad\&\quad \mathsf{g} = 2\,\Gamma\,\sin^2(\zeta - \chi),
\ee
where $\mathsf{g}$ again takes on the role of a cost function in the optimization (see \eqref{zeta-star-condition}). 
The conditional dynamics in our one--dimensional sub-manifold are given by 
\be 
\dot{\chi} = 2\,r\,\sqrt{\Gamma}\,\sin(\zeta - \chi). 
\ee
It should be clear at this point that our derivation led us to a Lindblad operator \eqref{Bell_lindop} whose parametric dependence $\zeta$ effectively corresponds to a rotation in the plane of the concurrence angle $\chi$.
We now have the base ingredients from which to assemble the CDJ stochastic Hamiltonian 
\be \begin{split}
\mathcal{H}^\star &= 2 \Gamma  \left([\Lambda_\chi  \sin (\zeta -\chi )+\cos (\zeta -\chi )]^2-1\right) \quad\text{with}\quad r^\star = 2 \sqrt{\Gamma } \left(\Lambda_\chi  \sin (\zeta -\chi )+\cos (\zeta -\chi )\right).
\end{split} \ee
As in our single--qubit example, $\mathcal{H}^\star$ depends strictly on $\zeta-\chi$, such that optimization \eqref{zeta-star-condition} will lead to the co-states $\Lambda_\chi$ conjugate to $\chi$ being conserved in the optimal dynamics. 
The solution moreover takes exactly the same form as in the single qubit case: We find
\begin{subequations} \be 
\chi^\star = \chi_i + (\chi_f - \chi_i)\tfrac{t}{T}, 
\ee generated by \be \label{zeta_chi_opt}
\zeta^\star = \chi_i + (\chi_f - \chi_i)\tfrac{t}{T} + \arctan\left( \frac{\chi_f - \chi_i}{4\,\Gamma\,T} \right)
\ee \end{subequations} 
for measurement--only dynamics. 
Lindbladian simulation according to this optimal schedule (see Fig.~\ref{fig-BellFid}) reveals the expected behavior: The measurement--based Zeno controls achieve asymptotically better fidelity and state purity as we approach the adiabatic limit.
STZ may again be applied via \eqref{h_2Qubit}, leading to perfect finite--time evolution. 

This example illustrates that for simple target evolution, we may still obtain simple analytical solutions embedded in larger Hilbert spaces. 
The proposed approach to entanglement generation is substantially different from other schemes that use measurement or dissipation to create or sustain coherent correlations (see e.g.~\cite{Chantasri_2016, martin2017optimal, Doucet_2020, doucet2023scalable, Lewalle:21, blumenthal2021, ZenoGateTheory, Greenfield_2024} and references therein).

%%%%%%%%%%%%%%%%%%%%%%%%%%%%%%%%%%%%%%%%%%%%%%%%%%%%%%%%%%%%%%%%%
%%%%%%%%%%%%%%%%%%%%%%%%%%%%%%%%%%%%%%%%%%%%%%%%%%%%%%%%%%%%%%%%%
\subsection{An Analytically--Solvable Class of Problems}

We remark on the properties of a class of problems to which both of our analytically--tractable examples belong, stemming from operations restricted to a planar rotation as in \eqref{h_planar-rotation}. 
If the target operation can be parameterized by a scalar $\zeta$, and restricted to a plane as in \eqref{h_planar-rotation}, then a single coordinate $\phi$ can be defined to parameterize the system's response to pure--state $\zeta$ rotations.
That is, it will be possible to reduce $\dot{\rho} = i[\dot{\mathsf{h}},\rho]$ to an equation $\dot{\phi}$ uncoupled from any other evolution. The same will apply to the pure--state conditional evolution arising from $\hat{L}(\zeta)$ constructed from the recipe in \secref{sec-get-L}.  

Regarding the existence of the desired coordinate $\phi$: Consider a pure state $|\phi_0\rangle$ under the weak measurement (with unit efficiency) of an observable $\hat{L} \propto |\phi_1\rangle \langle \phi_1|$. The post measurement state $|\phi_0^{\prime}\rangle$ is a linear combination of $|\phi_0\rangle$ and $|\phi_1\rangle$, i.e.
\begin{equation}
    |\phi_0^{\prime}\rangle \propto \hat{M} |\phi_0\rangle \in \mathrm{RSpan}\{|\phi_0\rangle, |\phi_1\rangle \}.
\end{equation}
The Kraus operator $\hat{M}$ is given by \eqref{Kraus-expand-0} with $\eta = 1$, and $\mathrm{RSpan}\{|\phi_0\rangle, |\phi_1\rangle \}$ is $\mathrm{Span}\{|\phi_0\rangle, |\phi_1\rangle \}$ with real coefficients, i.e.
\begin{equation}\label{eq_plane}
\begin{aligned}
    &\mathrm{RSpan}\{|\phi_0\rangle, |\phi_1\rangle \} = \left\{|\phi\rangle \, \bigg{|} \, |\phi\rangle = \mathrm{sin}(\zeta)|\phi_0\rangle + \mathrm{cos}(\zeta) |\phi_1\rangle, \, \zeta\in [0,2\pi) \right\}.
\end{aligned}
\end{equation}
Obviously $\mathrm{RSpan}\{|\phi_0\rangle, |\phi_1\rangle \}$ is a plane defined by $|\phi_0\rangle$ and $|\phi_1\rangle$, and any state in it can be described by a single parameter $\zeta$ as seen in \eqref{eq_plane}. 

This also holds when we are in the limit of continuous measurement $\Delta t \rightarrow 0$, where the dynamics are described by \eqref{ideal-hermitian} (or a stochastic Schrodinger equation) for pure states.  
When a pure state $|\psi(t)\rangle \in \mathrm{RSpan}\{|\phi_0\rangle, |\phi_1\rangle \}$ is driven by a Hamiltonian as in \eqref{h_planar-rotation}, the state will remain in $\mathrm{RSpan}\{|\phi_0\rangle, |\phi_1\rangle \}$.
Adding measurement, if $\forall \,t \in [0, T]$, we have the measurement observable $\hat{L}(t)$ be related to a pure state $|\phi(t)\rangle \in \mathrm{RSpan}\{|\phi_0\rangle, |\phi_1\rangle \}$ via $\hat{L}(t) \propto |\phi(t)\rangle \langle \phi(t)|$, then we will again have the system state $|\psi(t)\rangle \in \mathrm{RSpan}\{|\phi_0\rangle, |\phi_1\rangle \}$ under this continuous measurement. 
Therefore, when the dynamics include only the continuous monitoring and/or the Hamiltonian dynamics driven by $\dot{\mathsf{h}}$, one only needs a single parameter $\phi$ to describe the system dynamics, where the rate of unitary motion $\Omega$ and measurement axis position $\zeta$ both act solely within $\mathrm{RSpan}\{|\phi_0\rangle, |\phi_1\rangle \}$. 

When the above holds, one is then guaranteed that the conditional evolution (with or without STZ) can be expressed by a single coordinate as per $\dot{\phi} = \mathsf{A}_\phi(\phi-\zeta) + \mathsf{b}_\phi(\phi-\zeta)\,dW(t)/dt$.
If this is true, then in turn, one may immediately understand that \eqref{zeta-star-condition} guarantees that $\dot{\Lambda}_\phi = -\partial_\zeta \mathcal{H}^\star = 0$, such that the co-state $\Lambda_\phi$ conjugate to $\phi$ will be conserved under the optimized dynamics, as in the examples above. 
It should then be generically possible to solve both for the optimal schedule $\zeta^\star(t)$ and the optimal evolution $\phi^\star(t)$ explicitly. 

%%%%%%%%%%%%%%%%%%%%%%%%%%%%%%%%%%%%%%%%%%%%%%%%%%%%%%%%%%%%%%%%%
%%%%%%%%%%%%%%%%%%%%%%%%%%%%%%%%%%%%%%%%%%%%%%%%%%%%%%%%%%%%%%%%%
\subsection{Beyond Analytical Solutions: Outlook on Numerical Approaches \label{sec-numerical}}

Restricting ourselves to simple planar rotations inside a larger Hilbert space does limit the power and applicability of the CDJ--P method.
There will be many settings where it will either be i) impossible to impose such a simplification due to experimental constraints, and/or ii) be advantageous to allow the optimization procedure more freedom to choose the path between boundary states without restriction. 
These typical cases will likely have to be approached numerically. 
Fortunately, many existing numerical strategies might be adapted to the CDJ--P action, and thereby be used for measurement--driven control.  
Those strategies include e.g.~the Gradient--Ascent, Krotov, or PRONTO algorithms \cite{Boscain2021, Krotov_1996, Hauser_2002, Hauser_2003, Naris_2024} all of which have a basis in classical control \cite{Bryson_Control, Kirk_Control}, and have been successfully adapted to unitary quantum control \cite{Goerz_2019, Morzhin_2019, Boscain2021, Shao_2022, Shao_2024}. 
It appears relatively straightforward to adapt these algorithms to the pure--state CDJ action associated with \eqref{HCDJ-general}, simply by treating $\mathbf{r}$ as another auxiliary set of controls:
Joint numerical optimization of $\lbrace \mathbf{r},\bmath{\zeta} \rbrace$ for cost--function $\mathcal{G}$ \eqref{G-full}, under the constraint $\dot{\mathbf{q}} = \boldsymbol{\mathcal{F}}(\mathbf{q},\mathbf{r},\bmath{\zeta})$, and with some enforcement of the final boundary conditions, is a problem compatible with the structure of existing control algorithms. 
It is conceivable that the analytical $\mathbf{r}^\star(\mathbf{q},\boldsymbol{\Lambda},\bmath{\zeta})$ solutions might be adapted into versions of these algorithms specifically suited to measurement--based CDJ--P control in future work, but this does affect the way co-states and boundary conditions are handled. 

All of those algorithms just mentioned begin with a trial control trajectory, and aim to iteratively adjust the controller trajectory until it converges satisfactorily close to its optimum. 
As such, a good first guess $\bmath{\zeta}^\star_{(0)}(t)$ and $\mathbf{r}^\star_{(0)}(t)$ is generally beneficial, if not required. 
Fortunately, our simple examples above provide numerous insights that could ostensibly be leveraged to make such first guesses in a systematic and informed way.

%%%%%%%%%%%%%%%%%%%%%%%%%%%%%%%%%%%%%%%%%%%%%%%%%%%%%%%%%%%%%%%%%
%%%%%%%%%%%%%%%%%%%%%%%%%%%%%%%%%%%%%%%%%%%%%%%%%%%%%%%%%%%%%%%%%
%%%%%%%%%%%%%%%%%%%%%%%%%%%%%%%%%%%%%%%%%%%%%%%%%%%%%%%%%%%%%%%%%
\section{Discussion and Outlook \label{sec-conclude}}

In this work we have put forward two methodologies related to Zeno dragging (measurement--driven control) of quantum systems. 
First, we have shown that a ``Shortcut to Zeno'' (STZ) may be derived in the same spirit as all--unitary shortcuts to adiabaticity \cite{Guery_STA-RMP}. Second, we have demonstrated that the CDJ stochastic path integral \cite{Chantasri2013, Chantasri2015} offers a pathway towards performing a Pontryagin--style optimization of a measurement--driven controller. 
We applied both methods to the example of a single qubit \cite{ZenoDragging}, and found that they generate identical optimal solutions, revealing that a monitored unitary offers ideal performance in which one finds unit--probability success for a Zeno operation that provides dissipative stabilization of a subspace. 
The CDJ--P action offers insight into this point: It simultaneously represents control cost and quantum trajectory probability, and hence the controlled behavior converges towards occurring deterministically when the action is minimized all the way to zero.
We have further demonstrated that redundant unitary-- and measurement--driven dynamics have some resilience against controller errors, compared to unitary evolution alone. 
Finally, we have shown how the method can be adapted to larger systems, and included an explicit two--qubit example.

%%%%%%%%%%%%%%%%%%%%%%%%%%%%%%%%%%%%%%%%%%%%%%%%%%%%%%%%%%%%%%%%%
%%%%%%%%%%%%%%%%%%%%%%%%%%%%%%%%%%%%%%%%%%%%%%%%%%%%%%%%%%%%%%%%%
\subsection{Discussion}

We may state an informal definition of Zeno dragging, based on the situations considered in this work, as follows.
We conjecture that Zeno dragging is a viable approach to driving a quantum system from some initial $\ket{\psi_i}$ to a final $\ket{\psi_f}$, if and only if there exist parameter(s) $\bmath{\zeta}$ controlling the choice of measurement such that i) a continuous sweep in $\bmath{\zeta}$ is possible, and ii) that this generates a continuous deformation of a local minimum of $\mathsf{g}(\rho,\bmath{\zeta})$ that traces a path from $\ket{\psi_i}$ to $\ket{\psi_f}$. 
In the Zeno frame, any such local minimum would be rendered stationary.
For single--measurement situations this condition is clearly met, since by moving the measurement axis we necessarily control the trajectory of a root of $\mathsf{g}(\rho,\bmath{\zeta})$. However, we further expect 
the concept of Zeno dragging as a whole to extend in principle to more general situations that preserve the stated condition on $\mathsf{g}(\rho,\bmath{\zeta})$.

Fig.~\ref{fig-g-rotation} presents the simplest example of what we have just described. 
For a single qubit, by varying $\zeta$, we vary a single measurement axis and thereby have the freedom to rotate the roots of $\mathsf{g}(\mathbf{q})$ (which are the measurement eigenstates, or Zeno points) to any pure state in the $xz$ Bloch plane that we wish. 
In the event that we monitor this process with perfect efficiency, we will have a pure real--time state estimate in individual runs of the Zeno dragging protocol. 
We have used these pure--state conditional dynamics mathematically to derive our optimal control scheme. 
However, when performed well, Zeno dragging does not actually require monitoring at all; just controlling the dissipation without detection is adequate. 
This can be understood from the following considerations:
\begin{enumerate}
\item If we follow a measurement eigenstate perfectly as $\zeta$ changes, then there is never actually any diffusion away from that instantaneous eigenstate of $\hat{L}$ (or equivalently from the root of $\mathsf{g}$).
\item If there is no diffusion away from the eigenstate, then the dynamics have become deterministic. 
\item Equivalently, to 2., the dynamics resulting from such a deterministic process on an instantaneous eigenstate are guaranteed to provide a pure state on average (irrespective of the efficiency $\eta$), since there is no purity loss without averaging over a distribution of possible conditional states \cite{Ticozzi_2013, Benoist_2017}.\footnote{Such a distribution of conditional states could be calculated by a solving the appropriate Fokker--Planck equation \cite{GardinerStochastic, bookVanKampen}. Fig.~\ref{fig-g-rotation}(b) contains an illustration of this point, and see also \secref{sec-ExRobustness}.}
\end{enumerate}
Each of these statements is effectively equivalent, and can be taken as a description of a perfect Zeno dragging process, as implemented either with the globally optimal CDJ--P / STZ solution (i.e.,~with unitary assistance), or in the adiabatic limit of $\Gamma\,T \rightarrow \infty$ (with measurement only).
We have demonstrated this explicitly for the single qubit example in Fig.~\ref{fig-fidelities}, where we see that our globally optimal CDJ--P / STZ solution does in fact give perfect state fidelity on average, indicating that we have realized deterministic control. 
The fact that Zeno dragging can work well on average, i.e.,~without actually monitoring the results in individual runs, makes it a much more appealing protocol for experiments.
We see that our continuous measurements are effectively ensuring dissipative stabilization of the Zeno dragged state in the manner of \emph{autonomous} processes. 
From this perspective, our chosen example is similar to autonomous state stabilization protocols (see e.g.~Ref.~\cite{Kater_2012}), where we have now considered the optimal way to dynamically vary the point that is stabilized. 
We note that related analyses have also appeared in the literature on feedback control \cite{Cardona_2018, amini2023exponential}.

The connection to autonomous processes is not an accident, but is in fact deeply embedded in the construction of our CDJ--P method. 
Consider that we derive the CDJ action from statistical premises \eqref{cdj-CK} \cite{Chantasri2013}, such that action extremization is simultaneously the condition for optimized measurement--driven controls, and extremized event probability. 
While the CDJ optimal paths have in the past been used extensively to study properties of rare sequences of measurement events, we here have a confluence of optimal control and the most--probable events. 
This is arguably the most fundamental result of this manuscript: A good cost function for optimal measurement--driven control is the integrated log--probability density for sequences of readouts, such that controller optimization and statistical optimization are then necessarily performed together. The CDJ action \cite{Chantasri2013, Chantasri2015} here provides the link between the statistics and Pontryagin principle. 
Thus, the possibility of realizing deterministic and pure--state dynamics is a relatively general feature of the situation we consider, despite the intrinsic presence of stochasticity (or decoherence) in a measurement--driven (dissipation--driven) scenario. 

\subsection{Outlook}

The STZ and CDJ--P methods present different perspectives for further development.
For the CDJ--P optimization, while we have found an example with a straightforwardly tractable solution, we do expect the optimization procedure to be relatively more difficult to implement in larger systems. 
Furthermore, the coordinate parameterization of larger systems can become cumbersome in general (even with a restriction to pure states). 
The CDJ--P analysis in the single qubit example of \secref{sec-example} was made significantly easier because it was simple to intuitively guess a good coordinate system for the problem a priori; while we have shown that there exist situations in larger systems that retain this simplicity, it is unlikely to be easy to divine similarly helpful coordinates outside of those situations where simple target dynamics can be enforced. 
Moreover, the imposition of simplified dynamics substantially restricts the power of the CDJ--P method: Given more freedom, it is perfectly capable in principle of finding the optimal path between boundary states along with the accompanying schedule (see remarks in Ref.~\cite{Kokaew_2022}).
We expect that for many practical applications, numerical solution of the optimality conditions will be important, as discussed in \secref{sec-numerical}.

There are also some obvious criticisms of the STZ solution using a matched unitary. 
First, one may ask whether it is worthwhile to add dissipative stabilization if we are going to perform a unitary operation that would realize the target evolution on its own anyway. 
We have argued in \secref{sec-ExRobustness} that under simple error models, the redundancy adds robustness to the operation, provided the measurement motion and unitary motion are both relatively well--calibrated. Further advantages are possible when the Zeno effect is used for continuous error suppression (see further remarks below).
Second, for the simple qubit example of \secref{sec-example}, the STZ result is obvious enough that it can be written down intuitively, without 
requiring the theoretical formalities developed in this work.
However, we have not only shown that the STZ and CDJ--P methods agree and reproduce an intuitive result in a simple setting, but have also generalized their use well beyond the single--qubit problem where the answer is intuitive.

There are a number of potential ways to build on the tools we have established and connected above for more general applications.
Our STZ suggests a method for mapping dissipative stabilization or Zeno dragging operations onto adiabatic unitary evolution and vice versa; this may prove useful in and of itself. 
Additional extensions are suggested by consideration of recent results in which the Zeno effect has been used for control. 
In particular, is possible to engineer measurements that Zeno block the escape from a subspace without monitoring within that subspace, i.e.,~one may use the Zeno effect to define an effective decoherence--free subspace (DFS) \cite{Lidar_DFS} within a larger system \cite{Burgarth2020quantumzenodynamics, Facchi_2002, facchi2008quantum, Benoist_2017, blumenthal2021, ZenoGateTheory}. 
Such use of the Zeno effect allows one both to stabilize that subspace dissipatively and to alter the dynamics within the subspace in useful ways. 
This suggests a natural extension of the tools we have developed here. For example, one may optimize the schedule on which a DFS is dynamically varied over the course of a system's evolution, and have a systematic way of deriving a paired unitary that improves the probability of the Zeno blocking (DFS confinement) succeeding. 
This is one possible extension of the methodological foundation we have presented here which could be usefully explored in future work. 
We are aware of one experimentally--accessible example of a process like the one just described, namely the dissipatively--stabilized cat qubits encoded in bosonic modes \cite{Mirrahimi_2014, Touzard_2018, Guillaud_2019, Lescanne_2020, Gautier_2022, gautier2023designing, Regent2023heisenberg}. 
These already use a time--dependent dissipation channel paired with an optional ``feed forward'' Hamiltonian to perform some logical one-- and two--qubit gates.
The use of the Zeno effect to stabilize subspaces in these systems is a key feature in suppressing qubit errors, and provides an example of a setting where the presence of dissipation is extremely helpful (recall the first point in the preceding paragraph). 
Extension of these dissipative stabilization methods to grid states of an oscillator have also been proposed \cite{Sellem_2022, Royer_2020, sellem2023stability}; this setting is again of contemporary experimental interest \cite{Fluhmann_2019, PCI_2020}, including within the context of error correction \cite{deNeeve_2022, Sivak_2023, lachancequirion2023autonomous}.
Broadly speaking, the use of Zeno--like dissipators to suppress errors is referred to as Autonomous Quantum Error Correction \cite{Paz-Silva_2012, Dominy_2013, Cohen_2014, Lihm_2018, lebreuilly2021autonomous, Gertler_2021, shtanko2023bounds}; the cat qubit examples we have just mentioned are one experimentally--accessible example of this paradigm. 
Following this logic, application of the ideas in the present manuscript to a larger system may also have interesting points of contact with continuous quantum error correction \cite{Ahn_2002, Ahn_2003, Ahn_2004, Sarovar_2004, vanhandel2005optimal, Oreshkov_2007, Mascarenhas_2010, Atalaya_CQEC, Mohseninia2020alwaysquantumerror, Atalaya_2021_CQEC, Convy2022logarithmicbayesian, Livingston_2022, Convy_2022}. 
In other words, just as insights about the stabilizing properties of the dissipative (average) Zeno effect \cite{Facchi_2002, Ticozzi_2013} can be leveraged to correct measurement--driven quantum operations on average, so too should feedback be able to systematically improve these stabilizing properties \cite{Mirrahimi_2007, Ticozzi_2008, amini2011stability, Ticozzi_2013, Benoist_2017, Cardona_2018, liang2019exponential, Cardona_2020, amini2023exponential, liang2023exploring}, to implement real--time correction of those operations. Quantum measurements are invasive, and are thus never a passive element in a monitored unitary; well--engineered measurements for monitored operations offer the possibility of using that invasiveness constructively, to realize robustly correctable quantum control.

To summarize, in this work we have laid out some strategies to compute the optimal time--dependence for an evolving quantum dissipator, and illustrated that a ``shortcut to Zeno'' allows such time--dependent dissipation to be paired with optimally designed control unitaries to achieve perfect fidelity dissipatively stabilized quantum operations. 
This is expected to be a relatively general feature, because the CDJ path integral shows us that the controller cost that should be optimized is derived directly from the probability density for the measurement readout statistics.
It is apparent from the examples cited in the previous paragraph that the use of dissipation engineering to protect and manipulate quantum information is a diverse and increasingly active sub-field. 
We expect that the general theoretical results developed in this work will have use well beyond the simple examples we have presented to illustrate the optimal Zeno dragging approach, and will find wider use supporting ongoing efforts in quantum information science. 

%%%%%%%%%%%%%%%%%%%%%%%%%%%%%%%%%%%%%%%%%%%%%%%%%%%%%%%%%%%%%%%%%
%%%%%%%%%%%%%%%%%%%%%%%%%%%%%%%%%%%%%%%%%%%%%%%%%%%%%%%%%%%%%%%%%
%%%%%%%%%%%%%%%%%%%%%%%%%%%%%%%%%%%%%%%%%%%%%%%%%%%%%%%%%%%%%%%%%

\subsection*{Acknowledgements}

This material is based upon work supported by the U.S. Department of Energy, Office of Science, National Quantum Information Science Research Centers, Quantum Systems Accelerator. 
PL gratefully acknowledges discussions with Areeya Chantasri and Thiparat Chotibut about their work \cite{Kokaew_2022}, as well as with Arianna Cylke, Andrew Jordan, Howard Wiseman, and Areeya Chantasri about the interpretation of $\mathsf{g}(\mathbf{q})$ in connection with \cite{Cylke_2022inprep}. 
PL further acknowledges conversations with Justin Lane and Hugo Ribeiro about STA, which later inspired some of the results about STZ that were developed above. 
PL and YZ have benefited from discussions in which Torin Stetina shared a perspective about the connection to adiabatic approaches to quantum computing.
PL and KBW are grateful for discussions with Mazyar Mirrahimi, Pierre Rouchon, Alain Sarlette, Ronan Gautier, J\'{e}r\'{e}mie Guillaud, Fran\c{c}ois-Marie Le R\'{e}gent, R\'{e}mi Robin, and Lev-Arcady Sellem, that helped us to understand the connection to dissipatively stabilized qubits encoded in bosonic modes.
We furthermore thank Shay Hacohen-Gourgy, Archana Kamal, Josh Combes, Uwe Fischer, and Naoki Yamamoto for pointing out some relevant references.  
PL is grateful to the UMass Lowell department of Physics \& Applied Physics for their hospitality during part of this manuscript's preparation. 
This document was written without the use of AI. Simulations and calculations were performed with the help of Python and Mathematica.

%%%%%%%%%%%%%%%%%%%%%%%%%%%%%%%%%%%%%%%%%%%%%%%%%%%%%%%%%%%%%%%%%
%%%%%%%%%%%%%%%%%%%%%%%%%%%%%%%%%%%%%%%%%%%%%%%%%%%%%%%%%%%%%%%%%
%%%%%%%%%%%%%%%%%%%%%%%%%%%%%%%%%%%%%%%%%%%%%%%%%%%%%%%%%%%%%%%%%

\appendix
\section*{\centering Appendices}
%%%%%%%%%%%%%%%%%%%%%%%%%%%%%%%%%%%%%%%%%%%%%%%%%%%%%%%%%%%%%%%%%
%%%%%%%%%%%%%%%%%%%%%%%%%%%%%%%%%%%%%%%%%%%%%%%%%%%%%%%%%%%%%%%%%
%%%%%%%%%%%%%%%%%%%%%%%%%%%%%%%%%%%%%%%%%%%%%%%%%%%%%%%%%%%%%%%%%
\section{Solving Lindblad Dynamics for a Linear Dragging Schedule} \label{sec-LME-solve}

%%%%%%%%%%%%%%%%%%%%%%%%%%%%%%%%%%%%%%%%%%%%%%%%%%%%%%%%%%%%%%%%%
%%%%%%%%%%%%%%%%%%%%%%%%%%%%%%%%%%%%%%%%%%%%%%%%%%%%%%%%%%%%%%%%%
\subsection{A Liouvillian Formulation of Uni-Dimensional Lindblad Rotations \label{sec-OldAppA}}

In \secref{sec-get-L} we described how to derive a Lindblad operator that uses Zeno dragging to mimic a planar rotation \eqref{h_planar-rotation} between two boundary states. 
Here we make some extended comments about this particular construction. \eqref{DZ-example} is generically compatible with a measurement based on Gaussian pointer states and/or detectors ($e^{\Delta t \,\tilde{D}_\mathfrak{Z}}$ behaves like an un-normalized Kraus operator in the diagonal basis, where the normalization condition \eqref{povm} could then be applied), monitoring a Hermitian observable that leads to diffusive quantum trajectories in the time continuum limit. 
One common way to realize measurements of the type of above is with dispersive qubit cavity coupling \cite{Gambetta2008, Murch2013, Blais_CQED, LeighShay2020, Steinmetz_2022, Korotkov2016}, and quantum--limited amplifiers that behave much like optical homodyne or heterodyne detection. 
Many other physical implementations are also compatible with the above however. 
Generalization of $\tilde{D}_\mathfrak{Z}$ to more than two outcomes/subspaces is straightforward (and is likely to arise naturally in modeling the physics of specific measurement devices \cite{blumenthal2021, ZenoGateTheory}).

Confinement to a state or subspace should always work well even on average (i.e.~in the Lindbladian dynamics) in the adiabatic limit of slow rotation of the measurement axis. 
The construction of \textsc{eqs.}~(\ref{h_planar-rotation},\ref{DZ-example}) also makes clear that if outcomes can be grouped to mark a state or subspace, then escape can be detected and possibly corrected when true measurements are made (i.e.~if the experimentalist has access to the readouts $r$ with efficiency $\eta > 0$). 
Works on continuous quantum error correction have considered detection and correction of errors in closely--related scenarios  \cite{Atalaya_CQEC, Mohseninia2020alwaysquantumerror, Atalaya_2021_CQEC, Convy2022logarithmicbayesian, Livingston_2022, Convy_2022}. 

Given the autonomous nature of a near--optimal Zeno dragging process, we now look at the Lindblad dynamics for the simple construction of \secref{sec-get-L} more closely.
The Lindblad dynamics \eqref{lme} ($\eta = 0$) may be written $\kett{\dot{\rho}} = \mathscr{L}\kett{\rho}$ in terms of the Liouvillian superoperator
\be 
\mathscr{L} = \hat{L}^\top \otimes \hat{L} - \tfrac{1}{2}\,\hat{\mathbb{I}}\otimes\hat{L}^2 - \tfrac{1}{2}\,(\hat{L}^\top)^2 \otimes \hat{\mathbb{I}}
\ee 
(for column--major vectorization $\rho \rightarrow \kett{\rho}$). 
The dissipative Liouvillian is itself diagonalized by the operation that diagonalizes $\hat{L}$, i.e. 
\be \begin{split}
D_\mathscr{L} &= (\hat{Q}^\top \otimes \hat{Q}^\dag) \,\mathscr{L}\, (\hat{Q}^\ast \otimes \hat{Q}) \\ 
&= \hat{D}_L \otimes \hat{D}_L - \tfrac{1}{2}\,\hat{\mathbb{I}}\otimes \hat{D}_L^2 - \tfrac{1}{2} \hat{D}_L^2 \otimes \hat{\mathbb{I}} \\ 
&= \hat{D}_L \otimes \hat{D}_L - \Gamma\,\hat{\mathbb{I}}\otimes\hat{\mathbb{I}}.
\end{split} \ee
It follows that the Liouvillian gap \cite{Sarandy_2005, Campos_2016} is given by $|\Delta_\mathscr{L}| = 2\Gamma$ in the suggested construction \eqref{DZ-example}.

\subsection{A Solvable Liouvillian in the Zeno Frame \label{sec-OldAppB}}

We briefly describe how the idea of moving to the Zeno frame, as in \eqref{Zeno-frame-dynamics}, can be used to obtain quasi-analytical solutions to the Lindblad Master Equation \eqref{lme} for a Zeno dragging operation with a time--linear schedule. 
Recall that we defined $\varrho = \hat{Q}^\dag\,\rho\,\hat{Q}$ and $\hat{D}_L = \hat{Q}^\dag\,\hat{L}\,\hat{Q}$, with $\hat{Q} = e^{-i\,\hat{\mathsf{h}}}$ such that $\dot{Q}^\dag\,\hat{Q} = i\,\dot{\mathsf{h}}$ and $\hat{Q}^\dag\,\dot{Q} = -i\,\dot{\mathsf{h}}$. 
Let us apply this frame change to \eqref{lme}, such that
\begin{subequations} \label{lme-zeno-frame} \begin{align} 
\dot{\varrho} &= \dot{Q}^\dag\,\rho\,\hat{Q} + \hat{Q}^\dag\,\rho\,\dot{Q} + \hat{Q}^\dag\,\dot{\rho}\,\hat{Q} \\
&= \dot{Q}^\dag\,\hat{Q}\,\varrho + \varrho\,\hat{Q}^\dag\,\dot{Q} + \hat{Q}^\dag\left\lbrace  i[\rho,\hat{H}] + \hat{L}\,\rho\,\hat{L}^\dag - \tfrac{1}{2}\,\hat{L}^\dag\hat{L}\,\rho - \tfrac{1}{2}\,\rho\,\hat{L}^\dag\hat{L}\right\rbrace \hat{Q} \\
&= i[\varrho,\hat{Q}^\dag\,\hat{H}\,\hat{Q}-\dot{\mathsf{h}}] + \hat{D}_L\,\varrho\,\hat{D}_L - \tfrac{1}{2}\,\hat{D}_L^2\,\varrho - \tfrac{1}{2}\,\varrho\,\hat{D}_L^2,
\end{align} \end{subequations}
where we have applied $\hat{L} = \hat{L}^\dag$ in the last line. 
This equation is linear in $\varrho$, and our frame change has eliminated time--dependence from the dissipator, shifting it to the $\dot{\mathsf{h}}$ term that modifies the Hamiltonian.  
In the event that the schedule is linear in time (i.e.~$\dot{\mathsf{h}}$ is time--independent), and $\hat{Q}^\dag\,\hat{H}\,\hat{Q}$ does not have any time dependence remaining in this Zeno frame, then an analytic solution in terms of the eigenvectors of the Liouvillian may be constructed in this Zeno frame. 
Related comments about the Zeno frame have been made in the literature, (see e.g.~Refs.~\cite{Aharonov_1980, shea2023action}).

Let us apply this to compute the Fidelity of the average Zeno dragging dynamics (without STZ) under our optimal schedule \eqref{pre-compute-linear-gZeno}. 
We now have a (non-Hermitian) Liouvillian, such that
\be 
\kett{\dot{\varrho}} = \mathscr{L}\kett{\varrho}= \left( i\,\left\lbrace \hat{Q}^\dag\,\hat{H}\,\hat{Q}-\dot{\mathsf{h}} \right\rbrace^\top \otimes \hat{\mathbb{I}} - i\,\hat{\mathbb{I}}\otimes \left\lbrace \hat{Q}^\dag\,\hat{H}\,\hat{Q}-\dot{\mathsf{h}} \right\rbrace + \hat{D}_L \otimes \hat{D}_L - \Gamma\,\hat{\mathbb{I}}\otimes\hat{\mathbb{I}} \right) \kett{\varrho}
\ee
is equivalent to \eqref{lme-zeno-frame}. 
We may then diagonalize $\mathscr{L}$ via similarity transformation, i.e.
$
\mathscr{D}_\mathscr{L} = \mathcal{Q}^{-1}\,\mathscr{L}\,\mathcal{Q},
$
where $\mathcal{Q}$ is a square matrix whose columns are the right eigenvectors of $\mathscr{L}$.
Then a solution to \eqref{lme-zeno-frame} of the form
\be \label{lme-solved}
\kett{\varrho(t)} = \mathcal{Q}\,\exp(t\,\mathscr{D}_\mathscr{L})\,\mathcal{Q}^{-1} \kett{\varrho(0)}
\ee
exists so long as $\mathcal{Q}$ is invertible, and $\mathscr{L}$ is time--independent in the Zeno frame. 
(Equivalently, the solution above exists for any choice of time--independent parameters that do not form an Exceptional Point (EP) of $\mathscr{L}$ \cite{Ashida_2020}). 
The full solution procedure then reads: 
\begin{enumerate} 
\item Write the initial state in the Zeno frame, i.e.~$\rho_i ~\rightarrow~ \varrho(0) = \hat{Q}_0^\dag\,\rho_i\,\hat{Q}_0$
\item Apply \eqref{lme-solved} to solve the dynamics in the Zeno frame.
\item Return the solution back to the original frame $\varrho(t) ~\rightarrow~ \rho(t) = \hat{Q}_t\,\varrho(t)\,\hat{Q}_t^\dag$, where $\hat{Q}_t = \hat{Q}(\zeta(t))$. 
\end{enumerate}
We reiterate that the above requires i) that $\zeta$ depend linearly on time in the sense that $\dot{\mathsf{h}}$ is a constant, and ii) that $\mathcal{Q}$ be invertible for the parameters chosen.  

Following the example in the main text, let us use $\hat{H} = \tfrac{1}{2}\,\Omega\,\hat{\sigma}_y$, and 
\be 
\dot{\mathsf{h}} = \frac{\dot{\zeta}}{2}\,\hat{\sigma}_y = \frac{\theta_f - \theta_i}{2T}\,\hat{\sigma}_y = \frac{\Delta\theta}{2T}\,\hat{\sigma}_y, \quad\&\quad \hat{D}_L = \left(\begin{array}{cc}
\sqrt{\Gamma} & 0 \\ 0 & -\sqrt{\Gamma}
\end{array}\right),
\ee
which is the example of \secref{sec-example}. 
This is consistent with the use of the optimal schedule \eqref{linear-zeta-star-general}, which aims to Zeno drag the qubit state an angular distance $\Delta \theta$ over the time interval $T$, with a measurement strength $\Gamma$. 
We have $\hat{Q}^\dag\,\hat{H}\,\hat{Q} = \hat{H}$ because $[\hat{Q},\hat{H}] = 0$ for our chosen example. 
In this particular case, we find
\begin{subequations}\be 
\mathcal{Q} =\left(
\begin{array}{cccc}
 1 & 0 & \Omega -\dot{\zeta} & \Omega -\dot{\zeta} \\
 0 & -1 & \Gamma -\sqrt{\Gamma ^2-(\Omega-\dot{\zeta})^2} & \Gamma +\sqrt{\Gamma ^2-(\Omega-\dot{\zeta})^2} \\
 0 & 1 & \Gamma -\sqrt{\Gamma ^2-(\Omega-\dot{\zeta})^2} & \Gamma +\sqrt{\Gamma ^2-(\Omega-\dot{\zeta})^2} \\
 1 & 0 & \dot{\zeta}-\Omega  & \dot{\zeta}-\Omega  \\
\end{array}
\right), \quad\text{and}
\ee \be  
\mathscr{D}_\mathscr{L} = \left(
\begin{array}{cccc}
 0 & 0 & 0 & 0 \\
 0 & -2 \Gamma  & 0 & 0 \\
 0 & 0 & -\Gamma+\sqrt{\Gamma ^2-(\Omega-\dot{\zeta})^2}  & 0 \\
 0 & 0 & 0 & -\Gamma -\sqrt{\Gamma ^2-(\Omega-\dot{\zeta})^2} \\
\end{array}
\right).
\ee \end{subequations} 
We have an EP at $\Gamma^2 = (\Omega-\dot{\zeta})^2$. 
This marks a transition from an overdamped regime where solutions damp to the target state via the Zeno effect (for $\Gamma^2 > (\Omega-\dot{\zeta})^2$), and an underdamped regime (for $\Gamma^2 < (\Omega-\dot{\zeta})^2$) where unitary rotations win out over the Zeno dynamics (low purity oscillations take place). 
The solution method above works for all parameter choices \emph{except} $\Gamma^2 = (\Omega - \dot{\zeta})^2$, i.e.~at the critical damping point (or EP) marking the boundary between the overdamped and underdamped solution regimes. 

We use solutions based on this method to construct Fig.~\ref{fig-fidelities}, illustrating the average fidelity of a Zeno dragging operation on a qubit.  
The solutions obtained from the above process with our optimal schedule are sufficiently cumbersome that we do not reproduce them here in full (the expressions are impractical to handle without a computer algebra system). 
While the difficulty of doing even parts of this procedure analytically will quickly become prohibitive with an increase in system size, the above analysis can be the basis for a good numerical scheme in the adiabatic regime (which necessarily includes long evolution times), as long as $\mathscr{L}$ can be diagonalized numerically.
See e.g.,~\cite{Manzano_2020} for further context.

%%%%%%%%%%%%%%%%%%%%%%%%%%%%%%%%%%%%%%%%%%%%%%%%%%%%%%%%%%%%%%%%%
%%%%%%%%%%%%%%%%%%%%%%%%%%%%%%%%%%%%%%%%%%%%%%%%%%%%%%%%%%%%%%%%%
%%%%%%%%%%%%%%%%%%%%%%%%%%%%%%%%%%%%%%%%%%%%%%%%%%%%%%%%%%%%%%%%%
\section{STZ as an Optimal Feedback Protocol \label{sec-feedback}}

We here revisit the example of \secref{sec-example} from a third perspective (supplementing the STZ and CDJ--P procedures described in the main text). 
Specifically, we will here show that the proportional and quantum state--based (PaQS) procedure \cite{zhang2020locally, martin2019single}, which is a protocol for optimal feedback control, also leads us to the STZ solution \eqref{STZ-optimal}. 

PaQS, like our STZ procedure, encourages us to take the measurement schedule as a given, and then search for the optimal feedback unitary $\hat{U}(\phi) = e^{-i\,\phi\,\hat{\sigma}_y}$, where the control parameter $\phi$ may depend on the measurement records. 
Here we know the ``control Hamiltonian'' must be proportional to $\hat{\sigma}_y$ with the intuition that $\hat{U}(\phi)$ should be a real valued matrix. 
We denote the density matrix conditioned on measurement outcomes at time $t$ by $\rho_t$, and the (controlled) density matrix after the measurement as well as the feedback unitary by $\rho_t^c$. 
The It\^{o} stochastic master equation for continuous diffusive measurement reads 
\begin{subequations}\label{ito-sme}\be\begin{split} 
d\rho_t = \mathcal{L}(\hat{L},\rho_t)\,dt + \mathcal{K}(\hat{L},\rho_t)\,dW \quad&\text{for}\quad \mathcal{L}(\hat{L},\rho_t) = \hat{L}\,\rho\,\hat{L}^\dag - \tfrac{1}{2}\,\hat{L}^\dag\hat{L}\,\rho_t - \tfrac{1}{2}\,\rho_t\,\hat{L}^\dag\hat{L} \\
&\text{and}\quad \mathcal{K}(\hat{L},\rho_t) = \hat{L}\,\rho_t + \rho_t\,\hat{L}^\dag - \rho_t\,\tr{\hat{L}\,\rho_t + \rho_t\,\hat{L}^\dag}
\end{split}\ee
in general. Note that in constrast to the main text, we have switched to the It\^{o} formalism in this appendix, and write the measurement noise in terms of a Wiener increment $dW$.
For our qubit example with $\hat{L}(\zeta) = \sqrt{\Gamma}\left( \hat{\sigma}_x\,\sin\zeta + \hat{\sigma}_z\,\cos\zeta \right)$, \eqref{ito-sme} can be reduced to
\begin{equation}\label{1}
\begin{aligned}
    \rho_{t+dt} = \rho_t + d\rho_t 
    &=\rho_t + \Gamma \left\{\hat{L}\,\rho_t\,\hat{L} -\rho_t \right\}dt +\sqrt{\Gamma}\,\left\{\hat{L}\,\rho_t + \rho_t \,\hat{L} - 2\rho_t\, \tr{\hat{L}\,\rho_t} \right\}dW.
\end{aligned}
\end{equation} \end{subequations}
Applying feedback we have
\begin{equation}\label{2}
    \rho_{t+dt}^c = \rho_t^c + d\rho_t^c = \hat{U}(\phi)\,\rho_{t+dt}\,\hat{U}^{\dagger}(\phi).
\end{equation}
Here we will assume that the measurement axis rotations are not stochastic (i.e.~$d\zeta = \dot{\zeta}\,dt$), but assume that the feedback could be stochastic as per $\hat{U} = e^{-i\,\phi\,\hat{\sigma}_y}$ with $\phi = \alpha\,dt + \beta\,dW$ where $\alpha$ and $\beta$ are some (real) numbers specifying the feedback strategy. Expanding to $\mathcal{O}(dt)$ using It\^{o}'s lemma $dW^2 =dt$, one finds that
\begin{equation}\label{3}
    \hat{U}(\phi) = \hat{\mathbb{I}} - i\,\beta\,\hat{\sigma}_y \,dW - \left(i\,\alpha\,\hat{\sigma}_y + \tfrac{1}{2}\,\beta^2\, \hat{\mathbb{I}} \right)dt + \mathcal{O}(dt\,dW)
\end{equation}
Using \eqref{ito-sme} and \eqref{2} and expanding everything as per It\^{o} calculus, we find
\begin{equation}\label{4}\begin{split}
    \rho_{t+dt}^c &= \rho_t^c + \left\lbrace \mathcal{K}(\hat{L},\rho_t^c) + i \,\beta\,[\rho_t^c,\hat{\sigma}_y] \right\rbrace dW + \left\lbrace \mathcal{L}(\hat{L},\rho_t^c) + \mathcal{L}(\beta\,\hat{\sigma}_y,\rho_t^c) + i[\rho_t^c,\alpha\,\hat{\sigma}_y] + i\left[ \mathcal{K}(\hat{L},\rho_t^c),\beta\,\hat{\sigma}_y\right] \right\rbrace dt \\
    & = \rho_t^c + \CMcal{K}^{(c)}(\rho_t^c)\,dW + \CMcal{L}^{(c)}(\rho_t^c)\,dt. 
\end{split} \end{equation}
Continuing to follow the PaQS protocol \cite{zhang2020locally, martin2019single}, we formulate a cost function as an expectation value, defining $F(\phi) = \tr{\rho_t^c\,\hat{L}(\zeta)} = \tr{\hat{U}(\phi)\,\rho_t\,\hat{U}^\dag(\phi)\,\hat{L}(\zeta)}$. Then the locally optimal feedback should satisfy 
\be \label{fb_opt} \begin{split}
\frac{\partial F}{\partial \phi}\bigg|_{t+dt} &=  \partl{}{\phi}{} \tr{\hat{L}(\zeta+d\zeta)\,\hat{U}(\phi)\,\rho_{t+dt}\,\hat{U}^{\dagger}(\phi)} 
= \tr{\left( \hat{L}(\zeta) + \dot{\zeta}\,\partl{\hat{L}}{\zeta}{}\,dt \right) [\rho_{t+dt}^c,i\,\hat{\sigma}_y]}
= 0 \\ 
&= \tr{\left( \hat{L}(\zeta) + \dot{\zeta}\,dt\,\partial_\zeta\hat{L} \right)\left[\rho_t^c + \CMcal{K}^{(c)}(\rho_t^c)\,dW + \CMcal{L}^{(c)}(\rho_t^c)\,dt ,i\,\hat{\sigma}_y\right] } 
\\ &= \tr{\hat{L}(\zeta) [\rho_t^c,i\,\hat{\sigma}_y]} + \tr{\hat{L}(\zeta)[\CMcal{K}^{(c)}(\rho_t^c),i\,\hat{\sigma}_y]}\,dW + \tr{\hat{L}(\zeta)[\CMcal{L}^{(c)}(\rho_t^c),i\,\hat{\sigma}_y] + \dot{\zeta}\,(\partial_\zeta\hat{L})[\rho_t^c,i\,\hat{\sigma}_y]}\,dt,
\end{split} \ee 
where the It\^{o} rule has been used as needed to truncate expressions to $\mathcal{O}(dt)$.
We may solve this equation by individually finding the root of the three terms that are $\mathcal{O}(1)$, $\mathcal{O}(dW)$, and $\mathcal{O}(dt)$. 
We may understand that the $\mathcal{O}(1)$ term reduces to $z_t^c\,\sin\zeta = x_t^c \,\cos\zeta$, which is satisfied by the obvious constraint that the optimally--controlled dynamics $\rho_t^c$ should follow an eigenstate of $\hat{L}(\zeta)$, i.e.
\begin{equation} \label{rho_eigenstate}
    \rho_t^c = \tfrac{1}{2}\left(\hat{\mathbb{I}} \pm \hat{L}(\zeta) \right).
\end{equation}
This could equivalently be understood as a condition that the optimal control problem was previously solved at time $t$, while we now consider solving it for $t+dt$ by considering the next two terms.
We substitute this form of $\rho_t^c$ \eqref{rho_eigenstate} into the $\mathcal{O}(dW)$ and $\mathcal{O}(dt)$ terms of \eqref{fb_opt}, in order to find that
\begin{subequations} \begin{align} \label{solve_odw}
\mathcal{O}(dW) \quad\text{gives}\quad & -4\,\beta\,\sqrt{\Gamma} = 0 \quad\rightarrow\quad \beta = 0, \quad\text{and} \\ \label{solve_odt}
\mathcal{O}(dt) \quad\text{gives}\quad & 2 \sqrt{\Gamma}(\dot{\zeta} - 2\,\alpha ) = 0 \quad\rightarrow\quad \alpha = \tfrac{1}{2}\,\dot{\zeta}.
\end{align} \end{subequations}
We may now recognize that the PaQS approach \cite{zhang2020locally, martin2019single}, slightly modified to account for our time--dependent observable, has reproduced the results we derived as a ``shortcut to Zeno'' in the main text \eqref{STZ_example}, since $\alpha = \tfrac{1}{2}\,\Omega$, \eqref{solve_odt} reproduces exactly \eqref{STZ-optimal}.
The condition $\beta = 0$ \eqref{solve_odw} just confirms that our control can be open--loop, in the sense that it remains independent of the diffusive noise in individual measurement realizations, as discussed in \secref{sec-stz}. 

As a final remark, this feedback derivation was performed with a cost function $F = \ensavg{\hat{L}}$, while in the main text we used $\mathsf{g} = 2\ensavg{\hat{L}^2} - 2\ensavg{\hat{L}}^2$. 
It is easy to check that in this case there is fact no difference between the results these give however: Because our qubit example satisfies $\hat{L}^2 \propto \hat{\mathbb{I}}$, we have $\mathsf{g} = 2\left(\Gamma - \ensavg{\hat{L}}^2 \right)$. 
Following the process of \eqref{fb_opt}, it is then simple to verify that optimum of $F$ is also the optimum of $\mathsf{g}$, such that $F$ and $\mathsf{g}$ may be used interchangeably in the context of this feedback optimization. 
This property applies for any single observable of the form $\hat{L} \propto \hat{\mathbb{I}} - 2\,\hat{\Pi}$, where $\hat{\Pi}^2 = \hat{\Pi}$ is a projector. 

%%%%%%%%%%%%%%%%%%%%%%%%%%%%%%%%%%%%%%%%%%%%%%%%%%%%%%%%%
%%%%%%%%%%%%%%%%%%%%%%%%%%%%%%%%%%%%%%%%%%%%%%%%%%%%%%%%%
%%%%%%%%%%%%%%%%%%%%%%%%%%%%%%%%%%%%%%%%%%%%%%%%%%%%%%%%%
\section{A Minimal Parameterization of Two--Qubit States \label{app_2Q}}

Note that as a matter of convention, we will notate two--qubit state vectors in the basis
\be 
\ket{\psi} = \left( \begin{array}{c}
a \\ b \\ c \\ d
\end{array} \right) \quad {\color{white!50!black}
\begin{array}{c}
\ket{ee} \\ \ket{eg} \\ \ket{ge} \\ \ket{gg}
\end{array}
}
\ee
with the usual notation for the Bell basis
\be 
\ket{\Phi^\pm} = \tfrac{1}{\sqrt{2}}\left(\ket{ee} \pm \ket{gg} \right), \quad\&\quad \ket{\Psi^\pm} = \tfrac{1}{\sqrt{2}}\left( \ket{eg} \pm \ket{ge} \right).
\ee

In general, we may write a two--qubit density matrix as
\be \label{2Q_general_coord}
\rho = \frac{1}{4} \left(\hat{\mathbb{I}}_4 + \sum_{i,j} q_{ij}\,\hat{\sigma}_{ij} \right)\quad \text{for} \quad \hat{\sigma}_{ij} = \hat{\sigma}_i^{(A)} \otimes \hat{\sigma}_j^{(B)}.
\ee
There are in general $15$ real coordinates in the vector $\mathbf{q}$, corresponding to $15$ generalized Gell--Mann matrices $\hat{\boldsymbol{\sigma}}$, that are defined for all combinates $i,j = I,X,Y,Z$, excluding $i = I = j$, i.e.~excluding $\hat{\mathbb{I}}_2 \otimes \hat{\mathbb{I}}_2 = \hat{\mathbb{I}}_4$. 
This parameterization has the property that $q_{ij} = \tr{\rho\,\hat{\sigma}_{ij}}$ and $\dot{q}_{ij} = \tr{\dot{\rho}\,\hat{\sigma}_{ij}}$, such that dynamics can be easily expressed as a dynamical system of equations in these $\mathbf{q}$. 

We can however greatly reduce the size of the coordinate space by using some simplifying assumptions:
\begin{enumerate}
    \item our initial state is pure,
    \item our initial state has only real amplitudes / an all--real density matrix $\rho$, and
    \item $\eta = 1$ on any and all measurements (to retain purity of the conditional dynamics).
\end{enumerate}
Then we have only real amplitudes and pure states for all time. 
In this case, it is possible to express our dynamics in terms of only three real coordinates. 
We reduce the pure state parameterization put forth by \textcite{wharton2016natural}, which for real amplitudes reads 
\be 
\ket{\psi} = \begin{array}{rcl}
& \left[\sin \left(\theta/2\right) \sin \left(\chi/2\right) \sin \left(\vartheta/2\right)+\cos \left(\theta/2\right) \cos \left(\chi/2\right) \cos \left(\vartheta/2\right)\right] & \ket{ee}\\
+ & \left[\cos \left(\theta/2\right) \cos \left(\chi/2\right) \sin \left(\vartheta/2\right)-\sin \left(\theta/2\right) \sin \left(\chi/2\right) \cos \left(\vartheta/2\right)\right] & \ket{eg} \\
+ & \left[\sin \left(\theta/2\right) \cos \left(\chi/2\right) \cos \left(\vartheta/2\right)-\cos \left(\theta/2\right) \sin \left(\chi/2\right) \sin \left(\vartheta/2\right)\right] & \ket{ge} \\
+ & \left[\sin \left(\theta/2\right) \cos \left(\chi/2\right) \sin \left(\vartheta/2\right)+\cos \left(\theta/2\right) \sin \left(\chi/2\right) \cos \left(\vartheta/2\right)\right] & \ket{gg}
\end{array}.
\ee
In this parameterization, the concurrence \cite{Wooters_1998} is given by 
\be 
\mathcal{C} = \sin\chi,
\ee
while $\theta$ and $\vartheta$ are an angle on the $xz$--plane great circle of the local Bloch spheres for qubits $A$ and $B$ respectively (with $\theta = 0$ and $\vartheta = 0$ corresponding to $\ket{e_A}$ and $\ket{e_B}$, respectively). 

The time evolution in the coordinates $\lbrace \theta,\vartheta,\chi \rbrace$ can furthermore be obtained from the general parameterization \eqref{2Q_general_coord} as per 
\begin{subequations} \label{angular_dynamics_general} \be 
\dot{\theta} = \sec (\chi ) \left[\dot{q}_{YY}\,\cot (\theta ) \tan (\chi )- \dot{q}_{ZI} \csc (\theta )\right],
\ee \be 
\dot{\vartheta} = \sec (\chi ) \left[\dot{q}_{YY} \tan (\chi ) \cot (\vartheta )-\dot{q}_{IZ} \csc (\vartheta )\right],
\ee \be 
\dot{\chi} = -\dot{q}_{YY}\,\sec{\chi},
\ee \end{subequations}
when our assumptions about real amplitudes and pure states hold. 

%%%%%%%%%%%%%%%%%%%%%%%%%%%%%%%%%%%%%%%%%%%%%%%%%%%%%%%%%%%%%%%%%
%%%%%%%%%%%%%%%%%%%%%%%%%%%%%%%%%%%%%%%%%%%%%%%%%%%%%%%%%%%%%%%%%
%%%%%%%%%%%%%%%%%%%%%%%%%%%%%%%%%%%%%%%%%%%%%%%%%%%%%%%%%%%%%%%%%
%\newpage
\addcontentsline{toc}{section}{\bf References}
{\sloppy
\printbibliography
}

\end{document}